\documentclass[final,3p,times]{elsarticle}

\usepackage{amssymb}
\usepackage{amsmath} 
\usepackage{ulem}
\usepackage{bm}
\usepackage{deluxetable} 
\usepackage{makecell}
\usepackage{wrapfig}
\usepackage{placeins}
\usepackage{rotating,tabularx}
\usepackage{tabularx}

\newcommand{\Msun}{\ensuremath{\rm \,M_\odot}}
\newcommand{\Rsun}{\ensuremath{\rm \,R_\odot}}
\newcommand{\Zsun}{\ensuremath{\,Z_\odot}}
\newcommand{\Lsun}{\ensuremath{\rm \,L_\odot}}

\newcommand{\msun}{\ensuremath{\rm \,M_\odot}}

\newcommand{\myr}{\ensuremath{\,\mathrm{Myr}}}
\newcommand{\gyr}{\ensuremath{\,\mathrm{Gyr}}}

\newcommand{\ergs}{\ensuremath{\,\mathrm{erg}\,\mathrm{s}^{-1}}}
\newcommand{\msy}{\ensuremath{\Msun\mathrm{\; yr}^{-1}}}

\newcommand{\Ledd}{\ensuremath{\,L_{\rm Edd}}}
\newcommand{\Mdoted}{\ensuremath{ \,\dot M_{\rm Edd}}}

\newcommand{\Teff}{\ensuremath{\,T_{\rm eff}}}

\newcommand{\dash}{\,\hbox{--}\,}
\newcommand{\Rsph}{\ensuremath{\,R_{\rm sph}}}
\newcommand{\nulx}{\#ULX}
\newcommand{\nbhulx}{\#BHULX}
\newcommand{\nnsulx}{\#NSULX}
\newcommand{\ttt}[2]{\ensuremath{#1\times10^{#2}}}
\newcommand{\nergs}[2]{\ensuremath{#1\times10^{#2}\ergs}}
\newcommand{\zsun}{\ensuremath{Z_\odot}}
\newcommand{\lx}{\ensuremath{L_{\rm X}}}

\newcommand{\ledd}{\ensuremath{L_{\rm Edd}}}
\newcommand{\lxmax}{\ensuremath{L_{\rm X, max}}}

\journal{New Astronomy Reviews. Published: 2023, Volume 96. DOI:10.1016/j.newar.2022.101672}

\begin{document}

\begin{frontmatter}



\title{Ultraluminous X--ray Sources}


\author[1,2,3]{Andrew King}, 
\author[4,5]{Jean-Pierre Lasota} 
\author[6]{Matthew Middleton}

\address[1]{Astrophysics Division, School of Physics \& Astronomy, University of Leicester, Leicester LE1 7RH, UK}
\address[2]{Astronomical Institute Anton Pannekoek, University of Amsterdam, Science Park 904, 1098 XH Amsterdam, Netherlands}
\address[3]{Leiden Observatory, Leiden University, Niels Bohrweg 2, NL-2333 CA Leiden, Netherlands}
\address[4]{Institut d'Astrophysique de Paris, CNRS et Sorbonne Universit\'e, UMR 7095, 98bis Bd Arago, 75014 Paris, France}
\address[5]{Nicolaus Copernicus Astronomical Center, Polish Academy of Sciences, ul. Bartycka 18, 00-716 Warsaw, Poland}     
\address[6]{Department of Physics and Astronomy, University of Southampton, Highfield, Southampton SO17 1BJ, UK}

\begin{abstract}
The study of ultraluminous X-ray sources (ULXs) has changed dramatically over the last decade. In this review we first describe the most important observations of ULXs in various wavebands, and across multiple scales in space and time. We discuss recent progress and current unanswered questions. We consider the range of  current theories of ULX properties in the light of this observational progress.
Applying these models to
neutron-star ULXs offers particularly stringent tests, as this is the unique case where the mass of the accretor is effectively fixed. 

\end{abstract}







\end{frontmatter}



\section{Introduction}
\label{sec:introd}

\label{sec:ulxs}

Ultraluminous X--ray sources (ULXs) are usually defined by fluxes which, if assumed isotropic, give luminosities
\begin{equation}
    L_X > 10^{39}\, {\rm erg\, s^{-1}}.
\end{equation}
The implication is that ULXs appear to be above the Eddington critical luminosity
\begin{equation}
\Ledd = 1.3 \times 10^{38} \left(\frac{M}{\Msun}\right)\, {\rm erg\,s^{-1}},   
\end{equation}
for accretor masses $M \lesssim 10\msun$ {(see Eq. \ref{eq:Ledd} for the exact definition of the luminosity $\ledd$)}.  Early systematic X--ray surveys 
found significant numbers of these objects: \citet{Fabbiano8901} reported 16 sources
with $L_X > 10^{39}\, {\rm erg\,s^{-1}}$. By now, more than about {    1800} ULXs \citep{Walton0122} are known, including several with $L_X > 10^{41} - 10^{42}\, {\rm erg\,s^{-1}}$. 
ULXs are (almost) all in external galaxies, but not located in their nuclei, making them distinct from
accreting supermassive black holes

From the beginning it was clear that the brightest ULXs required a significant shift in the standard
paradigm for the then known accretion--powered sources, and this provoked a wide variety of speculations. 
An obvious possibility was that the accreting mass $M$ was unusually large.
\citet{Colbert9907} used a sample of $\sim 20$ ULXs with apparent X--ray luminosities in the range $10^{40} - 
10^{42}\, {\rm erg\,s^{-1}}$
to argue that the typical accretors were black holes with masses in the range $10^2\msun \lesssim M \lesssim 10^4\msun$.
The suggested objects soon acquired the name `intermediate--mass black holes', or IMBH \citep{Tanigushi0006}. 
Models for IMBH\footnote{Some dwarf galaxies have active nuclei powered by accretion on to black holes of masses $\sim 10^5\msun$, and so with similar luminosities to some ULXs. These black holes appear to obey the same $M - \sigma$ relation between black hole mass and host spheroid velocity dispersion as universally holds for higher mass nuclear black holes and their host galaxies (see \citealt{Baldassare2007, King2008} for a discussion). 
Unlike these objects, ULXs are not unique systems at the dynamical centres of their hosts. 
Accordingly, in this review we confine use of the term `IMBH' to objects {\it not} in the nuclei of dwarf galaxies (see Section \ref{sec:hlx}).} formation
include rapid merging of stellar--mass black holes in globular clusters \citep{Miller0202}, or of stars in dense 
clusters 
\citep{Portegies0209}. A problem for all these models was to ensure that the IMBH would find a companion star 
and then form a binary able to sustain mass transfer at the required rate (up to $\sim 10^{-5} - 10^{-4}\msun\, {\rm 
yr}^{-1}$).

A quite different idea was that ULXs were fairly standard X--ray binaries with $M \lesssim 10\msun$ in unusual 
states. \citet{Begelman0204} suggested that the accretion discs in standard black hole X--ray binaries might be 
unstable to the photon--bubble 
instability, and simply radiate at super--Eddington luminosities. This is limited to luminosities
$L_{\rm sph} \lesssim 3\times 10^{40} \, {\rm erg\,s^{-1}}$, and requires that mass loss from the disc surface is small. 
\citet{Kording0201} suggested that the high
luminosities of ULXs could result from relativistic beaming in jets from microquasars oriented towards the 
observer. This is again limited, to $L_{\rm sph} \lesssim 10^{40} \, {\rm erg\,s^{-1}}$.
Similarly, it was known that magnetic fields with the strength 
$\sim 10^{12} - 10^{13}\,{\rm G}$ found in accreting neutron stars reduced 
the electron scattering opacity for X--rays propagating perpendicular to the field 
\citep{Canuto0571}\footnote{The possibility of this effect is probably the reason why the
occasional $L_X \simeq 10^{39}\, {\rm erg\, s^{-1}}$ outbursts of the Be--star -- magnetic neutron star system A0538-66
\citep{White0478} did not attract more attention.},
 but this works only for $L_{\rm sph} \lesssim 3\times 10^{40} \, {\rm erg\,s^{-1}}$, and even then 
requires a magnetar--strength field $B \sim 10^{14}\, {\rm G}$ 
\citep{Mushtukov1502}, although these do not seem to appear in binary systems (see Section \ref{sec:nomagnet}). The limits for all three of these stellar--mass models
are incompatible with the luminosities $L_X > 10^{41} - 10^{42}\, {\rm erg\,s^{-1}}$ observed in some ULXs. \citet{Narayan_spectra2017}
suggested that radiatively inefficient accretion on to black holes could explain ULXs as the radiative output was tightly beamed. Here much of the accreting gas is swallowed by the hole without radiating. 

\citet{King0105} noted that the non--magnetic neutron star in the low--mass X--ray binary Cyg X--2 (not a ULX) 
was known from evolutionary arguments \citep[][see 
Section \ref{sec:evolulx} below]{King9910,Podsiadlowski0002} to have survived a previous 
phase where it was subject to mass transfer at rates $\sim 10^2 - 10^3$ times its Eddington accretion rate of $\sim 
10^{-8}\msun\, {\rm yr}^{-1}$.
The neutron star had evidently
ejected the vast bulk ($\sim 2 - 3 \msun$) 
of the transferred mass, retaining no more than a few $\times\, 0.1\msun$.
Ejection like this is expected in the simple picture of \citet{Shakura73} of mass
expulsion by radiation pressure from thin accretion discs fed at super--Eddington rates\footnote{Note that advection, neglected by \citet{Shakura73}, cannot change the accretion luminosity for a neutron star accretor.} (see Section \ref{sec:ss73} below), so
a plausible deduction was that the intense accretion disc wind might
block the emitted X--rays from many lines of sight, perhaps leaving only narrow escape routes near the disc's rotational 
poles. If these
had total solid angle $4\pi b$, with $b \ll 1$, an observer viewing the source along these directions but assuming that the 
emission was isotropic would ascribe a luminosity
\begin{equation}
\label{eq:beaming1}
    L_{\rm sph} = \frac{1}{b}L \gg L 
\end{equation}
to the source, where $L$ is its true luminosity\footnote{Although this is almost always called `beaming', as in this review, 
it is important to note that it is simply collimation, rather than more complex processes such as relativistic beaming.}.

Early discussion of this model was hampered by
the lack of a clear relation between the beaming factor $b$ and other physical variables, which left a spurious degree of 
freedom. This gap was later closed (see Section \ref{sec:windbeam}), but even before this the model made interesting and testable 
predictions. Simple estimates showed that the characteristic velocity of the accretion disc wind was $\sim 0.1c$ \citep[][compare Section \ref{sec:feedback}]{King1003}. If $b \lesssim 10^{-2}$, the X--ray fluxes of the early observed ULXs gave
$L = bL_{\rm sph} \lesssim 10^{39}\, {\rm erg\, s^{-1}}$, below the Eddington value for a $10\msun$ black hole.
\citet{King0105} also noted that for  $b \lesssim 10^{-3}$, the
deduced $L$ would be compatible with a neutron star accretor, as expected if Cyg X-2 is a survivor of a ULX phase, 
and with a white dwarf accretor in cases where the observed X--ray flux was soft (photon energies $\lesssim 0.1$\,keV).
The brightest ULXs would correspond to the same value $b = 10^{-3}$ of the beaming factor, but with a $10\msun$ black hole 
accretor.

This idea gave sensible answers for the likely populations and lifetimes of ULXs \citep{King0105}. It suggested that 
ULXs were an extremely common stage of high--mass X--ray binary evolution characterised by very large mass transfer rates. 

From the analysis of Cyg X-2's evolution, the best candidate for this stage was the 
preceding phase of super--Eddington mass transfer. This inevitably follows the standard wind--fed high--mass 
X--ray binary stage once the donor star fills its Roche lobe, as this is always more massive than the accretor. Then
mass transfer shrinks the binary to conserve orbital angular momentum, 
maintaining very high mass rates limited only by the companion's ability to expand rapidly, usually on its thermal timescale
(see Section \ref{sec:evolulx}). {   Begelman et al. (2006) noted that the extreme binary SS433 was probably a system of this type, but not viewed along the accretion disc axis, so was probably a ULX `viewed from the side'.}

The current value of the mass transfer rate $\dot M_{\rm tr}$ in any system like this is (as usual) hard to determine from direct observations. In particular measuring the rate $\dot P$ of change of the orbital period $P$ gives at best a wide upper limit $\dot M_{\rm tr} \lesssim |\dot P/P|M_2$
to the mass transfer rate (where $M_2$ is the mass of the donor star). This is clearly seen for high--mass X--ray binaries: \citet{Falanga15} measure $\dot P/P \sim - 10^{-6}\,{\rm yr}^{-1}$
for LMC X-4, Cen X-3, 4U1538-522 and SMC X-1, which would give maximum mass transfer rates $\sim 10^{-6}\msun\, {\rm yr}^{-1}$ and make all of these systems ULXs, but their observed luminosities are all sub--Eddington ($\dot M_{\rm tr} \lesssim 10^{-8}\msun\, {\rm yr}^{-1}$). 

We now know that a small group of high--mass X--ray binaries with Be--star
companions can occasionally produce very high mass transfer rates for short 
intervals because of dynamical effects on the Be--star disc -- see Section \ref{sec:bemodel}. These systems are unique in repeatedly making
transitions from a `normal' X--ray binary to a ULX and back. This property distinguishes very sharply between models for ULXs.   

The short lifetimes of both types of high--mass X--ray binary (pre--ULX, or Be--star) showed that the disc--wind beaming
model required an association of ULXs with 
star formation. This was already suspected, in particular following the discovery of 7 ULXs in {\it Chandra} observations of 
the Antennae, where the merger of two galaxies has provoked vigorous star formation
\citep{Fabbiano0106}. {\it Chandra} observations of the Cartwheel galaxy
\citep{Gao0310} soon strongly confirmed it. They revealed more than 20 ULXs with $L_{\rm sph} > 3\times 10^{39} \, {\rm 
erg\, s^{-1}}$,
four of them in the range $1 - 5.5\times 10^{41} \, {\rm erg\, s^{-1}}$,
in a spreading ring of star formation triggered some $3\times 10^8\, 
{\rm yr}$ ago when a smaller intruder galaxy crossed the plane of the Cartwheel disc close to its nucleus. 
None of these features were easily
compatible with the IMBH hypothesis for ULXs \citep{King0401}.
\begin{figure}[ht]
\begin{center}
\includegraphics[width=0.6\columnwidth]{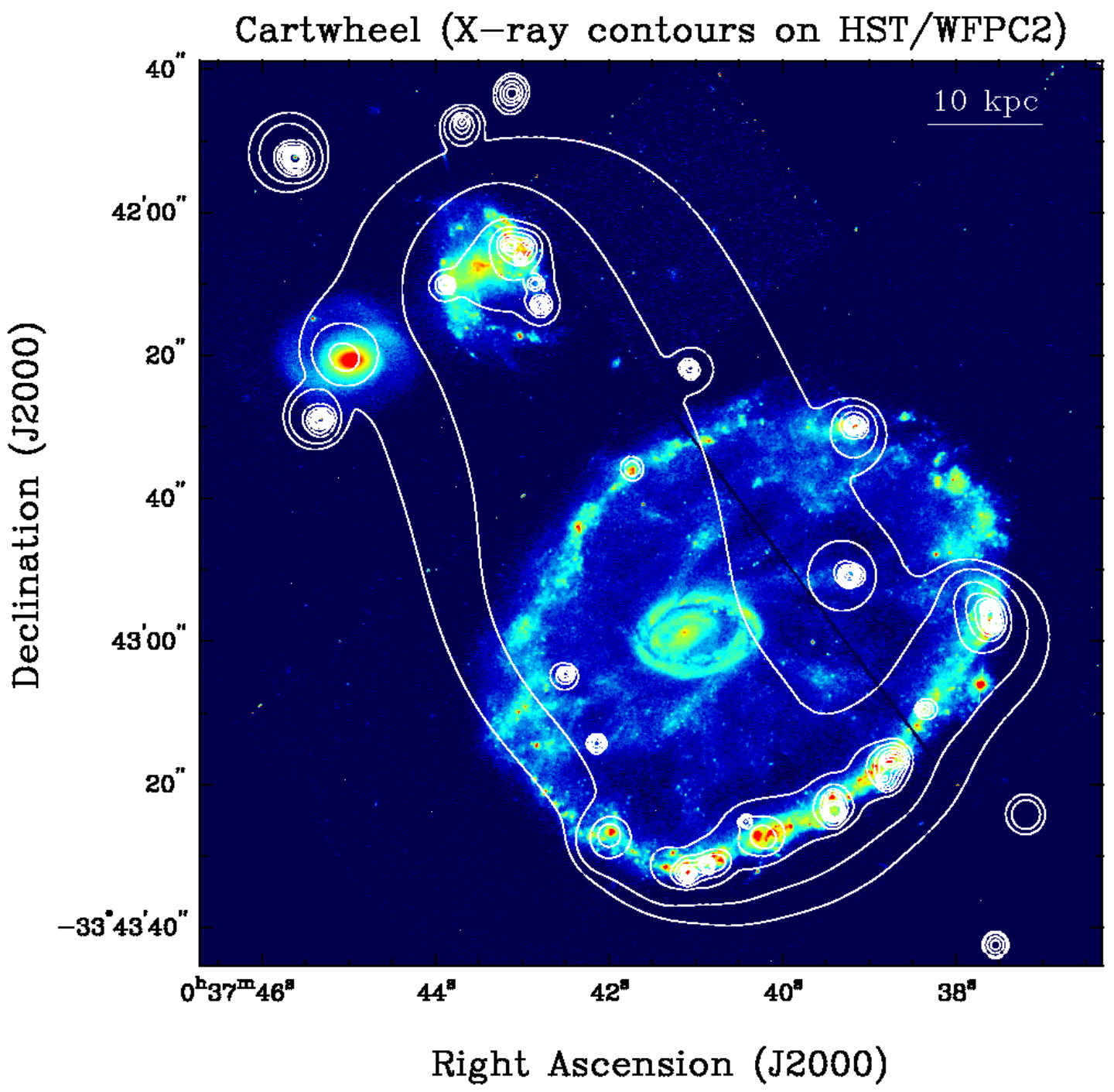} 
\caption{Broadband X-ray contours overlaid on the HST/WFPC2 optical image. The lowest contours are 0.0345, 0.0431 counts pixel$^{-1}$ (pixel=$\sim 0".5$)
and then increase successively by a factor of 2.\citep{Gao0310}.}
\label{fig:cartw}
\end{center}
\end{figure}
The proof that at least some ULXs had stellar masses, and indeed {neutron star accretors, as suggested by \citet{King0105} (see the discussion after Eq. \ref{eq:beaming1})},
finally arrived when \citet{Bachetti1410} found a coherent periodicity $P = 1.37\,{\rm s}$ in the ULX M82 X–2,
with $L_{\rm sph}  \simeq 1.8 \times 10^{40} \, {\rm erg\, s^{-1}}$. The 
only plausible interpretation, as the spin of a magnetic neutron star, required $M \sim 1\msun$, and showed 
that this object had an apparent X--ray luminosity $L_{\rm sph}$ more than 100 times its Eddington luminosity. {   Note that this observation did not identify the {\it cause} of the neutron star's super--Eddington luminosity -- this is still disputed, essentially between geometrical beaming as suggested by \citet{King0105} and the effects of very strong magnetic fields -- see the later part of the present review, and the reviews by \citet{Fabrika0121} and \citet{Mushtukov0422}. Models of this type suggest that a significant, if not dominant, fraction of ULXs are accreting NSs (see abstract in \citealt{Mushtukov1512}). It is interesting that only 6 out of the $\sim 1800$ non--Be ULXs have so far been observed to show pulsing, unlike what we might expect if this dominance holds. (We thank a referee for provoking this insight). A further problem here is that these very strong fields do not seem to be present in binaries which do not have high accretion rates. This picture then requires us to believe that strong--field neutron stars in binaries are unobservable unless the binary companion supplies them with mass, at a  super--Eddington rate.}

{The observation by \citet{Bachetti1410} removed the main argument motivating IMBH models, and discussion of ULXs is now tightly focussed.
The finding of ULXs with
$L_X > 10^{41} - 10^{42}\, {\rm erg\,s^{-1}}$, at least one of which contains a neutron--star accretor  is a stringent test of theoretical models. It is compatible with disc--wind beaming,
and this is also the only model naturally allowing standard Be -- X--ray binaries to become ULXs during bright outbursts.}

The main question in discussions of ULXs is whether they are all generically similar, or divide into subgroups with 
different physical models for each. We discuss this question in detail at the end of this review (Sect. \ref{sec:conclusions}), but
note here that in the absence of observations dividing ULXs into clearly distinct groups, Occam's razor\footnote{`Do not multiply hypotheses beyond necessity', or more simply, `don't invent two theories for the same thing.'}
leaves disc--wind beaming
as the only model open to test against {\it all} observations 
of ULXs. {   For reviews taking differing points of view, see \citet{Fabrika0121}, and \citet{Mushtukov0422}, the latter for pulsed ULXs containing very strongly magnetic neutron star accretors.}

\section{Observations}
\label{sec:observations}

Following the launch of NASA's {\it Chandra} (\citealt{Weisskopf2000}) and ESA's {\it XMM-Newton} (\citealt{Jansen2001}) satellites in 1999, the field of studying X-ray binaries in external galaxies was revolutionised. Both satellites carry X-ray CCD instruments sensitive in the 0.3-10 keV range, and have spectral, timing and imaging capabilities (with on-axis resolutions of $<$ 5"), the latter being vital for successfully disentangling an X-ray bright object from the centre of its host galaxy. 

\subsection{Catalogues \& X-ray Surveys}
\label{sec:catalogues}

After more than two decades of pointed observations, the number of individual sources now totals over 300,000 from {\it Chandra} (CSC 2.0: \citealt{Evans2010_CXC}) and over 500,000 from {\it XMM-Newton} (4XMM-DR9: \citealt{Webb2020_XMMDR9}) respectively. These extensive catalogues (and their previous iterations) have led to the formation of valuable ULX catalogues, most recently by \cite{Kovlakas2020} utilising {\it Chandra}, \cite{Earnshaw_ULX_cat} utilising {\it XMM-Newton} and {   \citep{Walton0122} using the combination of {\it Chandra}, {\it XMM-Newton} and {\it Swift}} (with previous catalogues by \citealt{Swartz_cat2004,Swartz_2011_Complete_sample_of_ULXs, Walton_cat2011, Wang_cat2016}).  

In these catalogues, ULXs ({   numbering $\sim$ 1800 in the most recent: \citealt{Walton0122}}) are located by cross-matching against local galaxy catalogues (e.g. {\sc HECATE}: \citealt{Kovlakas_HECATE2021}), with checks that sources lie within the host's optical D$_{25}$ isophote but are not coincident with the central region, and estimates applied for foreground and background (AGN) contamination. {   These catalogues enable quantification and searches for variability (e.g. propellor states)}, targeted follow-up observations, and the study of correlations with host type, including star formation rate and metallicity, thereby testing the predictions of binary population synthesis models (e.g. \citealt{Wiktorowicz1904, Kuranov2020}). 

\begin{figure}
  \begin{center}
    \includegraphics[trim=0 0 0 0, clip, width=0.5\textwidth]{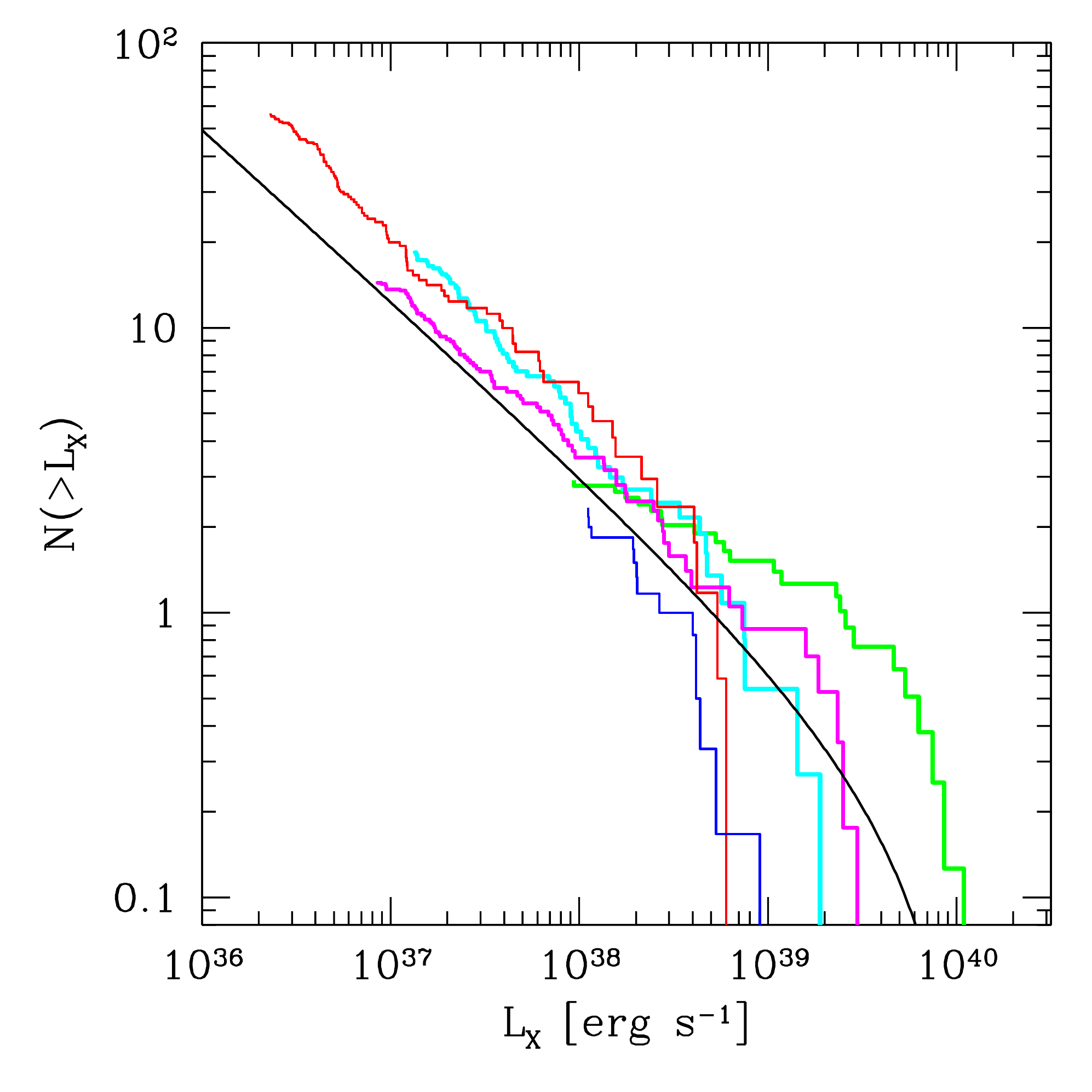}
  \end{center}
  \caption{From \cite{Mineo2012} showing the point source X--ray luminosity functions (XLFs) (normalised by star formation rate) created from {\it Chandra} observations of M101 (blue), NGC 3079 (red), M51 (cyan), Antennae (magenta), NGC 3310 (green).
    }
\end{figure}

Both of the most recent ULX catalogues find that spirals are the most common host galaxies -- consistent with ULXs being associated with regions of high star formation. An excess of ULXs is found in lower metallicity galaxies (see also \citealt{Soria0501, Mapelli1010, Prestwich1306, Brorby2014, Tzanavaris2016}), the numbers scaling with star formation rate as (\citealt{Kovlakas2020}): 

\begin{align}
    N_{\rm ULX} = 0.45^{+0.06}_{-0.09}\times{\rm SFR}_{M_{\odot}/{\rm year}} +3.3^{+3.8}_{-3.2} \times \frac{M_{\star}}{10^{12}M_{\odot}}.
\end{align}

It is important to note that, regardless of spatial coverage, the long timescale (months or longer) variability of ULXs (as discussed below) demands repeat observing to obtain a reliable census of the population. The all--sky survey by {\it eROSITA} (eRASS - \citealt{eROSITAonSRG}) will provide precisely these repeat scans, eight being taken over the space of four years (2020-2024) with a far greater sensitivity than the ROSAT all--sky survey (\citealt{Voges_RASS_1999}) and coverage from 0.3-10 keV (matching the nominal range of {\it Chandra} and {\it XMM-Newton}).

The production of such extensive catalogues provides unique statistical insights, notably the X-ray luminosity function (XLF, such as that shown in Figure 2) giving the combined numbers of low-- and high--mass X--ray binary populations and how these depend on metallicty (\citealt{Lehmer2021}). In turn these XLFs allow estimates for the impact of accreting binary populations on reionisation and galaxy evolution (\citealt{Fragos_2013_reionization,Kovlakas2022}). XLFs can also be directly compared with population synthesis calculations which include the evolution of systems through their various phases (including mass transfer, common envelope, SNe and gravitational inspiral). Recent attempts include those by \cite{Wiktorowicz1904} where the effect of geometrical beaming (following equation 55) is included, \cite{Khan2021}, where the observational impact of precession is considered (see below), and \cite{Kuranov2020} who obtain the XLF for magnetised neutron stars in the absence of beaming and precession of the wind-cone (see section 2.3.3). 
\begin{figure}
  \begin{center}
    \includegraphics[trim=0 0 0 0, clip, width=0.5\textwidth]{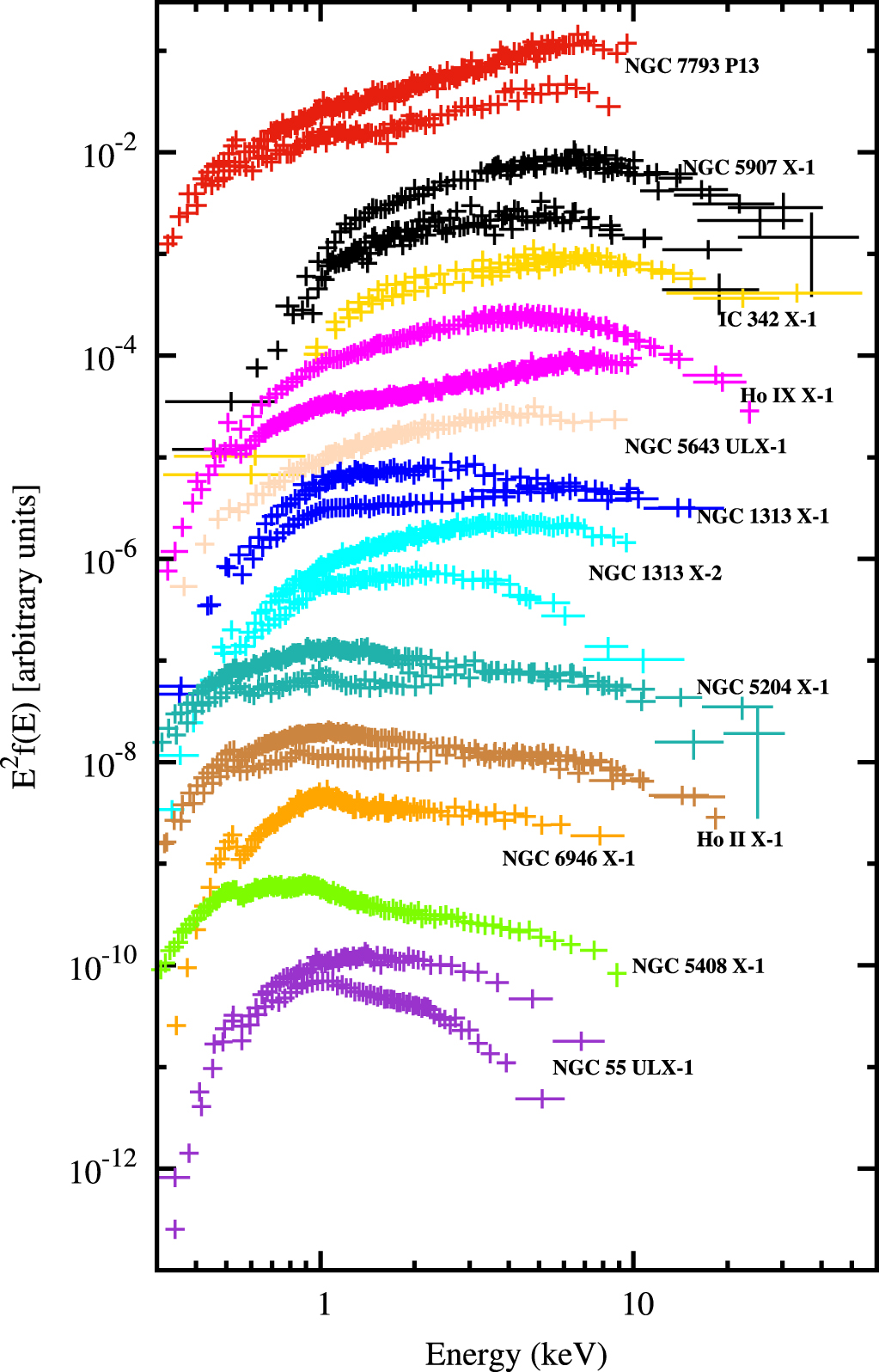}
  \end{center}
  \caption{From \cite{Pinto2017unified}, illustrating the variety of observed spectra of ULXs. Each dataset has been unfolded through the instrumental response only ({\it XMM-Newton} and/or {\it NuSTAR}) and show two spectral states for each ULX (one hard and one soft). The fluxes on the y-axis are arbitrary.} 
\end{figure}

\subsection{Spectroscopy}
\label{sec:spectro}

{\it Chandra} and {\it XMM-Newton} have made it possible to study ULXs with (relatively) high signal-to-noise and energy resolution, and have provided a `canonical' picture of ULX X-ray spectra. {  The improved data quality from these satellites allowed single component spectral models to be excluded, instead indicating the presence of a soft excess and signs of a break to a steeper spectral index at the highest energies, where instrumental effective area typically diminishes (e.g. \citealt{Kaaret2003Sci, Miller2004,Roberts2005,Stobbart2006}).}

If the soft excess is modelled as a thin disc (set to truncate at the ISCO), the characteristic temperatures are sufficiently low (0.1-0.2 keV) that IMBHs accreting at significantly sub-Eddington rates were proposed (\citealt{Miller2004b, Miller2004a}). But these accretion rates imply an upscattered power-law component extending unbroken to very high energies (around the electron's thermal energy, see e.g. \citealt{Done2007}). The presence of a break below 10~keV conflicts with this (as too do theoretical arguments - see \citealt{King0401}), supporting the alternative hypothesis of super--critical accretion (\citealt{Shakura73}). The first large--scale spectral analysis based on a picture of super-Eddington mass supply was performed by \cite{Gladstone2009}. Here the harder emission was created by inverse Compton scattering of seed photons (presumed in this case to originate from the soft excess). Since then the field has benefited from data from newer X-ray satellites, notably {\it NuSTAR}, whose ability to image in the 3-70 keV bandpass (\citealt{Harrison2013}) confirmed the $<$ 10 keV break with high signal-to-noise ratio (e.g. \citealt{Bachetti2013, Mukherjee2015}), and allowed for the most complete picture of the X-ray SEDs of ULXs (Figure 3). This wide band-pass has led to important advances in the field, including the discovery of the first ULX pulsar (\citealt{Bachetti1410}).

Current modelling of the X-ray spectra of ULXs remains to some degree phenomenological (although more physically motivated models have been developed: e.g. \citealt{Vinokurov2013}). Models used to describe the soft component variously include an advection dominated disc (typically {\sc diskpbb} -- with a radial temperature profile $T\propto r^{-p}$ with $p < 0.75$, e.g. \citealt{Mineshige1994}), or a standard disk blackbody (typically {\sc diskbb} -- \citealt{Mitsuda1984}). Each of these has been used to describe the wind, inclination and advection-modified emission from the super-critical disc or from an accretion curtain (depending on field strength and accretion rate -- \citealt{Mushtukov1705}). The harder X-ray emission (typically $>$ 1 keV) is variously modelled with Compton components (with empirical assumptions), broken power-laws and -- more recently -- the addition of emission from a neutron star accretion column (\citealt{Walton1804}). Such two- or three-component spectral fits are motivated by spectral timing methods (e.g. the covariance spectrum - \citealt{Middleton1503}) and pulse-resolved modelling (\citealt{Brightman2016, Walton2018P13}). Although multiple components are often used, simple modelling has encouraged the wide adoption of three simple, empirical categories: soft ultraluminous, hard ultraluminous, and broadened disclike (\citealt{Sutton2013} - Figure 4). It is important to note that these definitions are merely descriptive, they do not imply precise statements about the relative contributions of the underlying physical components but provide a useful shorthand for the shape of the spectrum and its evolution with time.

The simple phenomenological modelling described above can be connected to the theoretical picture of super--Eddington mass supply. {   The characteristic temperatures obtained from modelling the spectra can approximate the colour-corrected temperatures $T_{\rm c, ph}$, $T_{\rm c, sp}$ and $T_{\rm c, max}$, 
of the quasi-blackbody emission from the disc at characteristic radii (\citealt{Poutanen0705}): the outer photosphere of the wind ($R_{\rm ph}$), spherization radius ($R_{\rm sph}$ -- noting that the  assumed position of this radius in \citealt{Poutanen0705} slightly differs from Eq. 14), and innermost edge of the disc ($R_{\rm max}$):

\begin{equation}
  T_{\rm c, ph} = 0.8 f_{\rm col}\left(\frac{\zeta\beta}{\epsilon_{\rm w}}\right)^{1/2}m^{-1/4}\dot{m}^{-3/4} ~~ {\rm keV}
  \label{eq:2}
\end{equation}
\begin{equation}
  T_{\rm c, sp} = 1.5 f_{\rm col}m^{-1/4}\dot{m}^{-1/2} ~~ {\rm keV}
\end{equation}
\begin{equation}
  T_{\rm c, max} = 1.6 f_{\rm col}m^{-1/4}  ~~ {\rm keV}
\end{equation}
}

Associated values for the accretion rate then follow under the assumption of a fixed fraction of accretion energy deposited in the wind ($\epsilon_{\rm w}$), colour temperature correction ($f_{\rm col}$) as well as dynamical values associated with the wind (i.e. the wind velocity relative to the Keplerian velocity at $r_{\rm sph}$, $\beta$, and the cotangent of the wind opening angle, $\xi$ -- \citealt{Poutanen0705}). As an example, typical values of the soft X-ray component's temperature lie between 0.1 and 0.4 keV (\citealt{Middleton1503}); for values of $f_{\rm col} \sim$ 2 this indicates Eddington-scaled mass supply rates in the range $\sim$ 20-300 for a 10 M$_{\odot}$ black hole and 50-800 for a 1.4 M$_{\odot}$ neutron star. 

\begin{figure}
  \begin{center}
    \includegraphics[trim=0 0 0 0, clip, width=0.55\textwidth]{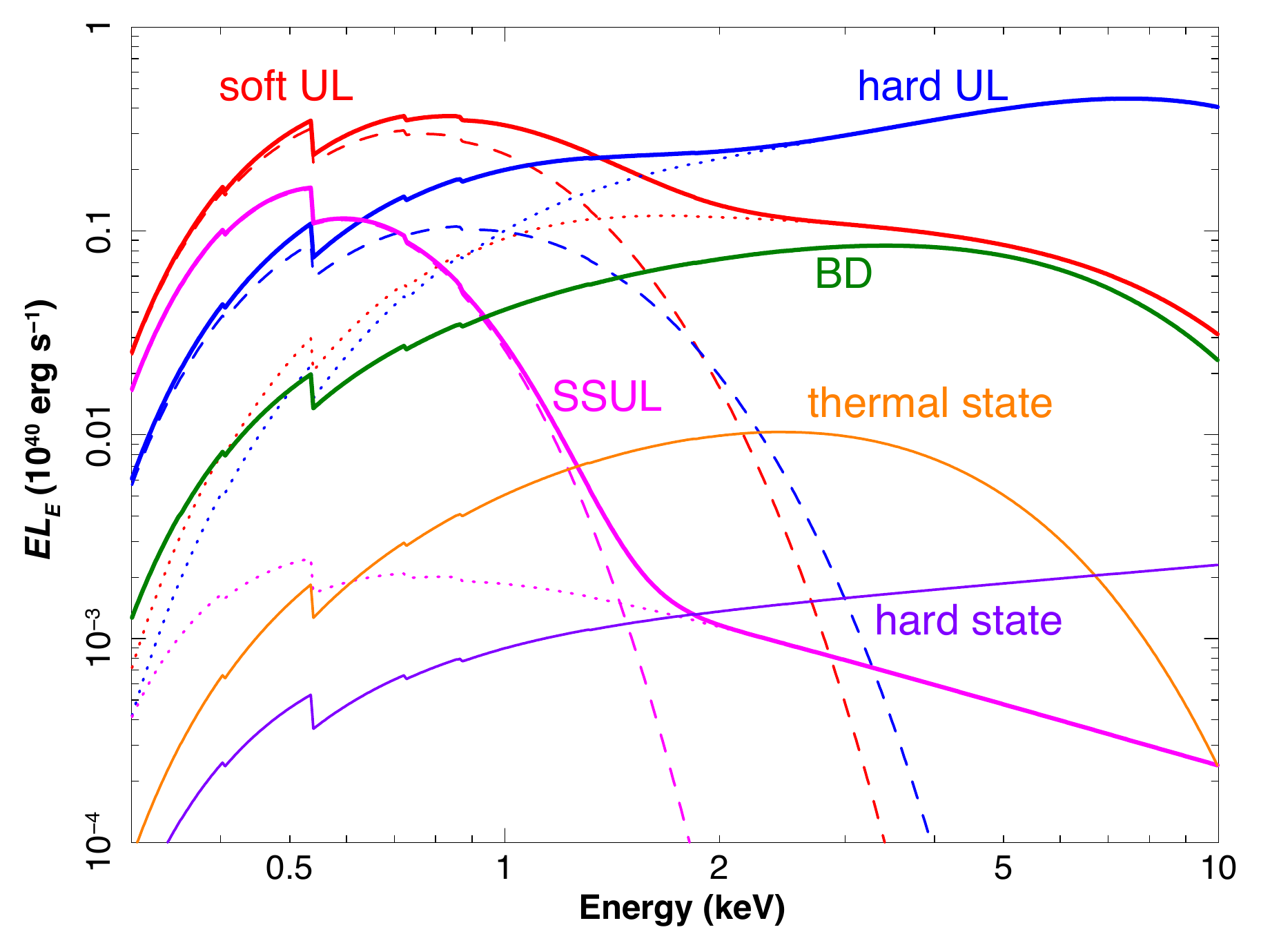}
  \end{center}
  \caption{From \cite{Kaaret1708}, showing the variety of phenomenological states of ULXs (SSUL: supersoft ultraluminous; BD: broadened disc; H/SUL hard and soft ultraluminous) as defined by \citealt{Sutton2013}) as compared to the standard canonical soft (thermal) and hard state spectra of XRBs. }
\end{figure}

One must be cautious when modelling ULX spectra, as descriptions are often degenerate (see Section 2.3.4), and the presence of a neutron star can have a major impact on the interpretation of characteristic temperatures (see Section 2.2 for more detail). As an example, strong dipole fields could truncate the {\it thin} disc (i.e. before the local Eddington limit is reached) which comes with the natural corollary that such a disc emits below the Eddington limit for a neutron star mass (see \citealt{Middleton_Brightman_2019_M51}). Additional uncertainty in the spectral fitting (regardless of the nature of the compact object) arises as Compton down-scattering within the wind itself can distort the temperature of the photons produced at various radii. It remains likely that the soft component is inclination-dependent, so that a more edge--on system appears increasingly softer until the temperature from the face of the wind dominates (and from which we expect a luminosity $\sim L_{\rm Edd}$ to be emitted: \citealt{Poutanen0705, Zhou2019, Qiu_Feng2021}). Together with changes in accretion rate, the inclination dependence provides a natural explanation for soft and super-soft (e.g. \citealt{Feng2016}) ULXs which fit well into the simplified unified model of ULX spectra (\citealt{Poutanen0705, Middleton1503, Soria_Kong2016,Urquhart_Soria2016a,Pinto2017unified}).

A notable absence from ULX spectra (with the exception of Swift J0243.6+6124: \citealt{Bykov1022}, see also the claim of a feature in NGC 7456 ULX-1: \citealt{Mondal2021}) is that of Fe K$_{\alpha}$ emission, with \cite{Walton2012} placing strong constraints on its presence in Ho IX X-1. This absence is remarkable, given the strong fluorescent yield of Fe and its typically high abundance. Importantly this may point towards an extremely highly ionized gas environment, consistent with an irradiated conical outflow (see also \citealt{Medvedev_Fabrika2010, Pinto2021}). It is quite likely that weak Fe K in reflection would be produced far out in the wind cone where the material is not fully ionized, in the disc beyond $R_{\rm sph}$ provided that this radius is small enough that the outer disc is irradiated by X-rays with energies $>$ 7 keV, and beyond the magnetospheric radius (so long as this is $> R_{\rm sph}$).

More elaborate models of emission from slim discs already exist (e.g. \citealt{Straub2013}), and models for fitting to data may now be obtained directly from 3D radiative MHD simulations for super--Eddington mass supply on to black holes (e.g. \citealt{Narayan_spectra2017}) and magnetised neutron stars (\citealt{Kawashima2002}). Still more elaborate spectral modelling may include self--consistent photoionization. Looking forwards, it is hoped that such models may bridge the gap between theory and observation.

\subsubsection{High energy-resolution X-ray spectroscopy}
\label{sec:HEres}

Phenomenological models can reasonably describe ULX spectra as observed by current instrumentation, although clear residuals to these models are found between 0.5 and 2 keV (\citealt{Stobbart2006}) in multiple ULXs (\citealt{Middleton2015_resids}), and seen in both {\it XMM-Newton} and {\it Chandra} spectra (ruling them out as instrumental artefacts - \citealt{Roberts2006, Pintore2015}). Initially thought to originate in diffuse star-forming plasma (\citealt{Strohmayer2007, Strohmayer2009, Miller2013}), the luminosities are too high for such an interpretation, and an association with mass-loaded outflows was instead made (\citealt{Middleton2014}). Both via high angular resolution studies with {\it Chandra} (\citealt{Sutton2015}), and by studying the correlated variability of the features with spectral hardness (\citealt{Middleton2015_resids, Pinto2020_1313}), it was determined that they must be associated with the ULX and not with the surrounding nebula (e.g. \citealt{Sathyaprakash2019_nebula}). 

Instruments with higher energy resolution (e.g. {\it XMM-Newton}'s Reflection Grating Spectrometer, RGS) have shown that the CCD-quality features are indeed composed of blueshifted absorption lines and rest--frame emission lines (\citealt{Pinto_2016Natur.533...64P, Pinto2017unified} -- but see \citealt{2018_Kosec_MNRAS.473.5680K} for apparently blueshifted emission lines in  NGC 5204 X-1), confirming the suggestion by \cite{Middleton2014}. In NGC 1313 X-1, counterpart Fe absorption lines have also been found in {\it NuSTAR} CCD-quality data with a blueshift consistent with that inferred from the lower energy resolved lines (\citealt{2016_Walton_ApJ...826L..26W}). The limit implied on the outflow velocity is $>$ 0.2 c, considerably faster than observations of thermal (or magneto-centrifugal) winds from sub--Eddington XRBs (e.g. \citealt{Ponti_winds2012}), but consistent with radiatively driven winds from a super-critical inflow (e.g. \citealt{King1003,Jiang1412, Kitaki0121}). 

Ultrafast outflows have now been located in several ULXs via individual searches of RGS spectra (NGC 1313 X-1 and NGC 5408 X-1: \citealt{Pinto_2016Natur.533...64P}, NGC 55 ULX: \citealt{Pinto2017unified}, NGC 247 ULX-1: \citealt{Pinto2021}) and notably in the transient NS ULXs, NGC 300 ULX-1 (\citealt{2018_Kosec_MNRAS.479.3978K}) and Swift J0243.6+6124 (\citealt{vandenEijnden2019}). Intriguingly this latter source simultaneously launches both a wind and a jet (see Section 2.4.1). {   In addition to these studies, a larger scale search within the data of 19 ULXs, using a faster algorithm by \citep{Kosec2021}, indicates that rest-frame emission lines and blue-shifted absorption lines are indeed a ubiquitous feature.}

The velocity of the wind, $v_{\rm wind}$, a measure of the ionisation state, $\xi$ (both obtained by using consistent models for photoionised plasma - \citealt{Pinto_2016Natur.533...64P}), and an assumption for the ionising luminosity, allows the kinetic luminosity of the wind to be estimated from: 

\begin{equation}
P_{\rm wind} = \frac{1}{2} \dot{m}v_{\rm wind}^{2} = 
 2\pi L_{\rm ion} \frac{\Omega}{\xi}v_{\rm wind}^{3}m_{\rm p} 
\end{equation}

The filling and covering factor of the wind ($\Omega$) in the above formula remain unknown, but the high projected velocities (which are therefore lower limits on the true velocity) imply the kinetic luminosities for these winds are substantial, and may even dominate over the observed radiated power from the source (\citealt{Pinto_2016Natur.533...64P}). It is worth noting that such observations are potentially in conflict with MHD models which predict lower fractions of the radiative power spent in launching the wind (\citealt{Jiang1412}). Questions remain over the nature of the rest-frame emission lines - conceivably these may be due to collisional ionization as a result of interaction with the wind of the companion star (e.g. \citealt{Oskinova2005, Mauerhan2010}) or may be associated with lower velocity winds driven from the outer, irradiated disc (\citealt{Middleton_thermal2021}). 

Studies have recently shifted to determining how the wind appears to change as a function of spectral hardness. As first reported in \cite{Middleton2015_resids}, the strength of the CCD residuals in NGC 1313 X-1 appears to anti-correlate with increasing spectral hardness (i.e. the features are weaker when the source appears spectrally harder). One interpretation has been that the wind or system is precessing, or the mass accretion rate changing (the optical depth along our line-of-sight through the wind changing with either/both: \citealt{Poutanen0705}). A more recent in-depth study of NGC 1313 X-1 based on a series of long {\it XMM-Newton} observations, has resolved the atomic features in emission and absorption (using the RGS) and made a direct comparison to changes in the X-ray spectrum (\citealt{Pinto2020_1313}). These data have allowed for important progress to be made, including the discovery of a slower, more neutral component of the wind (moving with a projected velocity of $\sim 0.08c$ rather than the fast wind of $\sim 0.2c$), pointing towards a complex outflow (as one might expect from simulations - e.g. \citealt{Takeuchi2013}). These high energy-resolution results broadly confirm the picture from the CCD-quality residuals (as does the larger sample result of \citealt{Kosec2021}), with the observation of an anti-correlation between spectral hardness and line strength, possibly driven by changes in accretion rate and/or inclination angle via precession (\citealt{Middleton_2015ULX_modelpaper, Middleton1512}). Importantly the correlated changes further confirm that the winds are associated with the ULX rather than some surrounding, more distant material (which would not respond on such timescales). The high quality data made available via long RGS exposures has allowed, for the first time, photoionization modelling of a ULX wind (\citealt{Pinto_winds_thermalbalance}, see also \citealt{Pinto2021} for a comparison to other ULXs), indicating that material in the innermost regions is potentially thermally unstable. In future, the launch of {\it XRISM} and {\it Athena} (with its X-ray Integral Field Unit) will allow for far deeper studies of ULX winds.

\subsubsection{Spectral studies of neutron star ULXs}
\label{sec:spectnsulx}

The discovery of several pulsating neutron star ULXs (\citealt{Bachetti1410,Israel1702,Israel1703,Furst1611,Tsygankov1709,Carpano1805,Sathyaprakash1909,Doroshenko1805} -- {   see Table \ref{tb:combined}}) has led to scrutiny of the X-ray spectrum in light of the expected differences from the standard super-critical model, as a consequence of the presence of the magnetosphere -- accretion curtain and column -- and surface. The truncation of the disc at the magnetospheric radius and free-fall of material onto the magnetic poles, leads to two likely configurations depending on the accretion rate. At very high accretion rates or low dipole field strengths, the spherization radius is reached at radii greater than the magnetospheric radius -- if the wind is optically thick (see \citealt{Vasilopoulos1910, Middleton_2019_Accretion_plane}) it will act to collimate some of the radiation within. Regardless, we expect thermal emission from R$_{\rm sph}$ down to R$_{\rm mag}$, which tends increasingly towards a blackbody as R$_{\rm sph}$ tends towards R$_{\rm mag}$. At radii larger than R$_{\rm sph}$, the emission is from an unmodified, approximately thin disc (thin below $\sim$ 30\% Eddington, e.g. \citealt{McClintock2006}), although this emission is probably modified and scattered by the wind covering those regions between R$_{\rm sph}$ and R$_{\rm ph}$. For strong magnetic fields or relatively low accretion rates, R$_{\rm sph} < R_{\rm mag}$ and the disc emission beyond R$_{\rm mag}$ originates from an Eddington--limited thin disc. In either accretion-rate regime, within R$_{\rm mag}$, the material falls freely along magnetic field lines until meeting the standing shock front in the accretion column. The key characteristic of the magnetospheric curtain is the optical depth $\tau$ which depends on the accretion rate (and therefore dipole field strength) and angle from the accretion disc plane. According to (\citealt{Mushtukov1705}) the optical depth is given by:

\begin{equation}
    \tau \approx \frac{70 L_{\rm 39}^{6/7} B_{\rm 12}^{2/7}}{\beta(\theta)}\left(\frac{\cos\theta_{0}}{\cos\theta}\right)^{3}
\end{equation}

\noindent where $\theta_{0} \approx \pi - (R/R_{\rm M})^{1/2}$ is the angle to the accretion column base and $\beta(\theta) = v(\theta)/c$ is the local dimensionless velocity along the magnetic field lines. This optical depth determines the shape of the thermal 
spectrum from the curtain (becoming optically thick for $L_{39}>B_{12}^{1/4}$) with a range of emergent photon temperatures due to inclination dependence, as well as potentially diluting variability \citealt{Mushtukov1903}) and the presence of cyclotron resonance scattering features (see the following section). A rough estimate for 
the temperature of the curtain is provided by \cite{Mushtukov1705} as $T_{\rm out} \sim 0.5L_{39}^{11/28}B_{12}^{-2/7}m^{-1/14}R_6^{-5/7}$~keV (where $R_{6} = R/10^{6}$ cm and $m = M/M_{\odot}$) which can be $\gtrsim$ 1 keV for typical ULX luminosities, and could therefore explain one of the hard components in the spectra of ULXs.

\begin{wrapfigure}{r}{0.6\textwidth}
  \begin{center}
    \includegraphics[width=6.5cm,height=7.0cm,angle=0]{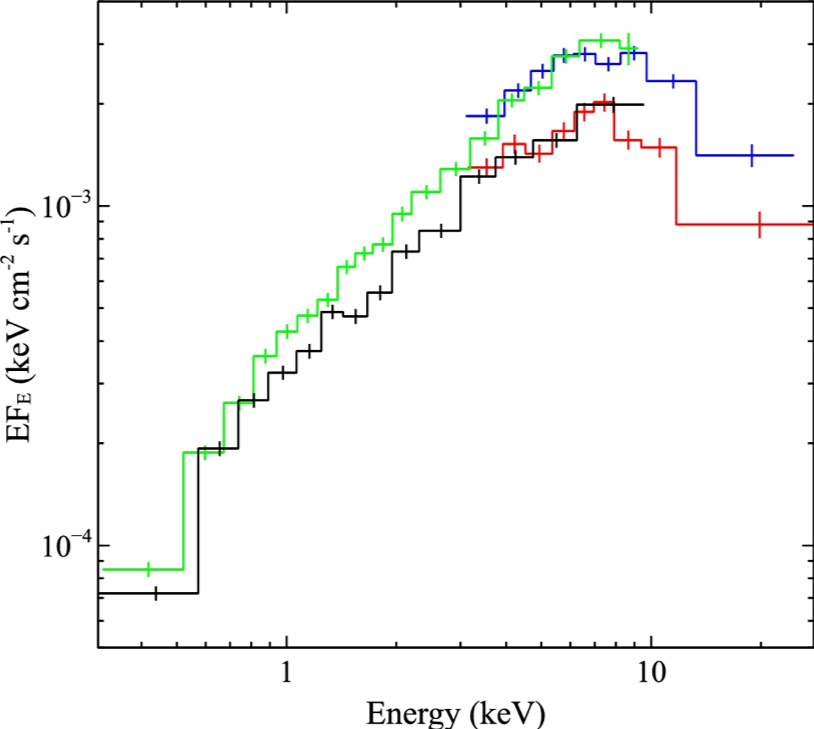}
\end{center}
  \caption{Pulse-on versus pulse-off spectra for NGC 7793 P13 from \cite{Walton2018P13}, using data from {\it NuSTAR}.}
\end{wrapfigure}

The final set of likely spectral components is then: at high accretion rates/low dipole field strength -- thin disc, thick disc and wind, curtain and column, and for low accretion rates/high dipole field strength --  thin disc, curtain and column. Using the latter description of the inflow, \cite{Koliopanos2017_sample}
explored the possible range of dipole magnetic field strengths for 18 ULXs (containing both pulsing ULXs and those without) under the assumption that R$_{\rm sph} < R_{\rm mag}$, finding values ranging from $\sim 10^{11} - 10^{13}$~G {   (although there is a tension between the assumption of a thin disc and observed super-Eddington luminosity from the soft component).}

The spectrum of the pulsed emission can also be extracted via selecting `off' versus `on' pulse times and obtaining the difference spectrum. The latter has been performed by \cite{Brightman2016} in the case of M82 X-2, and \cite{Walton2018P13} in the case of NGC 7793 P13 (Figure 5), who found the accretion column to be well-modelled by a power-law with an exponential cutoff (see also \citealt{Pottschmidt2016} for a similar finding for SMC X-3). The unambiguously spectrally hard nature of the pulsed emission has naturally led to suggestions that other, similarly hard ULXs (e.g. Ho IX X-1: \citealt{Walton2017HoIX} and IC 342 X-1:\citealt{Middleton1503}) may also harbour neutron stars (\citealt{Pintore2017_sample, Koliopanos2017_sample, Walton1804,Gurpide9221}). Showing whether the `hard excess' is indeed due to the presence of an accretion column will be valuable for identifying neutron star ULXs without the need for deep pulsation searches; this in turn will help reveal ULX demographics and test our grasp of binary evolution.

\subsubsection{Cyclotron resonance scattering features}
\label{sub:crcf}

In the presence of strong magnetic fields, electron and proton orbits are quantised into Landau levels; such particles can resonantly scatter incident photons of sufficient energy depending on the quantum mechanical cross section (see \citealt{Schwarm2017_CRSF_crosssections}) and leave cyclotron resonance scattering features (CRSFs) imprinted on the spectrum (although any subsequent scattering, for instance in an accretion curtain, can reduce their detectability - see \citealt{Mushtukov1705}). The transition energy of such lines is related to the magnetic field strength and particle type according to:

\begin{equation}
    \Delta E_{\rm e^{-}} = \frac{11.6}{(1+z)}\left(\frac{B}{10^{12} G}\right) ~~~ {\rm keV}
\end{equation}

\begin{equation}
    \Delta E_{\rm p^{+}} = \frac{6.3}{(1+z)}\left(\frac{B}{10^{15} G}\right) ~~~ {\rm keV}
\end{equation}

\noindent where $z$ is the gravitational redshift, given by:

\begin{equation}
    z = \left(1-\frac{2GM}{r_{\rm cyc}c^{2}}\right)^{-1/2} -1
\end{equation}

\noindent where $r_{\rm cyc}$ is the distance from the surface of the neutron star to where the line is formed. Studies of CRSFs in Galactic systems (e.g. \citealt{Tsygankov2006_CRSF, Jaisawal2016_CRSF}) have been useful for studying the regions in which these lines are formed (see the review of \citealt{Staubert2019_CRSF_review}).

\cite{Brightman1804} reported the first CRSF in a ULX, at an energy of $\approx$ 4.5 keV in a {\it Chandra} observation of M51 ULX-8 (see Figure 6); as this energy does not correspond to known instrumental features and cannot be readily explained as a highly blueshifted absorption line, a CRSF is a possible explanation. Given the line energy, the implications are either electrons orbiting in a field with a strength of $\sim5\times 10^{11}$~G (for z = 0.25) or protons orbiting in a field around 9$\times 10^{14}$~G. {   Given the narrow width of the line, it was concluded by \cite{Brightman1804} that the feature is a proton CRSF}. Combined with constraints from spectral fitting, it was concluded by \cite{Middleton_Brightman_2019_M51} that the dipole field is likely to be weak with the feature the result of a stronger, higher order multipolar (e.g. quadrupole) component closer to the neutron star (e.g. \citealt{Israel1702,Israel1703,Brice2021}), {   a situation now confirmed in the case of Swift J0243.6+6124 (\citealt{Kong2022_CRSF}).}

\begin{figure}
  \begin{center}
    \includegraphics[trim=800 300 800 300, clip, width=0.6\textwidth]{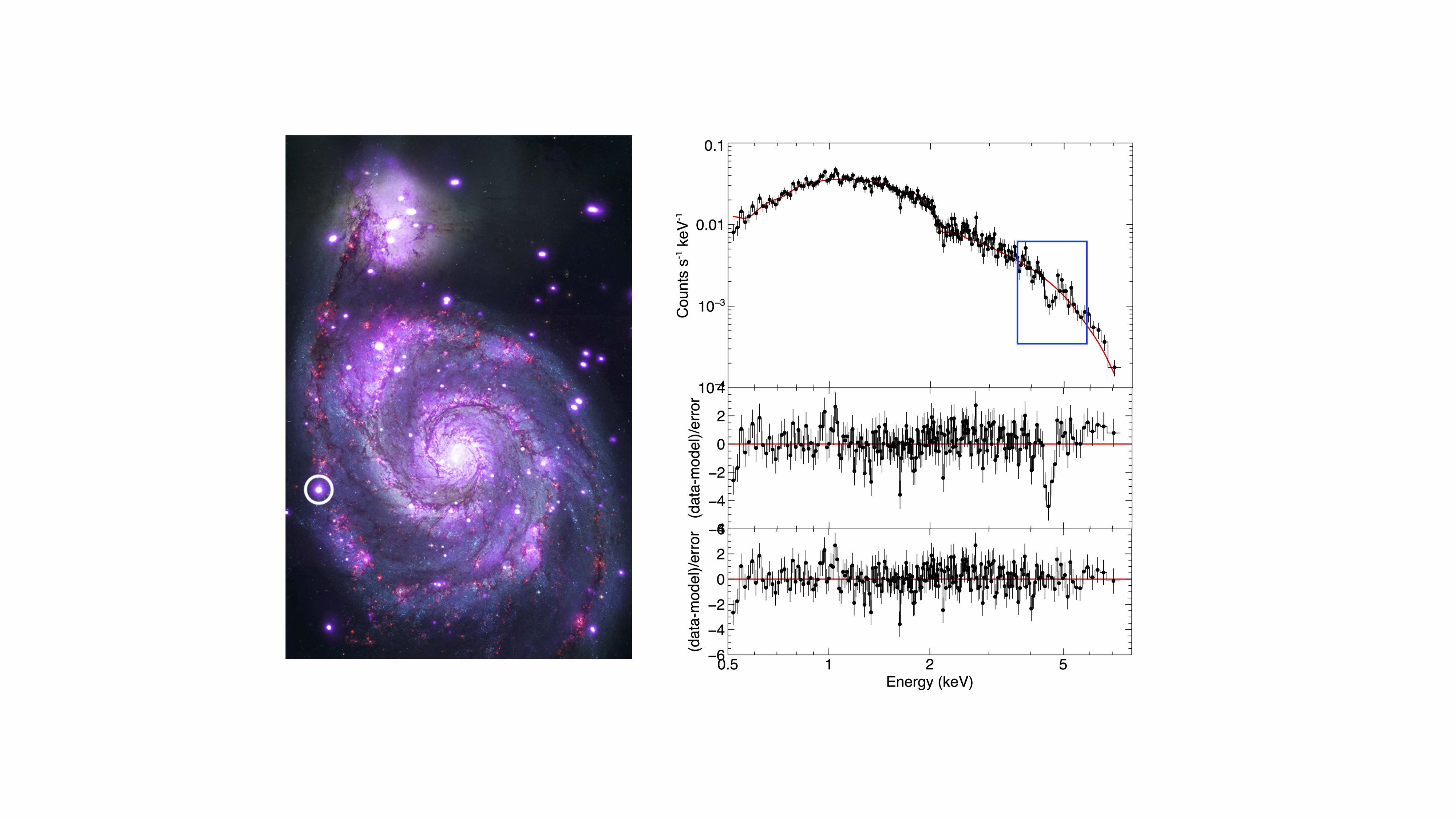}
  \end{center}
  \caption{Figure adapted from \cite{Brightman1804} showing the location of ULX-8 in M51 (left), and the CRSF identified at $\sim$4.5 keV in a {\it Chandra} observation (right).}
\end{figure}

At the time of writing, there is only a single other claim of a CRSF in a ULX, in the transient system, NGC 300 ULX-1, which was identified as a neutron star from its 31.6 second period (\citealt{Carpano1805}), with the presence of a CRSF inferred from spectral modelling of the pulsed component (\citealt{Walton1804b}). Although the line energy ($\approx$ 13 keV) appears consistent with the estimate of the dipole field strength from spin-up, doubts have been raised by \cite{Koliopanos_NGC300_CRSF} as to the presence of the line, as applying a spectral model based on emission from the accretion curtain and additional hard power-law tail, do not argue for its inclusion.

\subsubsection{Spectral evolution}
\label{sec:spectevol}

ULXs are well known to change in both brightness and in spectral shape over timescales of months or less. Most ULXs tend to follow a roughly predictable track with a given source becoming brighter as the hard and soft spectral components both increase in brightness, with the harder component tending to increase more than the soft (see e.g. \citealt{Middleton1503}) although this depends on how the spectrum is modelled (as this delineates the hard and soft components). As a result, for some of the brightest ULXs, there tends to be a clear correlation between the source becoming spectrally harder (with more flux emerging above $\sim$ 1 keV compared to that below) and brighter (\citealt{2018_Weng_ApJ...853..115W}) although this is not universal (see \citealt{Gurpide9221}). 

The coupled changes in X-ray spectrum and observed luminosity can be explained by a given ULX undergoing a change in accretion rate (at large radii), inclination angle (due to precession) or a combination of the two (\citealt{Middleton1503}). Under the condition that the neutron star dipole field is weak, so that $R_{\rm sph} < R_{\rm M}$, or the compact object is a black hole, increasing the accretion rate increases the spherization radius and the optical depth of the wind (as it is an integrated quantity -- see \citealt{Poutanen0705}) and decreases the temperature at both the spherization radius and outer photospheric radius (see equations 5 \& 6). Accordingly, we should expect to see an anticorrelation between the soft component's luminosity (which may be driven by radii interior to $R_{\rm sph}$) and the temperature at the spherization radius. Identifying and interpreting a correlation (or lack of one) is not necessarily straightforward, as increasing $R_{\rm sph}$, or moving to a more face-on orientation, leads to steadily increased beaming of the radiation interior to $R_{\rm sph}$  (\citealt{Walton2017HoIX,Walton_1313_2020}) which can make unambiguous spectral modelling challenging. Conversely, the hard X-ray emission is not expected to be particularly sensitive to accretion rate (see Eq. \ref{eq:2}) but is the region deepest within the wind cone and therefore experiences the largest amount of collimation and beaming (this is an important point when considering the effects on the local environment - see Section 2.6). An increase in accretion rate is only likely to affect the luminosity of the hard component if the wind opening angle is sensitive to accretion rate (c.f. \citealt{Jiang1412, Jiang_2019}). For a fixed inclination where sight-lines graze or intercept the wind, this picture is further complicated by obscuration and reprocessing (\citealt{Middleton1503}, although the effect of absolute collimation at any angle can be inferred numerically -- \citealt{Dauser1704}). 

Given the observed precession of SS433 (known to be highly super-critical and established to be a ULX seen edge-on - see Section 2.8.3), it is reasonable to expect precession to occur in other ULXs as well (for a variety of reasons, and with some support from observation e.g. \citealt{Pasham_M82_precessing_disc,Middleton2015_resids,2018_Weng_ApJ...853..115W}). It is therefore important to consider the effect such a changing effective inclination of the super-critical flow might have (see \citealt{Middleton1503} for details). Also, whilst changes in accretion rate and/or inclination lead to relatively predictable coupled evolution in spectrum and brightness, we also expect an impact on the timing signatures (see \citealt{Middleton1503}), and detected strength and speed of winds (see \citealt{Pinto_winds_thermalbalance, Pinto2021}). Such additional considerations are vital as they are independent of the spectral model used to describe the emission which can often be degenerate (see Section 2.3.4).

The {\it majority} of the spectral evolution observed in ULXs can be fitted into the above picture. Notably the model can broadly explain observations of super soft ULXs, which have extremely low temperature soft components ($kT_{\rm bb} \approx 50-150 eV$) and very little emission above 1 keV (e.g. \citealt{Kong0603, Urquhart_Soria2016a, Soria_Kong2016}) and which can be explained by an edge-on orientation and/or very high accretion rates (where the wind is highly obscuring). Conversely, the brightest and spectrally hardest ULXs, tend to become slightly softer as they brighten, consistent with increased beaming of emission within $R_{\rm sph}$ (\citealt{Walton2017HoIX,Walton_1313_2020}) and where the ULX is viewed at lower inclinations. In spite of the success of this simple model, a number of important open questions remain. 

One of the most obvious questions facing the unified inclination/accretion rate model of ULXs is how the presence of a strong dipole field affects the flow. For the most part, and as long as $R_{M} \ll R_{\rm sph}$, the expectation is that the spectrum will evolve along similar paths, as the position of the magnetospheric radius is expected to be independent of the mass-transfer rate for classical supercritical flows (see \citealt{Chashkina0619} for a discussion of how the location of $R_{M}$ can be affected by advection).
Rather more drastic deviations from the model above occur for those NS ULXs (note not necessarily ULX pulsars, as there are many reasons why a pulsation may be absent -- e.g. \citealt{Mushtukov2011}) which are close to spin equilibrium, as the onset of a propeller state can switch off the accretion curtain and column, leading to a quenching of the hardest spectral component (while the other components remain `on'). There are already hints of propeller states occurring in one ULX pulsar (\citealt{Tsygankov2016}) and such states might account for some of the more highly variable ULXs (e.g. \citealt{2018_Earnshaw_propeller_MNRAS.476.4272E, 2020_Earnshaw_ApJ...891..153E, Song2020_propellor}, although a propeller state is ruled out for recent, large-scale changes in X-ray flux observed in NGC 300 ULX: \citealt{Vasilopoulos1910}, and NGC 7793 P13: \citealt{Fuerst_P13_2021}).  

One of the most intriguing observations is that the shape and normalisation of the emission above 10 keV tends to stay remarkably constant in a number of bright ULXs as their spectra evolve (see \citealt{Walton_1313_2020}). This can be explained so long as the innermost regions are maximally beamed and any changes in accretion rate do not change the opening angle of the wind cone appreciably (although see \citealt{Jiang_2019} for indications that the cone's opening angle may well respond to larger changes in accretion rate). An alternative possibility is that the emission at these energies originates from a stable accretion column, and applying a phenomenological model for the column can indeed provide a good description of the hard emission (\citealt{Pintore2017_sample,Walton1804}).

There are important questions remaining over how the individual spectral components fit into the above picture, typically explored through use of the temperature-luminosity plane. Such analyses have a long history in the study of ULXs, with indications of a strong negative L-T correlation ($L \propto T^{-3.5}$) reported across a {\it sample} of ULXs (e.g. \citealt{Kajava_2008_Lsoft_T4, Kajava0909}). Given the expectation that the ULX population is heterogeneous, it is important to also study sources individually; this is becoming increasingly feasible with intense observing campaigns.

\cite{Walton_1313_2020} examined nine observations of NGC 1313 X-1 (combining {\it XMM-Newton} with {\it NuSTAR} data) and fitted the spectra with a two-component thermal model of a disc black-body at low energies and an advection dominated disc at high energies (and an additional component to account for the hard excess). From this they found a positive (and relatively steep) correlation between the temperature and luminosity of the hotter, advection dominated disc (the soft component has a flatter, possibly negative correlation). This correlation is further subdivided into high and low luminosity branches and may provide an intriguing test of what processes drive the changing appearance of ULXs. A similar analysis has been performed by \cite{Robba2021_1313X2} for the ULX pulsar NGC 1313 X-2, although a second luminosity-temperature branch is not as evident as for NGC 1313 X-1. As shown in several works (\citealt{Luangtip2016_HoIX, Gurpide9221}), the presence and nature of any correlation (at low or high temperatures) can be affected by the choice of spectral model, reinforcing the important role other methods can play in understanding the nature of the flow and the components in the X-ray spectrum (see Section 2.3.4).

\subsubsection{Open questions and future directions}
\label{sec:openquest}

Important questions remain regarding the modelling of ULX spectra, including: 

\begin{itemize}
    \item the apparent stability and similarity at energies above the spectral break (\citealt{Walton_1313_2020})
    \item the presence of fainter spectral components which may correspond to bremsstrahlung emission, as seen from SS433's jets (e.g. \citealt{Walton2015_HoII})
    \item the trends of luminosity versus temperature for the individually modelled components (\citealt{Kajava0909,Walton_1313_2020}).     
    \item the impact of spin and advection on the hard spectral emission 
    \item the prevalence (or lack-thereof) of CRSFs
\end{itemize}

Many of these issues may be resolved as self--consistent spectral models derived directly from 3D simulations of black hole and neutron star ULXs become available, (e.g. \citealt{Ohsuga1107, Jiang1412, Sadowski1502, Sadowski0316, Narayan_spectra2017, Takahashi0118, Kawashima2002}).

\subsection{Timing analysis}
\label{sec:timing}

X-ray spectroscopy has allowed for a broad categorisation of ULXs, but as with other accreting systems, the timing properties (over short and long timescales) are just as important for testing models and for revealing new behaviours. 

\subsubsection{Short timescales}
\label{sec:shortt}

Because of its large effective area, {\it XMM-Newton} has been the favoured instrument for studying the short (intra--observation) timescale variability of ULXs. The simplest analysis uses the Fast Fourier Transform (FFT) of the lightcurve to obtain the periodogram (or, if averaged, the power spectral density -- PSD). As this is typically normalised to be in (rms/mean)$^{2}$/Hz units, the integral of the PSD gives the fractional root mean square (rms), which, when Poisson noise corrected and plotted against energy, yields the rms spectrum (see \citealt{Uttley2014} for a comprehensive review). 

The PSDs extracted from single observations of ULXs are generally described by either a red--noise power law (power scaling as $\nu^{-1} \rightarrow \nu^{-2}$) or a broken power--law (e.g. \citealt{Strohmayer2003_M82QPO, Dewangan2006, Strohmayer2007, Heil2009_sample, Atapin2019}), sometimes designated as the broadband noise. Binning the rms reveals a tentative linear relationship with X-ray flux (\citealt{Heil2009_sample, Heil2010, HernandezGarcia2015}) as expected from multiplicative propagation of fluctuations in accretion rate (\citealt{Lyubarskii1997, Churazov2001, Arevalo_Uttley2006}). \cite{Heil2009_sample} observed that the variability of some ULXs appeared very low, while conversely others appeared extremely high (up to 30\% fractional rms - \citealt{Middleton1503}). This can be explained via either the dilution of intrinsic variability from a neutron--star accretion column as photons scatter inside the accretion curtain before eventually escaping (\citealt{Mushtukov1903}) or by mass loss in the wind, which may remove variability from the inflow (\citealt{Middleton1503}). The substantial amounts of short timescale ($< 10$~ks) variability in those spectrally softer sources -- in which there is strong evidence for mass-loaded winds (e.g. NGC 5408 X-1 and NGC 6946 X-1) -- could arise from stochastic obscuration by clumps in the wind (\citealt{Middleton2011, Sutton2013}), produced via radiative hydrodynamic (e.g. \citealt{Takeuchi2013}) or Kelvin Helmholtz instabilities.

In addition to the broadband noise, there have been several claims of quasi-periodic oscillations (QPOs -- with power concentrated over a small frequency range) at 10s of mHz (NGC 5408 X-1: \citealt{Strohmayer2009, Pasham2012_5408QPOs}, NGC 6946 X-1: \citealt{Rao2010_6946QPO}, NGC 1313 X-1: \citealt{Pasham2015_1313QPOs}, M82 X-1: \citealt{Strohmayer2003_M82QPO}, IC 342 X-1: \citealt{Agrawal2015_IC342_QPO}) as well as at higher ($\sim$Hz) frequencies in M82 X-1 (\citealt{Pasham2014_M82QPOs}). There have been various attempts to use QPOs to obtain the accretor mass, based upon a simple frequency scaling (e.g. \citealt{Strohmayer2009, Pasham2015_1313QPOs}), which tend to give masses typical of IMBHs. However, while QPOs are detected from super-critically accreting systems such as TDEs (e.g. \citealt{Pasham2019_TDEQPO}) and in GRS 1915+105 (the well known 67-68 Hz QPO: \citealt{Belloni2013}), it is not clear that there are well-studied objects with well-understood signals available for direct comparison. At present, the origin of ULX QPOs remains unclear, although models invoking modulations of accretion rate (\citealt{Okuda2009_SS433_QPO}) or precession of the radiation-pressure supported disc -- which can also explain the lag seen around the QPO frequency (\citealt{DeMarco2013} -- see Section 2.3.4) have both been put forward (\citealt{Middleton_2019_Accretion_plane}). With the high throughput of forthcoming instruments (namely ESA's {\it Athena}), studies of ULX QPOs will become considerably more revealing.

\subsubsection{Pulsations}
\label{sec:puls}

Following the discovery of an X-ray pulsation from M82 X-2 (\citealt{Bachetti1410}), one of the most important tools now regularly applied to ULX data is the accelerated pulsation search (see e.g. \citealt{Ransom2001, Ransom2002, Andersen2018_Jerk}). This can account for the orbit of the neutron star about the system barycenter, which acts to smear the power in the pulsations over neighbouring Fourier frequency bins. Accelerated searches within both new and archival data have led to the discovery of several more pulsating ULXs (NGC 5907 ULX-1: \citealt{Israel1702}; NGC 7793 P13: \citealt{Furst1611,Israel1703}; SMC X-3: \citealt{Tsygankov1709}; NGC300 ULX-1: \citealt{Carpano1805} -- first identified as a supernova by \citealt{Monard2010}; NGC 1313 X-2: \citealt{Sathyaprakash1909}; M51 ULX-7: \citealt{RodriguezCastillo2005}; RX J0209.6-7427: \citealt{Vasilopoulos0620, Chandra2020}) and Swift J0243.6+6124: \citealt{Kennea2017}).

\begin{figure}
  \begin{center}
    \includegraphics[trim=0 0 0 0, clip, width=0.48\textwidth]{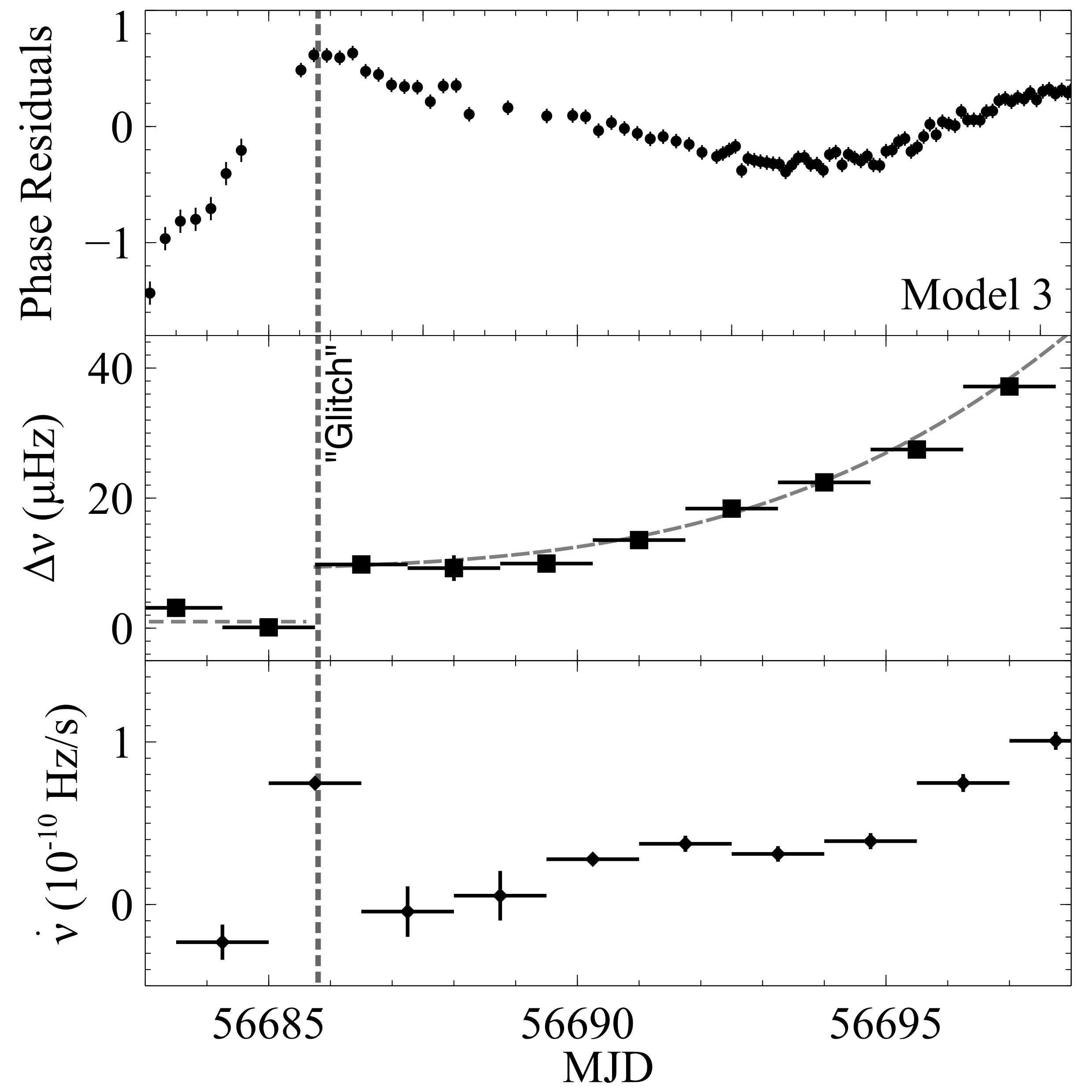}
  \end{center}
  \caption{From \cite{Bachetti2020} showing the positive glitch (with a change in frequency $\Delta\nu~\sim$ 8$\times$10$^{-11}$Hz/s) in M82 X-2, revealed by extensive monitoring with {\it NuSTAR}.}
\end{figure}

\begin{figure}[ht]
\begin{center}
\includegraphics[width=1\textwidth]{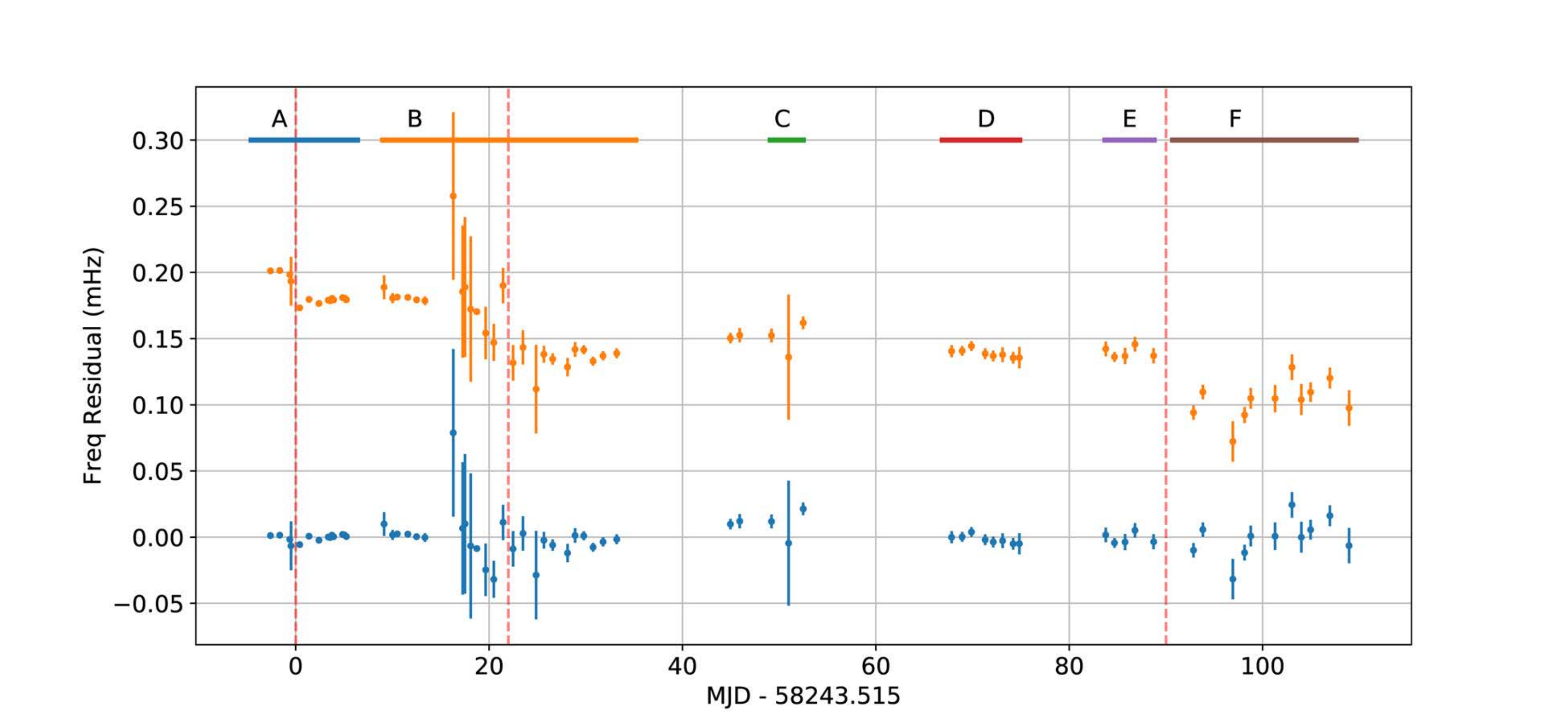}
\caption {Data from \cite{Ray2019}, highlighting the glitches in the ULX pulsar NGC 300 ULX-1. The upper (orange) points show offset residuals from the best-fitting model of the frequency evolution without glitches. These anti-glitches (with $\delta \nu/\nu > -10^{-4}$) are the largest yet observed in any pulsar system, and may be a result of the lag between the rotation of the crust and superfluid interior, driven by the enormous spin-up rate (Table \ref{tb:combined}).
}
\end{center}
\end{figure}

The observed pulsing ULXs typically have periods ranging from $\sim$ 1 - 30s (see Table \ref{tb:combined}). As is apparent from Tables \ref{tb:combined}, the secular $\dot{P}$ in these systems (i.e. $\Delta P/\Delta T$ where $\Delta T$ is typically months or longer -- except in the case of NGC 300 ULX-1) reveals a collection of objects which tend to spin up at extremely high rates (notably NGC 300 ULX: \citealt{Carpano1805}) when compared to Galactic X-ray pulsars (although M82 X-2 is observed to spin {\it down} between 2014 and 2016: \citealt{Bachetti2020}). The secular spin up/down rates may differ greatly from the `instantaneous' $\dot{P}$ obtained over the course of an observation due to the influence of orbital dynamics and changes in accretion rate between observations (\citealt{Israel1702, Bachetti2020}). In addition to these long timescale trends, glitches (perhaps associated with the coupling of differentially rotating inner crust -- and possibly core -- to outer crust: \citealt{Anderson_Itoh1975}) have now been observed in M82 X-2 (\citealt{Bachetti2020} -- Figure 7). These imply a change in spin frequency $\delta \nu/\nu \approx 10^{-5}$, similar in magnitude to glitches seen in magnetars and a small number of accreting neutron stars (\citealt{Galloway2004,Dib2009,Dib_Kaspi2014,Archibald2016,Serim2017}). Interestingly, there are two (or possibly three) anti-glitches in NGC 300 ULX-1 (Figure 8, \citealt{Ray2019}), where the neutron star undergoes a dramatic spin-{\it down}  $\delta \nu/\nu > -10^{-4}$. This may be the result of the enormous spin-up rate creating a sizeable lag between the rotation of the crust and superfluid interior (\citealt{Ray2019}) and is supported by the lack of corresponding radiative changes during the glitch. Only spin-up glitches have been observed in non-accreting pulsars, but accreting pulsars (\citealt{Klochkov2009}) and magnetars show both spin-up and spin-down glitches (e.g. \citealt{Dib_Kaspi2014}). The anti-glitches seen in NGC 300 ULX-1 are the largest yet-observed in any kind of pulsar system.

\begin{figure*}
  \begin{center}
    \includegraphics[trim=450 0 450 0, clip, width=1.0\textwidth]{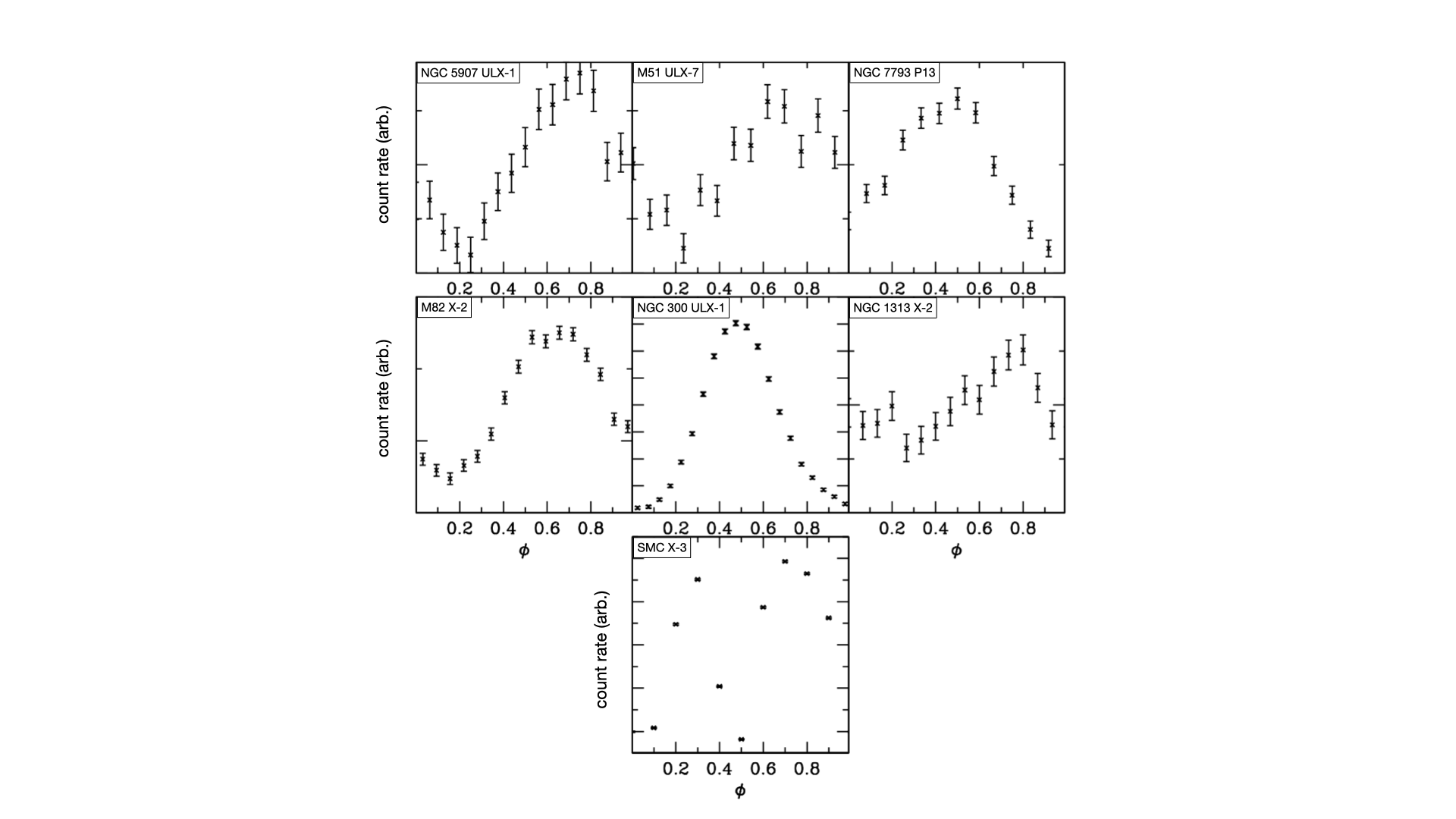}
  \end{center}
  \caption{Pulse profiles for the PULXs shown in Table \ref{tb:combined}, extracted from data taken by {\it XMM-Newton} (NGC 5907 ULX-1, M51 ULX-7, NGC 7793 P13, NGC 300 ULX-1, NGC 1313 X-1) and {\it NuSTAR} (M82 X-2 and SMC X-3). {Values for the count rates are not provided due to bandpass and detector differences, however, errors are provided for the purposes of comparing the level of constraints within each phase bin.}}
\end{figure*}

The known ULX pulsars have pulse fractions (defined as [Pulse max - Pulse min]/[Pulse max + Pulse min]) which increase with photon energy (as seen in Galactic X-ray pulsars: \citealt{Lutovinov2009}), and take values from $<$5-100 \% depending on the observation, source and energy band (see Table \ref{tb:combined} for more details). Notably, Figure 9 shows that several of the ULX pulsars have an a highly sinusoidal pulse profile (although both NGC 300 ULX-1 and SMC X-3 are notable exceptions), unlike that expected from a pencil beam geometry. This suggests a large emitting region, either a fan-beam geometry (\citealt{Basko_Sunyaev_1976}) or accretion curtain, the latter reprocessing (and therefore potentially diluting) the pulsed emission (\citealt{Mushtukov1705, Mushtukov1903}).

We can obtain a rough indication of the statistical significance of a given pulsed signal (with a width $\Delta nu$, fractional rms, $r$ -- related to the pulse fraction -- in an observation of length $T$, from a source of count rate $I$, with background count rate $B$) from $n_{\sigma} = 1/2~ [I^{2}/(I+B)]r^{2}\sqrt{T/\Delta \nu}$ (\citealt{Lewin1997}, noting that a true measure of the significance relies on considering both the noise and the number of independent free trials). Of the $\sim$ 1800 known ULXs (e.g. \citealt{Walton0122}), the vast majority do not have have sufficient data quality for deep pulse searches (down to small $r$). In those with sufficient data quality and not already identified as PULXs, pulsations appear to be absent down to the detection threshold (e.g. \citealt{Doroshenko2015, Israel1703}).  

There are several reasons why pulsations may be absent in a given ULX or in a given observation. Pulsations are observed to be transient in the known ULX pulsars (and indeed in Galactic accreting X-ray pulsars as well, e.g. \citealt{Kretschmar2000}) for reasons not yet well understood (see \citealt{Bachetti1410,Israel1702,Sathyaprakash1909,RodriguezCastillo2005, Bachetti2020}), and conditions during a given observation may not be favourable for detecting the pulsed emission from the column (and/or curtain). As an example, we may be prevented from observing the pulsations by dilution by a stronger stable component in the spectrum (\citealt{Walton2018P13}, e.g. if the disc/wind has precessed or accretion rate has changed \citealt{Pasham_M82_precessing_disc,Middleton1503,Middleton2015_resids,Middleton_2018_Lense_Thirring,Middleton_2019_Accretion_plane}). This can also happen if the neutron star has free-body precessed (\citealt{Fuerst2017_freebodyP}) so that the pulses are considerably weaker or beamed away from us. It may also be that we will {\it never} observe pulsations in some neutron star ULXs as we may never observe at suitable angles to view the accretion column. Alternatively the field may have been suppressed by accretion (e.g. \citealt{Urpin1995}), the neutron star and inner disc may be aligned (\citealt{King2003}), scattering in the wind cone may have suppressed pulsations on timescales shorter than the light-crossing time (\citealt{Mushtukov2011}), or some ULXs may simply harbour black holes (see \citealt{Middleton_King2017}. Binary evolution would seem to make the latter inevitable to some degree -- \citep{Wiktorowicz1904}.

\subsubsection{Long timescales}
\label{sec:longt}

It is well-accepted that LMXBs undergoing recurrent outbursts (see \citealt{Hameury1120}) may appear as ULXs for relatively short periods of time (e.g. \citealt{Middleton_2013_M31_microquazar,Carpano1805, vandenEijnden2018Nature}). Many bright ULXs  
are persistent (although we note the discovery of a new transient ULX in M51 reaching $10^{40}$ erg/s: \citealt{Brightman0620}), but their luminosities can fluctuate by orders of magnitude, periodically/quasi-periodically (on timescales of 10s of days: \citealt{Kaaret_2006_Modulation_M82,Kong_2016_Modulation_M82_X2,Walton_2016_Modulation_5907}) or seemingly at random. There are several possible explanations for such variations: changes in accretion rate, driven by effects in the outer disc {   (see Fig. \ref{fig:m51_xt}, Sec. \ref{sec:trans} and \citealt{Hameury1120,Middleton_thermal2021})}, an accreting neutron star entering a propeller phase, orbital variability as a neutron star passes through the decretion disc of a Be star, and precession of the disc/wind under external torques. In a number of the pulsating ULXs, the orbital period can be deduced from the Doppler modulation of the pulsations; this gives periods of a few days (see Table \ref{tb:combined}), with NGC 7793 P13 having a proposed period of 64 days (\citealt{Motch1410} although see \citealt{Fuerst_P13_2021} for details regarding the periodicities in this ULX). Other than NGC 7793 P13, the $\sim$ day orbital periods are considerably shorter than the `super-orbital' periods typically revealed by {\it Swift} monitoring but in some cases observed in the optical/UV (e.g. NGC 7793 P13: \citealt{Fuerst_P13_2021}). Both sets of periods are reported in Table \ref{tb:combined}.

Precession has been proposed as one means by which to explain the $\sim$ 10s of days super-orbital variability of ULXs by a number of authors. \cite{Pasham_M82_precessing_disc} suggested that the changes in the spectrum of M82 X-1 associated with a 62 day period (\citealt{Kaaret_2006_Modulation_M82, Kaaret_Feng2007}) could result from changes in the projected area of the inner disc (perhaps driven by a radiative warp: \citealt{Pringle_1996_warping_disc}). \cite{Middleton2015_resids} suggested that changes in the strength of atomic features, now known to be associated with the wind (\citealt{Pinto_2016Natur.533...64P,Pinto2020_1313}), could result from precession of the disc and wind in NGC 1313 X-1, and \cite{Luangtip2016_HoIX} appeal to a combination of precession and changes in accretion rate in order to explain the changing appearance of Ho IX X-1.

There are a number of possible mechanisms for driving a changing view of the inner regions of ULXs, including precession of a neutron star's magnetic dipole field (\citealt{1980_Lipunov_SvAL....6...14L}):

\begin{equation}
    t_{\rm mag,prec} \approx 1.5\times 10^{4} \frac{\mu_{30}^{-2}I_{\rm 45}R_{\rm M,8}^{3}P^{-1}}{\cos\phi(3\cos\zeta-1)}~ {\rm years}
\end{equation}

\noindent where $\mu_{30} = B_{12}R_{6}^{3}$, $I_{\rm 45}$ is the neutron star moment-of-inertia, $\phi$ is the angle between the normal to the disc plane and the spin axis, and $\zeta$ is the angle between the magnetic axis and the spin axis. To match observations of ULX superorbital periods of months or less, field strengths $> 10^{14} G$ are required (\citealt{Mushtukov1705, Vasilopoulos2020}).

Should the accretion flow be misaligned with the neutron star or black hole spin axis (which seems quite plausible: \citealt{King2003}) then frame-dragging of the fluid induces a torque. 
Similar to the hot inner flow of X-ray binaries, a thick super-critical disc/wind is expected to have an effective viscosity parameter $\alpha$ $< H/R$. The frame-dragging induced torque is then expected to be communicated as a bending wave from the inner edge of the flow out to the photospheric radius (\citealt{Middleton_2018_Lense_Thirring,Middleton_2019_Accretion_plane}) which may then induce solid body precession (\citealt{Fragile_2007_tilted_disc}) on a estimated timescale of:

\begin{equation}
    t_{\rm LT,prec} \sim \frac{GM\pi}{3c^{3}a_{*}}r_{\rm sph}^{3}\left[\frac{1-\left(\frac{r_{\rm in}}{r_{\rm sph}}\right)^{3}}{\ln{\left(\frac{r_{\rm sph}}{r_{\rm in}}\right)}}\right]~~{\rm s}
\end{equation}

\noindent for the disc and

\begin{equation}
    t_{\rm LT,prec} \sim \frac{GM\pi}{3c^{3}a_{*}}r_{\rm ph}^{3}\left[\frac{1-\left(\frac{r_{\rm in}}{r_{\rm ph}}\right)^{3}}{\ln{\left(\frac{r_{\rm ph}}{r_{\rm in}}\right)}}\right]~~{\rm s}
\end{equation}

\noindent for the wind (here $r$ is in units of gravitational radius -- $GM/c^{2}$ -- and $a_{*}$ is the dimensionless spin parameter). Both periods are sensitive to the mass accretion rate. This can be related to properties of the X-ray spectrum and may tell us the nature of the accretor (neutron star or black hole, \citealt{Middleton_2019_Accretion_plane}) without the requirement for pulsations which are themselves transient or may be diluted beyond detection (\citealt{Bachetti1410, Mushtukov2011}). 

If the neutron star is asymmetric about its rotation axis, free-body precession may also occur on a timescale of (\citealt{Jones2001}):

\begin{equation}
    t_{\rm free,prec} \approx \frac{2.32\times 10^{-5}}{\epsilon}\frac{P_{\rm NS}}{1s}\frac{0.5}{\cos\chi}~ {\rm days}
\end{equation}

\noindent where $\chi$ is the angle of the misaligned distortion of the neutron star relative to its rotation axis, and $\epsilon$ is the measure of distortion, related to the surface magnetic field strength through (\citealt{Lander2009}): 

\begin{equation}
\epsilon = 2\times10^{-11}\left(\frac{B_{\rm surf}}{10^{12}G}\right)^{2}
\end{equation}

\noindent which, notably, \cite{Vasilopoulos2020} used to estimate the dipole field strength in M51 ULX-7.

Besides the above torques, other effects may warp and precess the disc, e.g tidal torques from the secondary star (although these are unlikely to be strong in cases with known $\sim$ day orbital periods, e.g. \citealt{Bachetti1410}), magnetic warping (\citealt{Lai2003,Pfeiffer2004}), and radiative warping (\citealt{Pringle_1996_warping_disc}). In the latter case, where super-critical flows -- specifically a large-scale height disc and optically thick wind -- are present, it is not clear how efficiently the outer disc is irradiated (see also \citealt{Middleton_thermal2021}). But if the dipole field is strong and/or the accretion rate low, irradiation may well be possible, and similar warps and precession to that seen in Her X-1 are then quite plausible (see \citealt{Petterson_1977_Her_X1_Precession,Kotze2012}). Regardless of the mechanism driving the precession, it is possible to build a simplified model of the motion of the cone and then numerically extract the X-ray lightcurve. \cite{Dauser1704} developed a model which can produce a variety of lightcurve profiles (available within {\sc xspec}: \citealt{Arnaud1999} as {\sc ulxlc}), under the assumption that the X-ray source lies at the apex of a cone of outflowing, totally opaque, achromatically scattering plasma. Applying this model to the lightcurve of NGC 5907 ULX-1 implies a small total opening angle ($\sim 10-15\deg$) and substantial beaming (amplification by a factor 60-90) although it is as yet unclear whether this is in conflict with scattering reducing the observed pulse fraction (\citealt{Mushtukov2011}).

It is interesting to consider how the different physical origins for superorbital periods can be distinguished. A natural method would be to consider how the timescales respond to the accretion rate and spin-up/down of the neutron star. Equations 14 \& 15 are highly sensitive to the accretion rate (far more so than the spin) and so, where the accretion rate is variable (on long timescales), we would expect variability in the superorbital period (with correlated changes in the X-ray spectrum). Conversely, magnetic and free-body precession is more sensitive to the spin than accretion rate, and the superorbital period should therefore track changes in the neutron star's spin over time.

It is also worth noting that plotting the superorbital period versus the orbital period for Galactic systems and those ULXs with known periods (see Table \ref{tb:combined}, implies a potential correlation (see \citealt{Townsend_2020}), although care must be taken in interpreting any correlation between logarithmically scaled values which include systems known to be of very different types. It is plausible that the superorbital periods are in some cases responding to tidal interactions, while the mass supply to the compact object may correlate with the binary period, leading to a positive correlation through e.g. equation 15. This currently remains an open question but we can expect progress as the detected number of constrained superorbital periods increases.

In addition to the smooth, periodic/quasi-periodic variations seen in the X-ray and optical bands, there are a variety of changes in the lightcurves of ULXs which can help reveal the nature of the system. Dips are found in NGC 55 ULX-1 (\citealt{Stobbart2004_NGC55}); NGC 5408 X-1 (\citealt{Grise2013_NGC5408_dips}), NGC 247 ULX-1 (\citealt{Alston2021_dips_NGC247}), M51 ULX-7 (a known ULX pulsar, \citealt{Vasilopoulos0221}), and J125048.6+410743 (\citealt{Lin2013_dips}). Significantly, these ULXs tend to have relatively soft spectra (M51 ULX-7 is somewhat harder), possibly implying a correlation with sight-lines intercepting regions in the wind subject to instabilities and clumping (e.g. \citealt{Middleton2011, Middleton1503, Takeuchi2013}).

Whilst dips are seen in the lightcurves of a small number of ULXs, eclipses seem to be mostly absent from the currently known ULX population (\citealt{Middleton_King2016}). This would make sense if edge-on ULXs emit around the Eddington limit (\citealt{Poutanen0705}) implying sub-ULX luminosities in many cases (\citealt{Zhou2019}). However, there have been reports of eclipses in two ULXs in M51 (\citealt{Urquhart_Soria2016_eclipses, Wang2018_eclipses}). These give constraints on the angle between the companion star's orbit and the X-ray bright inner regions (noting that these do not have to be aligned in a single plane). If there is an estimate of the mass of the donor star this then constrains the compact object mass (\citealt{Wang2018_eclipses}). 

Neutron stars probably account for the majority of observed ULXs (\citealt{King1605, Middleton_King2017, King1702, Walton1804, Wiktorowicz1904}). Such systems are likely to approach spin equilibrium between gravity and centrifugal forces during their lifetimes, where the co-rotation radius approaches the magnetospheric radius. Whilst the effective torque vanishes when the two are close \citep{Dai0506}, should some process push the magnetospheric radius outside the corotation radius, a propeller state ensues where the emission from the accretion column and curtain -- but importantly not from radii down to the magnetospheric radius --  
 is switched off. 
 In systems where the magnetospheric radius is large, the result would be a severe drop in flux (although centrifugally driven winds from around the magnetospheric radius would remain). Such propeller states, widely observed in Galactic systems (e.g. \citealt{Cui1997,Campana2008, Tsygankov2016_galactic}) have been invoked to explain the large changes in luminosity seen in M82 X-2 (\citealt{Tsygankov2016}). However, in  NGC 7793 P13 (\citealt{Fuerst_P13_2021}) and NGC 300 ULX-1 (\citealt{Vasilopoulos1910}), the spin has continued increasing at the same rate when the source is considerably fainter in the X-rays, implying that some other mechanism -- possibly obscuration -- drives the change, rather than a propeller state. (In M51 ULX-7 it is at yet unclear whether a similar X-ray off state implies a propeller state or is associated with precession: \citealt{Vasilopoulos0221}). The predicted large-scale changes in X-ray flux in the propeller state have motivated searches for pulsar candidates among known ULXs. \cite{2018_Earnshaw_propeller_MNRAS.476.4272E} searched within those ULXs in the 3XMM DR4 catalogue (\citealt{Rosen2016_3XMM}) finding five objects which underwent a factor of 10 or more change in flux (with one propeller candidate: M51 ULX-4). \cite{Song2020_propellor} compiled {\it XMM-Newton}, {\it Swift} and {\it Chandra} flux lightcurves of 296 ULXs from the \cite{Earnshaw_ULX_cat} catalogue, and found that 25 out of 296 ULXs changed flux by more than a factor 10 (based on the available sampling), 17 of which show a bi-modal flux distribution, potentially consistent with sources at spin equilibrium.

\subsubsection{Spectral-Timing studies}
\label{sec:spectt}

As well as standalone studies focussing on either the spectral evolution or timing characteristics of ULXs, there are an increasing number which bring the two together. Such a spectral-timing analysis can be performed in the time or Fourier domain in a given energy band (e.g. the fractional rms or rms spectrum: \citealt{Revnivtsev1999_rmsspectra, Gilfanov2000M_rmsspectra}, and the covariance spectrum: \citealt{Wilkinson2009_CV}), or via lags between bands (via the CCF or Fourier lag spectrum) as a function of energy or frequency (see \citealt{Uttley2014} for a comprehensive review). Spectral-timing studies require high signal-to-noise or large amounts of variability (such that the Poisson noise is not dominant), making such analyses challenging for ULXs. However, there have been a number of successful attempts which have provided otherwise inaccessible insights. 

\begin{figure}[ht]
\begin{center}
\includegraphics[width=1\textwidth]{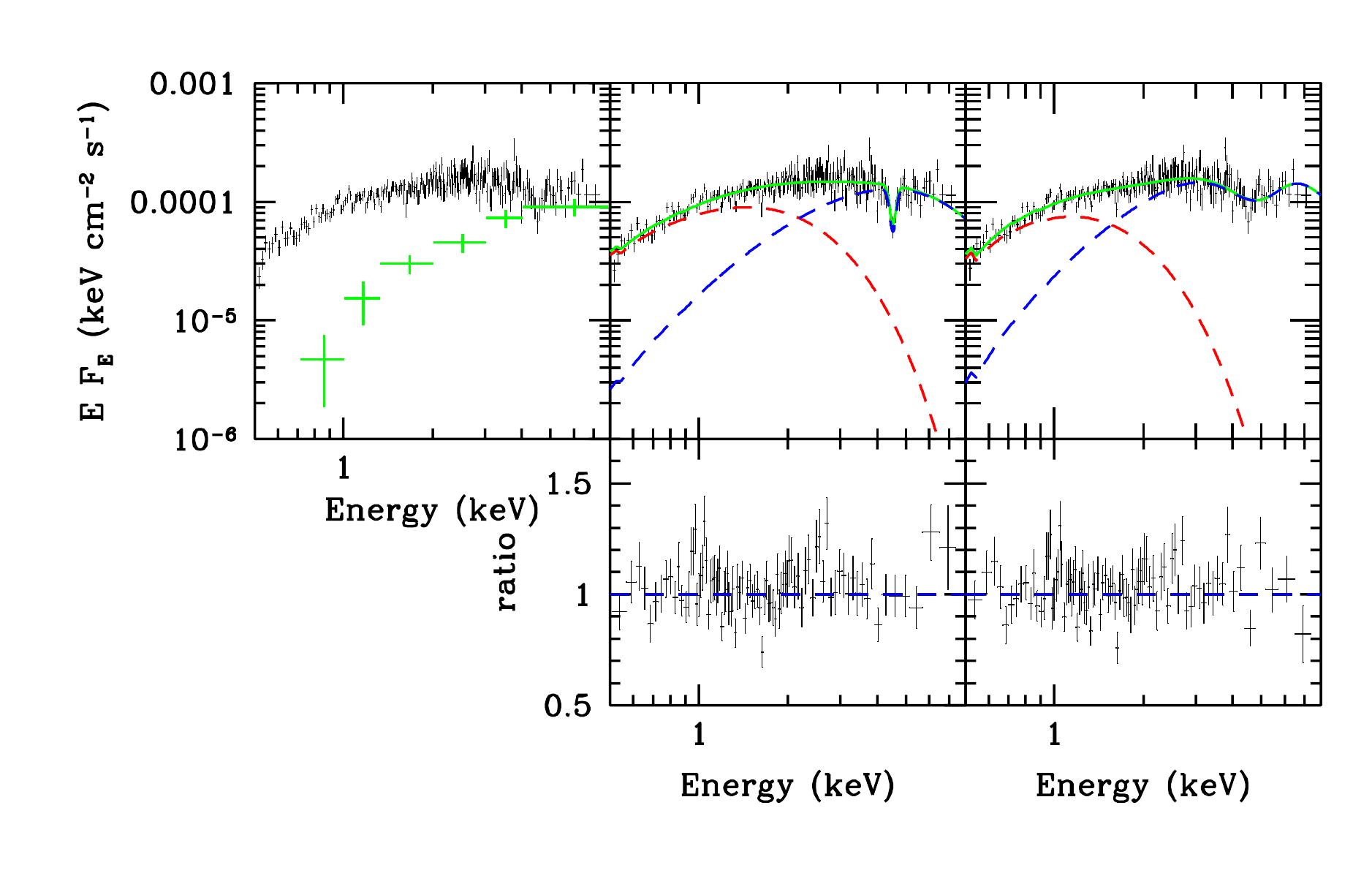}
\caption {From \cite{Middleton_Brightman_2019_M51} showing (left-hand panel) the covariance spectrum of M51 ULX-8 (in green) with the time--averaged {\it Chandra} data (in black). The covariance picks out the linearly correlated variability (see \citealt{Wilkinson2009_CV}) and allows the spectrum to be decomposed. In this case, the additional soft component (shown in red in the middle and right-hand panels) is above the Eddington limit for a neutron star and so inconsistent with a thin disc. This allowed \cite{Middleton_Brightman_2019_M51} to place indirect constraints on the dipole magnetic field strength (a higher-order multipole field is not excluded).
}
\end{center}
\end{figure}

Covariance spectra (\citealt{Wilkinson2009_CV}) have an advantage over rms spectra; as the Poisson noise between bands is uncorrelated, when the variability is weighted by the linear coherence between bands (see \citealt{Vaughan_Nowak1997,Uttley2014}), the constraints on the variability in each energy bin are improved. \cite{Middleton1503} presented the covariance spectra for a sample of variable ULXs with relatively high data quality and showed that the variability on relatively short ($\lesssim 1$ ks) timescales matched well to the shape of the hard component (seen also the studies of the covariance spectra of NGC 55 ULX-1: \citealt{Pinto2017unified} and NGC 1313 X-1: \citealt{Kara2020_1313}). This can be readily explained if the variability emerges from only this component on the timescales being studied. However, it is less clear why spectrally-harder ULXs (where this component is dominant) appear to be less variable on similar timescales (e.g. \citealt{Heil2009_sample}). One possible explanation is that the variability is extrinsic and being driven by obscuration by optically thick clumps of wind (\citealt{Middleton2011}), probably created due to fluid instabilities (\citealt{Takeuchi2013}). Clumps may also then be responsible for dips seen in the softer ULXs (e.g. NGC 55 ULX-1: \citealt{Stobbart2004_NGC55} and NGC 247: \citealt{Alston2021_dips_NGC247}) which in the standard paradigm are viewed at higher inclinations (\citealt{Poutanen0705,Middleton1503}). 

The covariance spectrum can also be used to isolate the {\it non-variable} component in ULXs. This can be modelled and better compared to theories for emission at soft energies (typically either a thin disc beyond the magnetospheric radius or emission from close to the spherisation radius). This can be complex, depending on the role of variable absorption, but it is at least possible to compare luminosities, and rule out the presence of a thin disc. This was demonstrated in \cite{Middleton_Brightman_2019_M51} (Figure 10), and allowed for independent constraints to be placed on the dipole field strength in M51 ULX-7 (see also discussions in \citealt{Brightman1804}, and \citealt{Christodoulou2019}). 

Time (or phase) lags between bands may be extracted in the time domain via the CCF (as shown to great effect when studying a variety of features in XRBs, e.g. lags corresponding to propagation in jets: \citealt{Gandhi2010}) or via the Fourier cross-spectrum (see \citealt{Uttley2014} for a detailed step-by-step guide). The lags are related to the nature and geometry of the accretion flow via the transfer function and are imprinted by propagation, and the light-travel time through reverberation or transmission (the latter affected by the amount and state of intervening gas). Lags in ULXs were first identified in a 2006 observation of NGC 5408 X-1 (\citealt{Heil2010}) between a soft (0.3 - 1 keV) and hard (1 - 10 keV) band. This `soft lag' is $\sim$ a few seconds in NGC 5408 X-1 (\citealt{DeMarco2013}) at frequencies of $\sim$ 10 mHz, is $\approx$ 150s at 0.1 mHz in NGC 1313 X-1 (\citealt{Kara2020_1313}), and is a few hundred seconds at around the same frequency in NGC 55 ULX-1 (\citealt{Pinto2017unified}). Soft lags have been interpreted as reprocessing in distant material (presumably the wind) and intriguingly appear around the QPO frequencies seen in NGC 5408 X-1 (\citealt{DeMarco2013}). Should the lag be imprinted as a consequence of resonant line scattering  or Compton down-scattering, the magnitude can provide an estimate of the optical depth and thickness of the wind. An alternative explanation is that hard photons created in the accretion column are Comptonised by a surrounding envelope (\citealt{Mushtukov1705}) and a lag is imprinted due to the light travel time across the curtain and the time taken to Compton scatter (\citealt{Mushtukov1903}). This picture would imply very large or dense curtains in the case of ULXs such as NGC 55 ULX-1.

\subsection{Multi-wavelength studies}
\label{sec:multiwave}

ULXs are defined (albeit empirically) by their X-ray luminosity, although much of the gravitational energy dissipated in the accretion flow may emerge in other wavebands through reprocessing 
or as kinetic luminosity in the wind and/or jet. It is therefore crucial to obtain as broad an SED as possible, and below we discuss insights obtained in several key bands (note that we discuss observations of bubble nebulae surrounding ULXs separately in Section 2.6.1).

\subsubsection{Radio}
\label{sec:radio}

Radio monitoring campaigns (\citealt{Koerding2005}) and cross-matching of X-ray catalogues with radio surveys (\citealt{Perez-Ramirez2011}) have returned limited numbers of ULX radio counterparts, perhaps suggesting that any jets are inherently weak or transient in nature. Targeted observations have tended to focus on the brightest ULXs (HLXs at $\sim 10^{41}$ erg/s) as these remain the most promising candidates for hosting IMBHs other than AGN in dwarf galaxies (\citealt{Farrell0907}). Their suggested sub-Eddington accretion rates would naturally lead to the expectation of flat spectrum radio emission associated with a compact jet (if in an analogous hard X-ray state as seen in X-ray binaries and AGN). Placing such HLXs on the radio--X-ray fundamental plane (\citealt{Merloni2003_FP,Falcke2004_FP,Gultekin2009_FP}) opens up the possibility of constraining the compact object mass, and this approach has been adopted for a number of ULXs (NGC 4088-X1: \citealt{Mezcua2011}, NGC2276-3c: \citealt{Mezcua2015}; N5457-X9: \citealt{Mezcua2013}, IC 342 X-1: \citealt{Cseh2012_IC342}).   

Confirming the presence of a jet in ULXs can be more difficult given the presence of a radio bright nebula in some cases. These can be up to 100s of pc across: \citealt{Mezcua2013_nebula}), potentially analogous to the W50 nebula surrounding SS433 
(e.g.  
\citealt{Pakull0601,Pakull_Grise2008,Russell2011,Cseh2012_IC342}), and energised by termination shocks between the surrounding gas and outflows. In only a single bright ULX -- Ho II X-1 -- has the nebula emission been resolved by using VLBI into discrete core and lobe emission (\citealt{Cseh_2014_HoII_jets,Cseh2015_HoII}, although there are also twin lobes either side of a bright radio core in the IMBH candidate NGC~2276-3c, see \citealt{Mezcua2015}). The fading of Ho II X-1's lobes over a period of around 2 years indicates that the jets are sporadic and may have more in common with the discrete ejections in near-Eddington LMXBs, placing a limit on the mass of the compact object in Ho II X-1 of $>$ 25 M$_{\odot}$ (\citealt{Cseh2015_HoII}). 

Jet-inflated nebulae may provide one of the few ways to locate edge-on ULXs. A key example is the radio-optical nebula in M83 (30 times more luminous than W50 at 5 GHz) and collimated jets in NGC 7793 (S26: \citealt{Pakull0710,Soria2010_S26}). In both cases, the estimated jet power lies above 10$^{40}$ erg/s, but the observed X-ray luminosity to-date is 3-4 orders of magnitude lower, and the sources would therefore not qualify as ULXs.
These systems are in many ways similar to SS433 (albeit with even higher jet kinetic luminosities) and provide ample evidence for similar edge-on ULXs being located in nearby galaxies. 

As well as jets associated with highly super-Eddington rates of mass transfer, transient jet ejections have been observed in some Galactic XRBs moving to less extreme, near-Eddington accretion rates (\citealt{Fender2009}). These have been successfully detected in nearby galaxies and can be associated with the appearance of a low luminosity ULX (e.g. M31: \citealt{Middleton_2013_M31_microquazar}). It is plausible that the transient jets detected from HLX-1 (\citealt{Webb0812}) suggest the source was moving to near-Eddington rates, which, at a luminosity of $>$ 10$^{42}$ erg/s would imply a BH mass of $\sim$10,000 M$_{\odot}$, consistent with the mass estimate from modelling the X-ray spectrum with relativistic disc models (\citealt{Davis2011_HLX_disc}). In future, with the advent of sensitive all-sky radio monitoring (SKA), the detection of such radio events should become commonplace. 

Finally, it is intriguing to note that a jet event has also been detected from the transient X-ray pulsar, Swift J0243.6+6124 -- which contains a neutron star with B $>$ $10^{12}$ G (\citealt{Doroshenko1805}) -- when accreting above its Eddington limit (\citealt{vandenEijnden2018Nature}). At peak radio brightness, the flux density from the jet was $<$ 100$\mu$Jy at 6 GHz (at an assumed  distance $\approx $ 7 kpc -- \citealt{vandenEijnden2018Nature}). Similar jets are therefore unlikely to be discovered in nearby galaxies (although there is the theoretical potential to drive more powerful jets through tapping the neutron star's angular momentum: \citealt{Parfrey2016}).

\subsection{Infra-red to ultraviolet}
\label{sec:iruv}

Generally speaking, observations of accreting systems at energies intermediate to those of X-rays and radio allow us to study the impact of reprocessing, the highest energy populations of electrons in jets and the nature of the companion star. In ULXs, such studies are of vital importance for constraining the true energy budget and age of the system, for understanding the impact on the surroundings and how such high rates of mass transfer actually occur.

\subsubsection{Infrared}
\label{sec:ir}

Campaigns studying ULXs in the IR have focused mostly on the nature of the donor star, using a combination of imaging and spectroscopy from ground-based instruments. Initial imaging studies by \cite{Heida2014} identified a number of possible red supergiant (RSG) companion stars (11/62 ULXs in 37 galaxies within 10 Mpc) from a combination of data from WHT/LIRIS, MMT/SWIRC and VLT/ISAAC. \cite{Heida2016} later performed NIR spectroscopy using Keck/MOSFIRE (Multi-Object Spectrometer for Infra-Red Exploration), identifying candidate counterparts for two ULXs (one in NGC 925 and one in NGC 4136) with RSGs. Building on the work of \cite{Heida2016}, a large study by \citealt{Lopez2017,Lopez2020} made a systematic search for NIR counterparts to 113 ULXs, through use of a combination of instruments (imaging via LIRIS/WHT, SOFI/NTT and WIRC on the Palomar Hale 5-m telescope, and spectroscopy with Keck/MOSFIRE). This study yielded candidate counterparts for 38 ULXs, 5 of which are identified as RSGs. The numbers obtained from ground-based studies match well to mid-IR studies performed using {\it Spitzer}, where \cite{Lau2019} located 12 candidate RSGs in a sample of 96 ULXs within 10 Mpc. These results point towards a highly probable association of RSGs with some ULXs, and an additional -- somewhat less considered -- form of mass transfer, i.e. that of wind-fed Roche-lobe overflow (e.g. \citealt{Copperwheat2007,ElMellah2019}) which has now been incorporated into models for binary population synthesis (\citealt{Wiktorowicz0221}).

There have been a number of focused campaigns to explore the brightest ULXs, and specifically the PULXs (determining the companion star in the latter being of obvious value when considering the evolution of these systems). Unfortunately for most PULXs the IR fields are extremely crowded, precluding the identification of a unique counterpart (\citealt{Heida2019a}). However the companion of the SN imposter, NGC 300 ULX-1 has been identified as an RSG (\citealt{Heida1910}).

As well as emission from the companion star, it is quite possible that synchrotron emission from jets and from circumbinary dust could both contribute to observed flux in the IR band. \cite{Lau2019} identify a mid-IR excess and a probable circumbinary disc in several ULXs (including Ho II X-1), and associate variable mid-IR emission from the X-ray spectrally hard ULX, Ho IX X-1, with a jet (as previously suggested by \citealt{Dudik2016HOIXjet} and \citealt{Sathyaprakash2022} in the case of the pulsating ULX, NGC 1313 X-2). 

Indirect constraints on the SED which photoionizes gas surrounding the ULX (see Section 2.6) can be extracted through modelling IR lines. \cite{Berghea2010a} identified Ho II X-1's nebula (thought to be inflated via jets) in {\it Spitzer} observations and, in \cite{Berghea2010b} later performed photoionization modelling to suggest that the lines are consistent with irradiation by the observed soft X-ray and UV emission, requiring little to no beaming (of these components). This is completely consistent with the picture where the soft emission is less beamed than the hard X-rays. The MF16 nebula surrounding NGC 6946 X-1 has also been detected by {\it Spitzer}, with photo-ionisation modelling by \cite{Berghea2012} indicating the possible presence of an O-type supergiant donor star.

 Roche lobe, this mass transfer mechanism known as "wind Roche lobe overflow" can remain stable even for large donor-star-to-accretor mass ratios

\subsubsection{Optical}
\label{sec:opt}

Identifying a unique optical counterpart to a ULX potentially gives a powerful probe of the system age via binary evolution arguments. If atomic lines can be isolated and associated with the companion, then subsequent studies may allow the mass of the compact object itself to be constrained by standard radial velocity techniques. There are a number of issues which make such investigations challenging. Given the distance to many ULXs, there may be several optical counterparts associated with the X-ray astrometric position (e.g. \citealt{Gladstone2013}). The distance can make lower mass companion stars undetectable, perhaps misleadingly favouring an association with any high mass star within the neighbourhood of the ULX. Further, the photometric signal of the star may be contaminated by direct emission from the wind photosphere depending on the accretion rate (\citealt{Poutanen0705}), and emission from the uncovered thin disc which may be increased by irradiation (should the geometry, advection and wind optical depth permit this -- \citealt{Sutton2014_irraddisc,Yao_Feng2019_discirrad}). Finally the spectroscopic signatures (e.g. He II emission lines) may also be contaminated or dominated by the wind (\citealt{Fabrika2015}).

Despite these challenges, there have been a number of attempts to find and study the optical counterparts of ULXs (besides the results discussed in the IR band above). Many studies (too many to recount in this review) over the last 20 years have targeted individual ULXs (see e.g. \citealt{Roberts2001_optical}), whilst others have aimed at describing the larger population. 
As important examples of the latter, the sensitivity and angular resolution of HST allowed \cite{Roberts2008} to locate counterparts to four out of six ULXs, whilst \cite{Tao2011} studied the optical counterparts of 13 ULXs, finding the optical flux and colour to vary on the timescale of days to years, ruling out changes in extinction as the origin. \cite{Gladstone2013} utilised HST photometry to study 33 ULXs with {\it Chandra} positions, locating 13 $\pm$ 5 counterparts and finding O--star companions to be rare. 

Where unique counterparts have been identified, there have been attempts to constrain the mass and species of compact object via dynamical arguments (most using the broadened He II $\lambda4686$ emission line):
\begin{itemize}
    \item Using {\it Gemini} observations, \cite{Roberts2011} attempted to constrain the mass of the compact object powering NGC 1313 X-2 (which is now known to be a neutron star: \citealt{Sathyaprakash1909}) but were unable to do so due to the absence of sinusoidal motion in the line velocities
    \item \cite{Cseh2013MNRAS} were unable to locate an orbital period, but placed a limit on the mass ($< 510 M_{\odot}$) of the compact object in NGC 5408 X-1 from a limit on the semi-amplitude in the absence of periodic modulations, and under the assumption that the He II line is produced in the irradiated outer accretion disc.
    \item \cite{Liu2013_M101} used the He II $\lambda4686$ line to place a lower limit 
    on the mass of the compact object in M101 ULX-1 of 5M$_{\odot}$.
    \item \cite{Motch1410} placed an upper limit on the mass of NGC 7793 P13 ($<$ 15 M$_{\odot}$) by modelling the $\sim$ 64 d period in the V and U band lightcurves as irradiation of the well-identified B9Ia companion star; this result is of course consistent with the discovery that a neutron star powers this ULX (\citealt{Furst1611,Israel1703}).
\end{itemize}
  
Use of the He II $\lambda4686$ emission line to determine the compact object mass via dynamical means is problematic, as \cite{Fabrika2015} show that this line in ULXs probably originates in the outer regions of an outflowing wind, as in SS433. There remains a question about the width of the He II lines in NGC 7793 P13 -- these cannot originate from the companion star (\citealt{Fabrika2015}) but also appear to be modulated on the same 64 d timescale as the optical flux (\citealt{Motch1410}); this may point towards the wind precessing as it does in SS433 (see Section 2.3.3 for a discussion of super-orbital periods).

\subsubsection{UV}
\label{sec:uv}

Models which consider the temperature dependence of the super-critical inflow (\citealt{Poutanen0705}) already imply that, for high mass accretion rates, the outer photosphere of the wind should emit in the UV around the Eddington limit for the compact object. Compton down-scattering of photons from the innermost regions could also produce extreme luminosities in the optical/UV band. This may explain the inference that SS433 emits at over 10$^{40}$ erg/s in the UV (\citealt{Dolan1997,Waisberg2019}).

UV observations of ULXs are limited by the difficulties inherent in observing in much of this band, but a particularly notable result is that of \cite{Kaaret2010_UV} who observed NGC 6946 X-1 with HST, finding a spectrally soft ULX showing strong variability and QPOs, surrounded by a photoionized and shock-heated optical nebula (MF16 -- see below). Spectral fitting carried out by these authors implied a UV luminosity in excess of 10$^{39}$ erg/s, consistent with the luminosity needed to produce the observed HeII $\lambda$4686 emission-line luminosity (\citealt{Abolmasov2008}). 

More recently, monitoring of ULXs with {\it Swift} has started to provide hints of how the X-rays (via the XRT) and UV (via UVOT) may be correlated (or anti-correlated). As shown in \cite{Fuerst_P13_2021}, the $\sim$ 64 d optical period observed in NGC 7793 P13 (\citealt{Motch1410}), is also seen in the X-rays (with a precise value of 65.21$\pm$0.15, see also \citealt{Hu2017}), in the UV (with a period of 63.75$^{+0.17}_{-0.12}$ d), and in the fits to the changing pulse period, yielding an orbital period of 64.86$\pm$0.19 d. It is clear from Figure 11 that the UV brightness is not in phase with the X-rays (and indeed, the UV modulation is considerably stronger when the X-rays are dim).

\begin{figure}
    \includegraphics[width=\columnwidth]{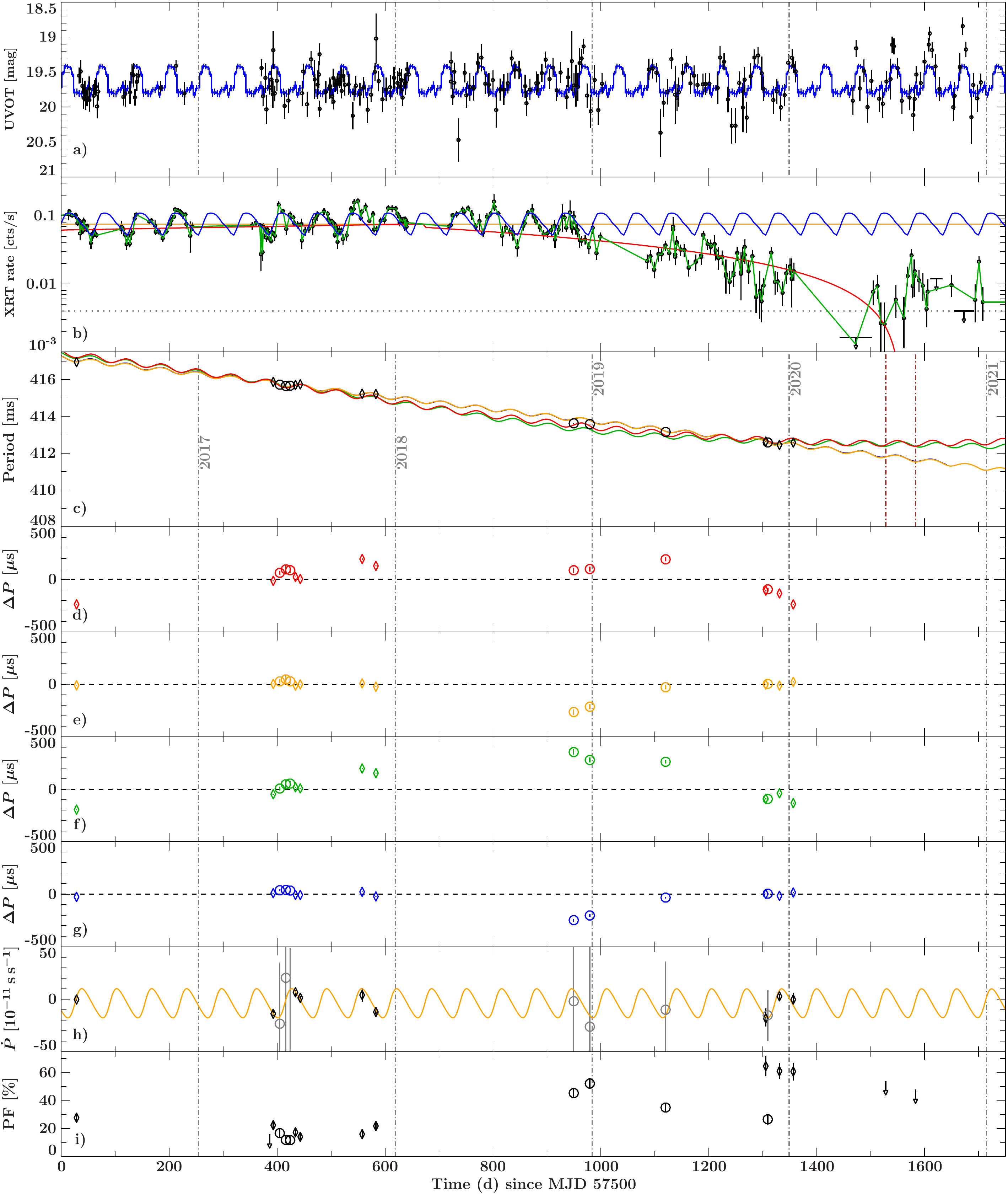}
    \caption{from \cite{Fuerst_P13_2021} showing the long term {\it Swift} X-ray (XRT) and UV (UVOT, U band) lightcurve of the ULX pulsar, NGC 7793 P13, with modulations around a $\sim$65 d period. The drop in X-ray flux in 2019 does not appear to result in a substantial decrease in the spin-up rate as indicated by the lower panels (see \citealt{Fuerst_P13_2021} for details). It is also intriguing that the UV and X-ray modulations appear to be -- to some extent -- anti-correlated.     }
    \label{fig:temperature}
\end{figure}

\subsection{The environment of ULXs}
\label{sec:envin}

ULXs are a product of their host environment, and interact with it through kinetic and radiative feedback. In this section we discuss the small and larger scale environments of ULXs and the insights these provide.

\subsubsection{ULX nebulae}
\label{sec:small}

As first reported by \cite{Pakull0202}, a substantial number of ULXs are observed to be surrounded by a bubble of warm gas (although the phenomenology is in reality somewhat varied), visible in the optical via emission lines (and in some cases detected in the radio as well). As indicated in Table 1, these `ULX  nebulae' are typically $\sim$10 - 100s pc across (the largest presently known being $\sim$500 pc in diameter) and tend to be considerably larger than typical SNRs (e.g. \citealt{Asvarov2014}) although there is sometimes disagreement on the actual size (e.g. \citealt{Moon2011} versus \citealt{Kaaret2004_HoII_nebula}). Bearing this point in mind, the dimensions provided in Table 1 do not necessarily account for the entire region influenced by the ULX (e.g. there is a reported 800 pc zone of weakly ionised gas surrounding NGC 1313 X-1: \citealt{Pakull0202}). 

Analogous to the W50 nebula of SS433, we expect gas from the original SNR to be inflated by jets and winds from the inner accretion flow (e.g. \citealt{Feng_Kaaret2008,Pakull_Grise2008}) and be photoionized by the UV--X-ray radiation field (which we broadly expect to be anisotropic). Indeed, ULX nebulae show a mixture of photoionized (notably He II 4686$\lambda$) and shock ionized lines ([OI] 6300$\lambda$ and [OIII] 5007$\lambda$) -- some nebulae appearing to be dominated by one process over the other (although both probably occur in any given source e.g. \citealt{Abolmasov2007}). ULX nebulae have conclusively been shown to not be inflated by O star associations (i.e. a superbubble) and are unlikely to be due to a hypernova explosion (\citealt{Pakull_Grise2008}). 

Applying a model where the ULX nebula is inflated via the mechanical action of outflows (i.e. a shock dominated rather than photo-ionised dominated nebula) gives interesting estimates for the age of the system. This in turn provides clues as to how the binary may have evolved, and for how long super-critical accretion can be maintained. From \cite{Weaver1977} 
the radius of the bubble is (compare Eq. (\ref{nebularradius}))

\begin{equation}
    R = \left(\frac{125}{154\pi}\right)^{1/5}\left(\frac{L_{\rm tot}}{\rho_{0}}\right)^{1/5}t^{3/5}
    \label{neb1}
\end{equation}
where R is the radius of the bubble at time $t$, $L_{\rm tot}$ is the mechanical luminosity of the wind and/or jet, and the ISM density (assumed to be uniform) is $\rho_{0} = \mu m_{\rm p}n$, where $\mu$ is the mean atomic weight of the gas, $m_{\rm p}$ is the proton mass and $n$ is the hydrogen number density. The characteristic age of the bubble is then given by $\tau = 3R/5v_{\rm exp}$.

These equations allowed \cite{Cseh2012_IC342} to determine the energy content and age of several ULX nebulae, the results of which are shown in Table 1. 
Assuming $L_{\rm k} = 1/2 \dot{m}_{\rm wind}v_{\rm wind}^{2}$, and that the wind provides the entire energy reservoir of the bubble 
(i.e. provides $L_{\rm tot}$), then equation 18 may alternatively be written in terms of the wind properties (e.g. \citealt{Pinto_winds_thermalbalance}):

\begin{equation}
    R = 0.76\left(\frac{\dot{m}_{\rm wind}v_{\rm wind}^{2}}{2\rho_{0}}\right)^{1/5}t^{3/5}
\end{equation}

\noindent where $\dot{m}_{\rm wind}$ is the mass loss rate in the wind and $v_{\rm wind}$ is the wind velocity. For measured wind velocities in ULXs (e.g. \citealt{Pinto_2016Natur.533...64P,2018_Kosec_MNRAS.479.3978K}), a kinetic luminosity of a few $\times$ 10$^{39}$ erg/s is plausible which gives bubble radii $\sim$100 pc, consistent with observation (Table 1). We note however that the radius depends on the driving luminosity (radiative and mechanical) to the power 1/5, so that even a difference of a factor 100 in luminosity alters the radius by only a factor 2.5.

\begin{table}
\begin{center}
\begin{tabular}{ |c|c|c|c|c|} 
\hline
ULX &   Dimension(s) & \makecell{Inferred age \\ (years)} & \makecell{Radio/IR (jet) \\ counterpart?} & \makecell{$L_{\rm kin}$ \\(10$^{39}$ erg/s)}  \\ \hline
NGC 1313 X-1  &  240 pc &   & & \\ \hline
NGC 1313 X-2 & 350x500 pc$^{\rm a}$  & $\sim$1$\times$10$^{6}$ (b) & & $\sim$10\\ \hline
IC 342 X-1  &  110 pc$^{\rm c}$ & $\sim$6$\times$10$^{5}$ & Y & $\sim$0.7 \\ \hline
Ho II X-1 &  45 pc$^{\rm d}$ & & Y &\\  \hline
Ho IX X-1  &  250 pc$^{\rm e}$ & $\sim$1$\times$10$^{6}$  & Y & $\sim$10\\ \hline
M81 X-6 & 115 x 42 pc$^{\rm f}$ &    &  & \\ \hline
IC 2574 X-1 &     &  &  & \\ \hline
NGC 2403 X-1 &   300 pc$^{\rm g}$ &    &  & \\ \hline 
NGC 4559 X-7 &   &    &  & \\ \hline
NGC 4631 H7 &  &    &  & \\ \hline
NGC 4861 X-1 &  &    &  & \\ \hline
NGC 4861 X-2 &  &    &  & \\ \hline
NGC 5204 X-1 &  360 pc &    &  & \\ \hline
NGC 5408 X-1 &  60 pc$^{\rm h}$  & $\sim$1$\times$10$^{5}$ (i)   & Y & 0.7 (i)\\ \hline
NGC 5885 ULX &  200 x 300 pc$^{\rm g}$ &    6$\times$10$^{5}$ (j) & Y & $\sim$20 (j)\\ \hline
NGC 6946 X-1 &  20$\times$34 pc$^{\rm k}$   &    & Y & \\ \hline\hline
NGC 7793 S26 &    185 $\times$ 350 pc (l) &  4$\times$10$^{5}$ (l) & Y & $\sim$10-100 (l,m)\\ \hline
SS433 (W50) &  50-70 pc$^{\rm n}$  & $\sim$2$\times$10$^{5}$ (b)   &  Y & $>0.1$\\ \hline
M83 MQ1 &  13~pc$^{o}$ & 1.3-2.0$\times$10$^{4}$ (o)   & Y  & $\sim$ 30 (n)\\ \hline
\end{tabular}
\end{center}
\caption{\label{tb:pulse3} The properties of ULX nebulae, with non-ULX (i.e. hidden ULX) systems included at the bottom in a detached table. Objects and values from \cite{Pakull0202} unless indicated otherwise. Dimension is diameter unless additional dimension is given. Corresponding references are: a: \cite{Ramsey2006}; b: \cite{Cseh2012_IC342}, c: \cite{Roberts2003}; d: \cite{Kaaret2004_HoII_nebula}; e: \cite{Miller1995}; f: \cite{Moon2011}; g: \cite{Pakull_Grise2008}; h: \cite{Grise2012_5408}; i: \cite{Soria2006_5408_jet}; j: \cite{Soria2021_5585}; k: \cite{Abolmasov2007}; l: \cite{Dopita2012} m: \cite{Pakull0710}; n: \cite{Fabrika_2004_SS433}; o: \cite{Soria2014_MQ1}.}
\end{table}

The He II 4686$\lambda$ recombination line is more easily created via photoionization than by shock-excitation (e.g. \citealt{Berghea2010b}), so using this line allows us to treat the nebula as a UV--X-ray bolometer, and infer the irradiating luminosity by using photoionization codes such as {\sc cloudy} (\citealt{Ferland1998_CLOUDY}). This has been performed for a number of ULXs, notably Ho II X-1 (\citealt{Pakull0202} and \citealt{Berghea2010b} who used the OIV line in the IR band) and NGC 5408 X-1 (\citealt{Kaaret_Corbel2009_5408}), with the finding that the required luminosities are a good match to the X-ray luminosities observed from the source. 

Such consistency is used as an argument against strong geometrical beaming in these ULXs as the nebula should see the isotropic emission (\citealt{Pakull0202}). However, the ULX wind photosphere emits a luminosity $\sim L_{\rm Edd}$ which -- depending on the mass transfer rate -- can have a temperature
$\sim 10^6\,{\rm K}$, implying strong near-isotropic soft X-rays and EUV (see the discussion of Eqs. (\ref{rphot}, \ref{Teff} below).

In agreement with this, both Ho II X-1 and NGC 5408 X-1 are spectrally {\it soft} and potentially viewed at larger inclinations than some of the spectrally harder ULXs. Should this be true then it is not surprising that the irradiating luminosity inferred from photoionisation modeling is consistent with the observed luminosity, as we would naturally expect the more obscured (at such angles) hard X-ray emission to be the most geometrically beamed (rather than the more isotropic soft emission).

In principle, the {\it number} of ULX bubble nebulae can also be used to explore the role collimation/beaming plays in the wider population based on the reasoning that, for every ULX observed with a given beaming factor, there should be many more which are unbeamed and may not even be detectable as a ULX. As the nebula's photoionisation responds mostly to relatively unbeamed soft X-rays and EUV photons, such a premise would hold only if the nebula was energised through shocks (which would indeed seem to be the case: \citealt{Abolmasov2007, Gurpide2022}) or if the accretion rate was high enough such that the photosphere emitted in the UV (with any substantial X-ray emission obscured or reprocessed). This then implies we should detect bubble nebulae with X-ray faint (non-ULX) counterparts; indeed there is already support for such observations both in other galaxies (e.g. \citealt{Soria2014_MQ1}) and within our own (SS433: see Section 2.8.3). Searches for X-ray quiet bubble nebulae are made somewhat harder as ULX nebulae with strong photoionisation signatures (NGC 5408 X-1 and Ho II X-1) appear to be smaller (10s of pc rather than 100s of pc: Table 1), precluding detection of such relatively small structures in the absence of high angular resolution, narrow filter instruments (\citealt{Russell2011}). However, \cite{Pakull_Grise2008} searched VLT images of NGC 1313 for additional bubble nebulae {\it without} ULX counterparts and found only one possible candidate. As shown by \cite{Wiktorowicz1904}, should geometric beaming be strongly related to accretion rate (\citealt{King_2009_Beaming}), then binary population synthesis would imply that (in the absence of precession) the difference between the observed and underlying ULX population should be a factor of 5-15 unless black holes are the dominant population (in which case the observed is very similar to the underlying population). This would imply $\sim$10-30 ULXs should be present in NGC 1313 (assuming neutron stars dominate the underlying population), around 1/3 of which would have large bubble nebulae (\citealt{Pakull_Grise2008}). This is not significantly inconsistent with observation and any tension is relaxed if the population demographic is not 100\% neutron stars or if the beaming is not as extreme a function of accretion rate as assumed

A number of ULXs have radio nebulae (see Table 1) of a similar size to their optical counterparts (\citealt{Lang2007,Cseh2012_IC342}). In Ho II X-1 (which shows a clear photo-ionisation dominated optical nebula), the radio nebula has been resolved into a discrete lobe-core-lobe structure (\citealt{Cseh_2014_HoII_jets}). The presence and variability of the radio structure in Ho II X-1 implies that repeat ejections also energise the surrounding gas (\citealt{Cseh2015_HoII}) in a similar manner to SS433 (W50), the inflated bubble nebula in M83 S2 (\citealt{Soria2020ApJ_M83_S2}), M82 Q1 (\citealt{Soria2014_MQ1}) and the collimated jet source in NGC 7793 (S26: \citealt{Soria2010_S26}).

\subsubsection{ULX host environment}
\label{sec:large}

There is a trend favouring increased numbers of ULXs with lower host metallicity (e.g. \citealt{Mapelli1010}) and higher host star formation rate (see Section 2.1). This makes it important to explore the sub-galactic environment where ULXs are located, as this can provide powerful constraints on formation channels. 

Besides the ionised environment (which can be extremely large, e.g. the 800 pc region around NGC 1313 X-1: \citealt{Pakull0202}), it has been widely observed that ULX counterparts tend to be coincident with young stellar clusters or OB associations (e.g. \citealt{Zezas2002,Goad2002,Gao0310,Grise2006,Grise2008,Grise2012_5408,Swartz2009,Liu2007,Abolmasov2007_7331}). Although ionizing, the stellar environment provides insufficient flux to explain the emission line luminosities seen from the bubble nebulae (and the expected kinetic input due to supernovae is also not enough to inflate them: \citealt{Ramsey2006}) -- indeed, in the case of IC 342 X-1, there are no O stars within 300 pc of the ULX (\citealt{Feng_Kaaret2008}).

ULXs are also found (albeit rarely) in globular clusters.
Relatively few (17 in total) are known at the time of writing: three in NGC 1399 (\citealt{Shih2010,Irwin2010,Dage2019a}), two in NGC 4649 (\citealt{Roberts2012_GCULX,Dage2019a}), five in NGC 4472 (\citealt{Maccarone2007M, Maccarone2011}) and a further seven in M87 (\citealt{Dage2020}). 

ULXs located in globular clusters are expected to differ markedly from those located in the wider population. Given the dense environment of a globular cluster, such ULXs likely have a dynamical origin (\citealt{Ivanova2010}) -- distinctly {\it unlike} those ULXs located in star forming regions (e.g. \citealt{Brightman2020_M82}), and the companion stars are also considerably older (and therefore less massive, e.g. the ULX may be feeding from a white dwarf: \citealt{Steele2014}, potentially in a short period, ultracompact binary \citealt{King1105}). In terms of brightness, the typical X-ray luminosity is $\sim$10$^{39}$ erg\,s$^{-1}$ with the only globular cluster ULX found to exceed 4$\times$10$^{39}$ erg\,s$^{-1}$ being GCU8 in NGC 1399 (\citealt{Dage2019a}). 

\subsection{`Hyperluminous' X-ray sources?}
\label{sec:hlx}

The term `hyperluminous' has sometimes been used to refer to sources with 
inferred (isotropic) luminosities in excess of 10$^{41}$ erg\, s$^{-1}$. However there is little evidence to support the implicit suggestion that
these bright ULXs form a physically distinct class. They certainly do not need to contain unusually high mass black holes or IMBH: the ULX pulsar  NGC 5907 ULX-1 reaches such luminosities (e.g. \citealt{Israel1702}) and evidently contains a neutron star. The well known extreme source SS433 is probably a ULX
viewed `from the side' (i.e. from outside the X-ray beam: see Section (\ref{sec:ss433}) and inferred on evolutionary grounds (see \citealt{King2000}) to be supplied with mass at a
rate $\sim 10^{-5} \rightarrow 10^{-4}\msun\, {\rm yr}^{-1}$. From the beaming formula (\ref{eq:beam}),
if viewed from within this beam,
SS433 would have an inferred isotropic luminosity $L_{\rm sph} \sim 10^{41}{\rm erg\, s}^{-1}$ if the accretor is a $10\msun$ black hole. We discuss the source HLX-1 in Section (\ref{sec:hlx2}) below.

\subsection{Galactic ULXs}
\label{sec:galactic}

A number of Galactic sources are known to reach or exceed their Eddington limit for extended periods of time (i.e. discounting type I bursts etc). Their proximity allows for more detailed tests of how super-Eddington mass supply operates. Here we report those sources which would qualify as ULXs if placed in another Galaxy, with observed (or inferred) X-ray luminosities $\gtrsim 10^{39}\, {\rm erg s^{-1}}$.

\subsubsection{Low mass X-ray binaries}
\label{sec:beX}

Several low mass X-ray binaries (LMXBs) containing black holes are known to exceed their Eddington limit during outbursts mediated by the classical disc instability (\citealt{Lasota0106}), recently modified to account for large discs powering super-Eddington outbursts (\citealt{Hameury1120}). GRS 1915+105 reaches rates close to its Eddington limit (and has been in outburst for at least 30 years: \citealt{Castro-Tirado1992_1915}), GRO J1655-40 may have entered a super-Eddington state in its 2005 outburst (\citealt{Neilsen2016_GROJ1655}), V404 Cygni is thought to have exceeded its Eddington limit in its recent 2015 outburst (\citealt{Motta2017_V404, Miller-Jones2019_V404}, and potentially in the earlier outburst if not all of the radiation was visible to us) in a similar fashion to V4641 Sgr in 1999 (\citealt{Revnivtsev2002_V4641}). In addition to these confirmed black hole systems, neutron star LMXBs certainly exceed their own Eddington limit and may even approach the empirically defined ULX threshold of $\sim10^{39}$ erg/s (for instance the bursting pulsar, GRO 1744-28: \citealt{Sazonov1997_Burstingpulsar}). It is notable that all of these systems have fairly long orbital periods (2.6 - 31 d), which results in a large outer disc radius and the creation of a large reservoir of material for accretion (see \citealt{Hameury1120} for details).

In addition to large X-ray and optical luminosities, and powerful mass-loaded outflows, optically thin, flaring jet emission is observed to coincide with outbursts of those sources approaching their Eddington limits (\citealt{Fender_Pooley1997}). Such radio-bright, highly variable events provide another way of detecting extra-galactic systems entering a near-Eddington/super-Eddington state  (and may favour the detection of black hole over neutron star systems: \citealt{Middleton_2013_M31_microquazar,vandenEijnden2018Nature}).

\subsubsection{High mass X-ray binaries}
\label{sec:hmxb}

A number of accretion powered pulsars fuelled by high mass donor stars (i.e. high mass X-ray binaries: HMXBs) are known to reach 
$\times10^{39}\, {\rm erg s^{-1}}$, including the X-ray millisecond pulsar AO538-66 (\citealt{Skinner1982}), SMC X-1 (e.g. \citealt{Bonnet-Bidaud1981}) and SMC X-3 (mentioned already in this review).

A recent important observation of Galactic HMXBs in the context of ULXs is that of Swift J0243.6+6124 (\citealt{Kennea2017}). This magnetised neutron star (with an inferred dipole field of $\ge$ 10$^{12}$G: \citealt{Doroshenko1805}) accretes from the decretion disc of an O9.5Ve star (\citealt{Reig2020}), reaches a (0.5-10 keV) luminosity of 10$^{39}\, {\rm erg s^{-1}}$, shows pulsations on a timescale of $\sim$ 10s (\citealt{Jenke2017}), shows evidence for fast outflows (\citealt{vandenEijnden2019,Tao2019}), and produces radio jets (\citealt{vandenEijnden2018Nature}). These jets are relatively weak radio emitters, with a maximum flux density of 100 $\mu$Jy at 6 GHz (at an assumed distance of around 7 kpc: \citealt{vandenEijnden2018Nature}), and are therefore unlikely to be detected in other galaxies in the absence of substantial Doppler boosting. However, detections of jet events from Galactic HMXBs may become more widespread with the advent of sensitive all sky radio monitoring (SKA) and will permit tests of neutron star jet models (e.g. \citealt{Parfrey2016}).

\subsubsection{SS433}
\label{sec:ss433_1}

Discovered by \cite{Stephenson1977_SS433}, the A type supergiant companion star in SS433 (Gies et al 2002) is transferring mass to the compact object at a rate $\sim 10^{-5} \rightarrow 10^{-4}\msun\, {\rm yr}^{-1}$ (\citealt{Shklovsky_ss433,King2000,Fuchs2006_SS433}). 
This would imply $\dot{m} \sim 100$ and 1000 for a 10 M$_{\odot}$ black hole or 1.4 M$_{\odot}$ neutron star respectively. This strongly super-critical mass transfer rate suggests that SS433 is `intrinsically' a Galactic ULX similar to those with luminosities $\ge 10^{40} {\rm erg s^{-1}}$ seen in other galaxies (\citealt{Begelman0607}). However it does not qualify as a ULX on the grounds of {\it observed} luminosity -- see Section \ref{sec:ss433}.

One of the most dramatic features of SS433 is the baryon-loaded jet (\citealt{Fabian_Rees1979_SS433,Margon1979,Kotani1994_SS433,Marshall2013})
which is ejected at $\sim$0.26c and is twisted into a helical shape by the 162 day precession of the surrounding disc and wind. The baryon-loading, low velocity ($v/c \simeq 0.25$), and 162.4 d precession period of the jet might result from a collision with the precessing outer disc (\citealt{Begelman0607}) or could simply be due to the jet being the highly collimated wind from the innermost regions (which matches simulations, e.g. \citealt{Jiang1412, Jiang0819}) tracing the precession.

We limit ourselves here to a discussion of the similarities and differences between SS433 and other ULXs, and direct the reader to \citep{Fabrika_2004_SS433} for a comprehensive review. 

\begin{wrapfigure}{R}{0.5\textwidth}
  \begin{center}
    \includegraphics[width=18.0cm,height=10cm,angle=0]{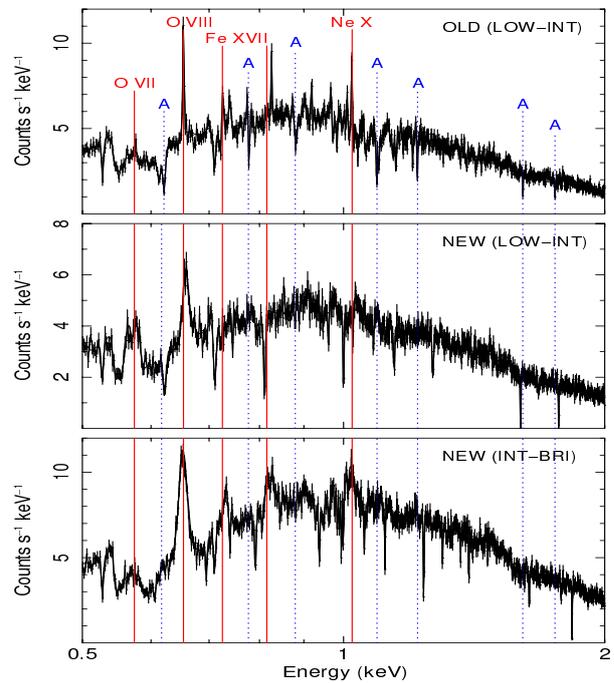}
  \end{center}
  \caption{From \cite{Pinto2020_1313}, showing simulations which highlight the spectral quality of observations of ULXs (in this case NGC 1313 X-1) which will be made possible by the launch of {\it Athena} with its X-ray integral field unit (X-IFU). The simulations utilise absorption and emission models derived from existing observations (see \citealt{Pinto2020_1313} for more details). 
    }
\end{wrapfigure}

SS433 can be regarded as a key example of a ULX in our own Galaxy, merely inclined such that we do not see directly into the wind-cone (\citealt{Begelman0607, Middleton2021_SS433}). As indicated in the previous sections, besides the similarities of the broad He II recombination lines (\citealt{Fabrika2015}) to those seen in other ULXs, there are a number of sources which look distinctly like SS433, viewed at similar edge-on inclinations. Notably the inflated nebula in M83 and collimated jet source in NGC 7793 (\citealt{Pakull0710,Soria2010_S26,Dopita2012}). In both of these cases, the jet power appears to be greater than in SS433 for similar or slightly higher apparent X-ray luminosities. Indeed, very many ULXs must also be oriented in such a manner such that we only observe a relative proportion of the entire ULX population (see \citealt{Middleton_King2017,Wiktorowicz1904}). Whilst it is rather hard to study the innermost regions of SS433 (although we can study the reflected emission and infer the intrinsic flux: \citealt{Middleton2021_SS433}), it provides a unique opportunity to explore how such systems are fed, its feedback into the local ISM, the nature of the binary (e.g. probes of circumbinary structure: \citealt{Waisberg2019}), and the structure of the X-ray obscuring regions (\citealt{Middleton2021_SS433}).

\subsubsection{Future ULX observations} 
\label{sec:future}

It is apparent that the study of ULXs has made dramatic progress as instrumentation has improved (e.g. the discovery of pulsations, winds and jets). In the near future, a number of developments will see the field evolve in new, exciting ways.

\subsection{X-rays}

Much of our understanding of ULXs derives from the X-ray band (even if this only actually grants us a restricted view of the full energetics of the system). Several new X-ray satellites are anticipated in the next two decades, including the {\it Astro-H} recovery mission, {\it XRISM} (dedicated to high energy-resolution X-ray imaging: \citealt{Xrism2020}), and ESA's flagship {\it Athena} mission (scheduled to launch in the 2030s: \citealt{Nandra2013_Athena}), comprising a wide field imager (WFI) and X-ray integral field unit (X-IFU). Both instruments onboard {\it Athena} represent a significant improvement in effective area for their respective specialisms: the WFI is around an order of magnitude more sensitive than {\it XMM-Newton}'s EPIC pn detector (from $\sim$0.5-2 keV), and the X-IFU is between 45 (at 1 keV) and 6 (at 7 keV) times more sensitive even than {\it XRISM}. This improvement in sensitivity will allow considerably deeper searches for pulsations in known ULXs, placing important constraints on the underlying population demographic. The sensitivity will also allow new searches for QPOs and reveal new details in the lag spectra (\citealt{Heil2010,DeMarco2013,Kara2020_1313,Middleton2021_SS433}). Finally, the impact on studying ULX outflows and how they vary with spectral state will be dramatic, allowing results previously obtained in $\sim$day exposures to be obtained in only a couple of hours (see Figure 12 which shows a simulation of the atomic lines detected in NGC 1313 X-1 as seen by {\it Athena}: \citealt{Pinto2020_1313}).

\subsection{Multi-wavelength}
\label{sec:multifut}

As we have discussed in the preceding sections, multiwavelength observations of ULXs have been highly revealing, allowing the discovery of jets, studies of optical nebulae and likely donor stars, and have provided insights into the nature of the accretion flow (irradiation and emission from the outer photosphere).

The future of optical observing of ULXs -- as with other variable astronomical objects of interest -- is poised to change dramatically with the introduction of all--sky, high angular resolution, high--throughput observing via the Vera Rubin LSST, which is due to start operations in 2023. LSST will provide highly sensitive snapshots (e.g. a limiting r-band magnitude of 24.7 in a single 30s exposure at SNR = 5) of $\sim$18,000 deg$^{2}$ of the Southern sky (DEC = -65 $\rightarrow$ +5). Combined with a median seeing--limited angular resolution of 0.7", this will allow deeper probes of possible optical counterparts and explore optical variability for signs of irradiation or precession (e.g. \citealt{Motch1410,Middleton_2019_Accretion_plane}). LSST is not equipped with narrow filters (nor any spectroscopic instrumentation), precluding locating new bubble nebulae, but the MUSE IFU on the VLT is set to transform our understanding of ULX feedback by spatially resolving different regions of ULX bubble nebulae (e.g. \citealt{Gurpide2022}). Finally, the introduction of the ELT (sometime after 2027), will also provide a step change in signal-to-noise, allowing deep spectroscopic studies of ULX counterparts. By identifying lines from the companion star, mass functions will no doubt follow. 

At longer wavelengths, there will be a great deal of interest in the JWST view of ULXs, as this will provide opportunities to locate further red super-giant companions (e.g. \citealt{Heida2016}), study the surrounding nebulae (e.g. using the OIV line: \citealt{Berghea2010b}) at a resolution even higher than can be achieved by MUSE, search directly for IR bright ejecta or indirectly search for the presence of jets via IR variability (e.g. \citealt{Dudik2016HOIXjet}). 

At the longest wavelengths, SKA will transform our view of the variable radio sky. In relation to ULXs, the enormous sensitivity of SKA will allow us to locate LMXB ULXs entering outburst in nearby galaxies (\citealt{Middleton_2013_M31_microquazar}), search for variability in radio bubble nebulae, and potentially resolve jet structures (\citealt{Cseh_2014_HoII_jets,Cseh2015_HoII}). In terms of the impact on studies of Galactic ULXs, SKA will allow the deepest searches for jet ejections associated with accreting pulsars (\citealt{vandenEijnden2018Nature}) and allow models for their production to be tested (e.g. \citealt{Parfrey2016}).

\begin{table}
\scriptsize
\begin{center}
\begin{tabular}{|c|c|c|c|c|c|c|c|} 
\hline
ULX & P$_{\rm pulse}$ (s) &  \makecell{Spin up rate (ss$^{-1}$) \\ (baseline)} & Pulse fraction (E range) &  P$_{\rm orb}$ & P$_{\rm s-orb}$ (d) & \makecell{L$_{\rm max}$ \\ ($\times 10^{39}$ erg s$^{-1}$)} & \makecell{M$_{2}$\\ ($\rm M_{\odot}$)} \\ \hline \hline

M82 X-2$^{\rm a,b,c}$ & 1.4 & 1 $\times$ 10$^{-11}$ (2.5 years) & \makecell{5-13\% (3 - 30 keV) \\ 8-23\% (10 - 30 keV)}  & 2.5 & $\sim$60 & 20 & $\ge$ 5.2\\ \hline

NGC 5907 ULX-1$^{\rm d,e}$ & 1.1 & -8$\times$10$^{-10}$ (11 years) & \makecell{12\% (0.2-2.5 keV) \\ 20\% (7-30 keV)} & 5.3 & 78 & 200 & ? \\ \hline

NGC 7793 P13$^{\rm f,g,h}$ & 0.4 & 4$\times$10$^{-11}$ (4 years) & 18-22\% (0.1-12 keV) & 64 & 64 & $\sim$ 10 & 18-23 (B9I) \\ \hline

NGC 300 ULX-1$^{\rm i,j}$ & 31.6 & -6 $\times$10$^{-7}$ (3.6 days) & \makecell{$\sim$60\% (0.2-10 keV) \\ $\sim$ 70\% (3-20 keV)} & $>$ 290 & ? & 5 & 8-10 (RSB?)\\  \hline

SMC X-3$^{\rm k,l,m,n}$ & 7.8 & $\sim$-4$\times$10$^{-9}$ (110 days) & 30-100 \% (3-70 keV) & $\sim$45 & ? & $\sim 3$ & $>$3.7 (Be?) \\ \hline

M51 ULX-7$^{\rm o,p}$ & 2.8 & -1$\times$10$^{-9}$ (13 years) & $<$5 - 20\% (0.1-12 keV) & 2 & 38 & $\sim$ 10 & $>$ 8 \\ \hline

NGC 1313 X-2$^{\rm q}$ & 1.5 & 1.2$\times$10$^{-12}$  (98 days) & $\sim$5\% (0.3-10 keV) & $<$ 4 & ? & 20 & ?  \\ \hline

NGC 2403 ULX$^{\rm r}$ & 18 & 3.4$\times$10$^{-10}$ (?) & ? & 60-100 & ? & 1.2 &  Be?\\ \hline

Swift J0243.6+6124$^{\rm s}$ & 9.86 & 2.2$\times$10$^{-10}$ (?) & ? & 28.3 & ? & $\ge$ 1.5 & Be?\\ \hline

RXJ0209.6-7427$^{\rm t}$ & 9.3 & 1.165$\times$10$^{-10}$ (?) & ? & $>$50 & ? & 1-2 &  Be \\ \hline


M51 ULX-8$^{\rm u}$ & ? & ? & ? & 8-400 & ? & 2 &  40(?)\\ \hline

NGC 1313 PULX$^{\rm v}$ & 765.6 & ? & 38\% (0.3-7 keV) &  & ? & 1.6  &  Be?\\ \hline

\end{tabular}
\caption{\label{tb:combined} Key parameters of interest for the ULX pulsars with values taken from those papers indicated by the reference attached to the ULX name. The period over which the spin-up/down rate is calculated, is given in parentheses, as is the energy range for the measured pulse fraction. a: \cite{Bachetti1410}, b: \cite{Bachetti2020}. It is worth noting that in the case of M82 X-2, there are periods of both spin-up and spin-down -- see \cite{Bachetti2020} for details. c: \cite{Brightman2019_SOP_M82}; d: \cite{Israel1702}; e: \cite{Walton_2016_Modulation_5907}; f: \cite{Furst1611}; g: \cite{Israel1703}; h: \cite{Fuerst_P13_2021}; i: \cite{Carpano1805}; j: \cite{Heida1910}, k: \cite{Edge2004}; l: \cite{Townsend1711}; m: \cite{Tsygankov1709}; n: \cite{Corbet2004}; o: \cite{RodriguezCastillo2005}; p: \cite{Brightman0620}; q: \cite{Sathyaprakash1909}; r: \cite{Trudolyubov0707}; s: \cite{Doroshenko1805}; t: \cite{Vasilopoulos0620};  u:\cite{Brightman1804}; v: \cite{Trudolyubov0806}.}
\end{center}
\end{table}

\section{Theory}
\label{sec:theory}

\subsection{Definitions and notation}
\label{sec:defnot}

The Eddington luminosity is
\begin{equation}
\label{eq:Ledd}
\Ledd = \frac{4\pi GM m_p c}{\sigma_T} = 1.26 \times 10^{38} \left(\frac{M}{\Msun}\right)\ergs, 
\end{equation}
where $m_p$ is the proton mass, $\sigma_T$ the Thomson scattering cross-section. 
{   In this review $\ledd$ defined by Eq. (\ref{eq:Ledd}) is a unit of luminosity depending on accretor's mass only.}

The Eddington accretion rate is defined as
\begin{align}
\dot{M}_{\rm Edd} & = 1.4\times 10^{18}\eta_{0.1}^{-1} \left(\frac{M}{\Msun}\right)\,{\rm g\,s^{-1}} \\
& = 2.2 \times 10^{-8}\, \eta_{0.1}^{-1} \left(\frac{M}{\Msun}\right) \rm M_{\odot}\,\rm yr^{-1},
\end{align}
where $\eta=0.1\eta_{0.1}$ is the radiative efficiency of accretion, $\kappa_{es}$ the electron scattering (Thomson) opacity.
We will often use accretion rate measured in units of Eddington accretion rate and stellar mass in solar masses:
\begin{equation}
\label{eq:mdot}
\dot m=\frac{\dot M}{\dot{M}_{\rm Edd}}\, \ \ \ \ \ \ \ \ \ {\rm and}\ \ \ \ \ \ \ \  m = \frac{M}{\Msun}.
\end{equation}

\subsection{Discs with super-Eddington mass supply rates}
\label{sec:sediscs}
At high local accretion rates $\dot m \gtrsim 1$, the radial structure of a stationary ($\dot M= const.$) disc around a compact object can be divided into three parts \citep{Shakura73}.
In the outermost regions the pressure is dominated by gas and the electron scattering contribution to the opacity can be neglected. For radii 
\begin{equation}
\label{eq:es}
R_{\rm es} < 5 \times 10^3 \dot m^{2/3} R_g,
\end{equation}
electron scattering is the main source of opacity, but gas pressure is still dominant. Finally, for radii
\begin{equation}
\label{eq:rradpres}
R _{\rm pr}< 1.9 \times 10^2 \dot m^{16/21}m^{-8/21}R_g,
\end{equation}
the pressure is dominated by radiation, while the opacity is mainly due to electron scattering.
{   $R_g\equiv GM/c^2$ is the gravitational radius.}

The standard accretion disc model \citep{Shakura73,Frank0201} applies only to geometrically thin discs, i.e., discs whose height satisfies the condition $H \ll R$.  When the luminosity approaches the Eddington value this approximation breaks down. 

The approximate vertical dynamical equilibrium equation
\begin{equation}
\label{eq:vertieq}
\frac{H}{R}\approx \frac{c_s}{v_{\rm K}},
\end{equation}
where $c_s=\sqrt{P/\rho}$, is the isothermal sound speed, $P$  and $\rho$ are the {   (total, gas + radiation)} pressure and density respectively, and $v_{\rm K}=\sqrt{GM/R}$, can be written as
\begin{equation}
\label{eq:rhedd}
 \frac{H}{R} \approx \frac{L(R)}{\Ledd}.
\end{equation}
To derive Eq. (\ref{eq:rhedd}), we have used the expression for the radiative pressure $P_r=4\sigma T_c^4/3c$ and the solution for the local {   vertical} radiative transfer equation $T_c^4=(3/8)\tau \Teff^4$ \citep{Lasota16}. 

Therefore already at $L \sim 0.1 \Ledd $ the thin-disc approximation ceases to be valid and for $L\gtrsim \Ledd$ one has $H \gtrsim R$ so that the accretion flow is more spherical than disc--like. 

The point at which the local flux takes the Eddington value is called the spherization radius \citep{Shakura73} and can be seen as the boundary separating the thin disc from a more spherical inner accretion flow.

\subsubsection{Spherization radius}
\label{sec:rsph}

In a stationary disc ($\dot M = const $), the local radiative flux generated by the disc's (anomalous) viscosity becomes equal to the local Eddington flux at the {\sl spherization radius} $\Rsph$.\\
The viscous flux \citep{Frank0201} can be written as 
\begin{equation}
\label{eq:fvis}
F_{\rm visc}=\frac{3}{8\pi}\frac{R_g}{R^3}\dot Mc^2\,\left[  1 - \left( \frac{R_{\rm in}}{R}\right)^{1/2}\right],
\end{equation}
and the Eddington flux is
\begin{equation}
\label{eq:eddflux}
F_{\rm Edd}= \frac{L_{\rm Edd}}{4\pi R^2}.
\end{equation}
From  $F_{\rm visc}= F_{\rm Edd}$, one obtains
\begin{equation}
\label{eq:resph}
R_{\rm sph}= {15}\dot m \,R_g,
\end{equation}
We have neglected the boundary term in square brackets since $R_{\rm sph} \gg R_{\rm in}$ for $\dot m \gg 1$ and $R_{\rm in}\approx 6R_g$.
The spherization radius defined in Eq. (\ref{eq:resph}) is 1.1 times larger than the spherization radius in \citet{Shakura73}. 
\citet{Ohsuga0914} assume $R_{\mathrm{sph}} \equiv\left(\dot{M} c^{2} / L_{\mathrm{Edd}}\right) R_{\mathrm{S}}$, where $R_{\mathrm{S}}=2R_g$\footnote{In the literature both units of length, $R_{\mathrm{S}}$ and $R_g$ are used.}, which gives a value of $R_{\rm sph}$ 1.3 times larger than in Eq. (\ref{eq:resph}).

The outer regions of accretion discs around compact objects can be strongly X--ray self--irradiated. Then the condition for $\Rsph$ is
\begin{equation}
F_{\rm Edd}= F_{\rm irr} + F_{\rm visc},
\end{equation}
where we put \citep[see][]{Dubus9902}:
\begin{equation}
F_{\rm irr} = \mathcal{C} \frac{\eta \dot{M_{\rm in}} c^2}{4 \pi R^2}, 
\label{eq:firr}
\end{equation}
where $\mathcal{C}$ is a constant containing all the information about the disc irradiation process and $\dot M_{\rm in}$ is the accretion rate onto the accretor. The spherization radius becomes 
\begin{equation}
\label{eq:resphirr}
R_{\rm sph}= {15}\left( 1 - \mathcal{C}\frac{\dot m}{2} \right)^{-1}\dot m \,R_g.
\end{equation}
For ``standard''  values of $ \mathcal{C} \sim 10^{-2}$ \citep{Dubus0107} disc irradiation can be neglected in the definition of the spherization radius for $\dot m \ll  200\, (0.01/\mathcal{C})$. {   One should notice, however, that at very high accretion rates it is nor clear how self--irradiation can proceed.
Even at sub--Eddington accretion rates when applying irradiation models to observed systems, one concludes that scattering in the wind is still not sufficient to produce the observed disc heating, even in combination with direct illumination \citep{Tetarenko0720}. On the other hand, until now, the disc considered in this context were rather smaller than those supposed to be present in many ULXs so it this latter case a component of order $\ledd$
emerging from the photosphere of the spherical wind, might irradiate the outer parts of the disc. Fortunately, as argued above, this effect is presumably not important at most luminosities of interest. }

High accretion rates implying large optical depth and high radial velocities in the inner regions of the flow can result in photons being trapped in the accretion flow.

\subsubsection{Trapping radius}
\label{sec:trapp}

The \textsl{trapping radius} is defined as the place where the photon diffusion (escape)  time $H\tau/c$ equals  the
viscous infall time $R/v_r$
\begin{equation}
\label{eq:trapp}
R_{\rm trapp}= \frac{H\tau\,v_r}{c}= \frac{H\kappa \Sigma}{c}\frac{\dot M}{2\pi R \Sigma}=20 \frac{H}{R}{\dot m}R_g,
\end{equation}
where $\Sigma=2 \int \rho dz$ is the disc surface density.
For $R < R_{\rm trapp}$ photons trapped in the flow are entrained (advected) faster than they can escape, so that if the accreting body is a black hole, most of the accretion energy is accreted by the black hole and adds to its mass. In this state there is in principle no upper limit to the accretion {\it rate} $\dot M$ but the accretion {\it luminosity} grows only logarithmically with $\dot M$ (see Eq. \ref{eq:ladv}). In other words, very super--Eddington mass supply rates do {\it not} imply (very) super--Eddington luminosities.

Eq. \ref{eq:trapp} assumes that the accretion flow moves only in the radial direction, and that there are no outflows. \citet{Shakura73}  proposed an alternative to such an advection--dominated solution (which they considered ``improbable'') where for $R\leq \Rsph$ the local flux is everywhere equal to its Eddington value.

\subsection{Shakura-Sunyaev (1973) second (windy) solution}
\label{sec:ss73}
To keep $F_{\rm vis}= F_{\rm Edd}$, everywhere for $R \leq R_{\rm sph}$, one must arrange that $\dot M\sim R$ (see Eqs. \ref{eq:fvis} and \ref{eq:eddflux}). Hence
\begin{equation}
\dot{m}(R) \simeq \dot{m}_{0} \frac{R}{\Rsph} 
\end{equation}

Using $\sigma T_{\text {eff }}^{4}\equiv F_{\rm visc}= F_{\rm Edd}$ one can calculate the luminosity of disc described by the windy SS73 solution
\begin{equation}
\label{eq:lss}
L_{\text {thick}}= 2\int_{R_{\mathrm{in}}}^{R_{\mathrm{sph}}} \sigma T_{\mathrm{eff}}^{4} 2 \pi R d R \approx  L_{\mathrm{Edd}} \ln \dot{m}.
\end{equation}
The total disc luminosity is then
\begin{align}
L_{\text {total }}=& L_{\text {thin }}+L_{\text {thick}}= \nonumber\\
& 4 \pi\left(\int_{R_{\text {in }}}^{R_{\text {sph }}} \sigma T_{\text {eff }}^{4} R d R+\int_{R_{\text {sph }}}^{R_{\infty}} \sigma T_{\text {eff }}^{4} R d R\right) \approx L_{\text {Edd }}(1+\ln \dot{m}).
\label{log}
  \end{align}

In this SS73 `windy' solution, maintaining the radiation flux at the local Eddington value requires the accretion rate to decrease with radius as $\dot M \sim R$. This requires that matter is expelled from the disc flow as a wind. But the trapping radius is close to the spherization radius, so one can also legitimately consider discs in which the
``excess'' radiation is trapped in the inflow and advected on to the accretor, while the accretion rate remains constant. We examine these
advective solutions in the next two subsections. {They have played a very important role in the understanding of high--rate accretion onto compact bodies but, as we will see, they do not give suitable models for ULX accretion.}

\subsection{Slim discs}
\label{sec:slimdiscs}

 For a black hole accretor, the advected energy disappears behind the event horizon (growing its mass) and we can regard advection as an additional cooling mechanism.  
In contrast, for a neutron star accretor, the advected energy must be radiated from the stellar surface {   or the magnetosphere}. Then the accretion luminosity violates the Eddington limit and the assumption of steady radial infall.

The energy conservation equation in a stationary disc is
\begin{equation}
\label{eq:encons1}
F_{\rm visc} = F_{r} + F_{\mathrm{adv}},
\end{equation}
where $F_{r}$ is the radiative flux and $F_{\mathrm{adv}}$ is the advective term \citep{Lasota16}
\begin{equation}
\label{eq:fadv}
F^{\mathrm{adv}}=\frac{\Sigma T \varv_{r}}{R} \frac{d s}{d \ln R}=\frac{\dot{M}}{2 \pi R^{2}} c_{\mathrm{s}}^{2} \xi_{a}.
\end{equation}
Here $\Sigma$ is the surface density, $T$ is the midplane temperature, $\varv_{r}$ the radial velocity, $c_{\mathrm{s}} = \sqrt{P/\rho}$ the speed of sound,
$s$ the specific entropy and $\xi_a \sim 1$ is a slowly varying function related to the entropy gradient.
From Eqs. \ref{eq:fvis}, \ref{eq:vertieq} and \ref{eq:fadv} one obtains
\begin{equation}
\label{eq:fhr}
\frac{F_{\mathrm{adv}}}{F_{\rm visc}} \sim \frac{c_{s}^{2}}{\Omega_{K}^{2} R^{2}} \approx\left(\frac{H}{R}\right)^{2},
\end{equation}
so that for thin ($H/R \ll 1$) discs the advective term can be neglected. But at luminosities approaching $\Ledd$ this term become dominant, leading to 
\begin{equation}
\label{eq:encons2}
F_{\rm visc} \approx F_{\mathrm{adv}}.
\end{equation}
Solutions of this equation represent advection dominated accretion flows (ADAFs), which for high accretion rates are called ``slim discs"
\citep{Abramowicz8809,Sadowski0809,Sadowski0811,Sadowski0311}.

Solutions representing vertically--integrated slim-disc structure have the form
\begin{equation}
\label{eq:slimeq}
\dot{m}\sim \kappa_{\mathrm{es}}  r^{1 / 2} \alpha \Sigma
\end{equation}
on the $\dot M(\Sigma)$ plane. {   $\kappa_{\rm es}$ is the electron-scattering opacity and $\alpha$ the disc viscosity coefficient.}

For these pure advection solutions, one has $H/R \approx 1$, {\it independent of accretion rate and radius} \citep[see Eqs. \ref{eq:fhr}, \ref{eq:encons2} and][]{Lasota16}.
But since $F_{\mathrm{adv}} \sim {\dot m}^2$ while $F_r \sim{ \dot m}^{1/2}$ \citep{Lasota1603}, below $\dot m \lesssim 1$, radiative cooling overtakes advection and the disc structure switches to the standard radiation-pressure--dominated SS73 disc:
\begin{equation}
\dot{m} \sim \kappa_{\mathrm{es}}^{-1}  r^{3 / 2}(\alpha \Sigma)^{-1}.
\end{equation}
Using Eq. (\ref{eq:rhedd}) the radiative flux of the advection-dominated flow becomes
\begin{equation}
F_{r}=\sigma T_{\mathrm{eff}}^{4} \approx\frac{L_{\mathrm{Edd}}}{4\pi R^{2}},
\end{equation}
so that the luminosity of this part of the accretion flow is
\begin{equation}
\label{eq:ladv}
L_{\text {slim }}= 2\int_{R_{\text {in }}}^{R_{\text {trans }}} \sigma T_{\text {eff }}^{4} 2 \pi R d R \approx L_{\text {Edd }} \cdot \ln \frac{R_{\text {trans }}}{R_{\text {in }}} \approx L_{\text {Edd }} \ln \dot{m}.
\end{equation}
Equations (\ref{eq:lss}) and (\ref{eq:ladv}) both result from the mechanical equilibrium equation of a radiation-pressure dominated disc
\begin{equation}
F_{r} = \frac{c}{\kappa_{es}}g_z  \approx \frac{c}{\kappa_{es}}\frac{GM}{R^2},
\end{equation}
for $H/R \approx 1$, and $g_z \approx (GM/R^2)H/R$ \citep{Paczynski0182,Poutanen0705}. But although the luminosities are similar, in the first case only $\dot m \approx 1$ is accreted on to the compact object while in an advection dominated flow the whole of $\dot m \gg 1$ arrives at the accretor's surface.

Since $H/R$ in slim discs is independent of the accretion rate, one cannot get so--called "thick discs" or "Polish doughnuts" by increasing the rate at which matter is supplied to the compact object. Contrary to assertions found in the literature \citep[e.g.,][]{Poutanen0705} slim discs are not the same objects as  Polish doughnuts. As we have seen, they do not significantly increase the true emitted luminosity above $L_{\rm Edd}$, and do not collimate or beam it such that the apparent luminosity may be super--Eddington in a restricted solid angle. Accordingly they do not offer an explanation for the defining feature of ULXs. However for a black hole ULX they may describe the fate of some of the super--Eddington mass supply. 

\subsection{Polish doughnuts}
\label{sec:donuts}

Unlike slim discs, for which $H/R \lesssim 1$, Polish doughnuts \citep{Kozlowski0278,Abramowicz0278,Jarosztnski0180,Paczynsky0880} 
have by construction $H/R \gg 1$, with long central funnels along which most of the radiation is emitted. They were devised in the late 1970s and early 1980s to try to explain the high luminosities and jets of QSOs by assuming that these objects are fed at highly super-Eddington rates \citep[see, e.g.,][]{Sikora0780}. They are often confused with slim discs and incorrectly assumed to be advection dominated accretion flows.
They have been invoked as solutions to the ULX problem, but we will see that this is not correct.

Polish doughnuts are models of stationary and axially symmetric accretion structures around black holes. All their properties 
are obtained from a single function $\ell(R)$  describing the specific (per unit mass) angular momentum distribution at the doughnut's photosphere. 
There is no explicit assumption about the doughnut's interior. One assumes only that a.) the photosphere coincides with an equipressure surface. b.) the specific angular momentum at the photosphere is assumed to be given by some function $\ell(R)$ (one often assumes $\ell(R) = const.$), c.) radiation is emitted at the photosphere at the local Eddington flux, that is, the local flux ${\bf F}_r$ is given by
\begin{equation}
\textbf {F}_r=\frac{c}{\kappa}\, \textbf{g}_{\mathrm{eff}},
\end{equation}
\citep{Paczynski0182},where boldface symbols denote vectors.
The specific angular momentum is assumed to be Keplerian at the inner and outer boundaries: $\ell\left(R_{\mathrm{in}}\right)=\ell_{\mathrm{K}}\left(R_{\mathrm{in}}\right) \text { and } \ell\left(R_{\mathrm{out}}\right)=\ell_{\mathrm{K}}\left(R_{\mathrm{out}}\right)$, and the inner radius lies between the innermost stable circular orbit (ISCO) and the innermost bound orbit (IBCO), and $R_{\mathrm{IBCO}}<R_{\mathrm{in}}<R_{\mathrm{ISCO}}$. Because of the strong radial pressure gradients (negligible for thin discs) needed to have $H/R \gg 1$, the inner flow edge is pushed inside the ISCO. 

The doughnut shape is obtained from
\begin{equation}
\frac{\mathrm{d} H}{\mathrm{d} R}=-\left(\frac{g^{R}_{\mathrm{eff}}}{g^{\mathrm{z}}_{\mathrm{eff}}}\right)_{\mathrm{z}=H} \equiv f(R, H),
\end{equation}
where $R,z$ are the cylindrical coordinates, and the radial component of the effective gravitational acceleration $g^{R}_{\mathrm{eff}} $is a function of $\ell(R)$.

The inner part of the doughnut forms a funnel with an opening angle obeying $\tan \alpha=1 / \chi \sim \sqrt{8 \rho}$, where  $\rho=(R_{\mathrm{in}}-R_{\mathrm{IBCO}})/{R_{\mathrm{S}}}$ ($0 \leq \rho \leq 1$) and $\chi =(H/R)_{\textrm max}$ \citep[][]{Wielgus0316}.

The luminosity is mainly emitted from the funnel, i.e.,
\begin{equation}
\label{eq:beamingPD}
L_{\textrm{apparent}}\approx \frac{L_{\textrm{funnel}}}{1- \cos \alpha }\equiv 
\frac{1}{b}L_{\textrm {funnel}}.
\end{equation}
With $b= (1- \cos \alpha)\sim \rho$, the funnel can be very narrow, but this does not lead to an increase of the apparent luminosity.
Indeed, 
\begin{equation}
L_{\textrm{apparent}}\approx \frac{1}{b}L_{\textrm{funnel}}= \frac{\epsilon_{\textrm{rad}}}{b}\dot Mc^2 \approx const.\, \dot Mc^2,
\end{equation}
because the accretion radiation efficiency $\epsilon_{\textrm{rad}}\sim \rho$, since by definition the binding energy at $R_{\textrm{IBCO}}$ is equal to zero.

So although Eq. (\ref{eq:beamingPD}) at first sight resembles Eq. (\ref{eq:beaming1}), in Polish doughnuts one does not have $L_{\textrm{apparent}} \gg L_{\textrm{true}}$.
This means that, contrary to initial appearances, Polish doughnuts are not relevant to BH ULXs \citep{Wielgus0316}, and we already know that they are not applicable to NS ULXs. Together with the similar conclusions we arrived at in Subsection (\ref{sec:slimdiscs}) for slim discs, this severely narrows the range of potential explanations for ULXs. 

We are left with only one possibility for making an accretion flow 
{\it appear} super--Eddington while simultaneously ensuring that the mass supply rate very close to the accretor is {\it not} significantly super--Eddington. In the next Subsection we ask if the disc winds required by the Shakura--Sunyaev `windy' solution can collimate the modestly super--Eddington intrinsic luminosity $L_{\rm Edd}(1 + \ln\dot m)$ (cf \ref{log}) given by a super--Eddington mass supply. We note that Eq. (\ref{log}) is an approximation since the spherization radius is not a uniquely defined quantity, which we should not see as a rigid limit defining the boundary of the wind.
Eq. (\ref{log}) is valid in both the SS73 and advection--dominated cases, but when these effects are simultaneously present one gets a slightly different formula (see e.g., \citealt{Poutanen0705}). In the following we therefore use $L=L_{\rm Edd}(1 + \ln\dot m)$ as a ``universal" reference value for unbeamed luminosity.

\subsection{Beaming by accretion disc winds}
\label{sec:windbeam}

We have seen that in the SS73 `windy' solution, the accretion flow carries only the local Eddington mass rate at each accretion disc radius, while the excess is blown away in winds at each radius near the disc centre. These winds must be quasispherical, but leave open funnels around the central disc axis (since they cannot achieve zero angular momentum). The funnels offer a suitable way of collimating the accretion luminosity, as all other photon escape routes have high optical depth. We will see in Section (\ref{sec:numer}) that numerical solutions of such flows are still some way from providing a picture which is easy to apply to observed ULXs. 

Fortunately,
soft X--ray observations of ULXs provide a remarkable insight into 
the beaming by disc winds. 
The soft components of ULXs 
can be fitted with blackbody spectra, giving the temperature $T$ and apparent (assumed isotropic) luminosity 
$L_{\rm soft}$. \citet{Kajava0909} show that the brightest soft X--ray components 
(above $3\times 10^{39}\, {\rm erg\,s^{-1}}$) in 9 ULXs cluster around the relation
\begin{equation}
L_{\rm soft} \simeq 7\times 10^{40}T_{\rm 0.1\, keV}^{-4}\, {\rm erg\, s^{-1}},
\label{softX}
\end{equation}
where $T_{\rm 0.1\, keV}$ is the blackbody temperature expressed in units of 0.1 keV/{\it k}, where $k$ is Boltzmann's 
constant. {   However, not all ULXs show such a relationship (cf \citealt{Walton_1313_2020, Gurpide9221}) and -- even in the absence of additional spectral components in the presence of a neutron star accretor -- the combination of anisotropy/beaming, precession and the lack of a suitable model for extracting L$_{\rm soft}$ unambiguously (see, e.g. \citealt{Robba2021_1313X2}), must result in complexity and a range of observed slopes.}

At first sight (\ref{softX}) seems paradoxical, as one would expect a blackbody source of fixed size to vary as
$L \propto T^4$ instead\footnote{Indeed one does find $L \propto T^4$ for sources which are permanently {\it below} 
$3\times 10^{39}\, {\rm erg\,s^{-1}}$ (\citealt{Kajava0909}). It is likely that
these sources are black holes with masses large enough that their luminosities are sub--Eddington.}.
But the relation (\ref{softX}) would instead imply  that the radius of the blackbody source varies along with the luminosity
and temperature. This is of course what we should expect for emission from the central part of accretion discs fed at 
super--Eddington rates, progressively ejecting the excess accreting matter from inside the spherization radius, 
as the latter varies $\propto \dot m$
(see Section \ref{sec:defnot}). We can see this explicitly by parametrizing the true emitted luminosity (pre--collimation) as
\begin{equation}
    L = 4\pi p R^2\sigma T^4
\end{equation}
where $R = r R_g \propto M$ with $r \simeq 15\dot m$ (see Eq. \ref{eq:resph}), and $\sigma$ is the Stefan--Boltzmann constant. 
Setting $L = lL_{\rm Edd} \propto M$, where $l \sim 1$, and $p$ allows for geometrical projection, we can write
\begin{equation}
L \propto R^2T^4p \propto M^2T^4r^2p \propto L^2T^4\frac{r^2 p}{l^2}, 
\end{equation}
where we use $R \propto rR_g$ at the
first step and $M \propto L/l$ at the second. The final form implies $L\propto T^{-4}$.
An observer assuming that the flux is isotropic with the observed value, rather than collimated, now infers a 
total blackbody 
luminosity $L_{\rm sph} = b^{-1}L$, so that, inserting the constants,
\begin{equation}
L_{\rm sph} = \frac{4\pi c^6}{\sigma \kappa^2}\frac{1}{T^4}\frac{l^2}{pbr^2} = 
\frac{2.8 \times 10^{44}}{T_{\rm 0.1\, keV}^{4}}\left(\frac{l^2}{pbr^2}\right)\, {\rm erg\,s^{-1}}.
\label{LT}
\end{equation}
The fact that observation gives $L \propto T^{-4}$ (cf Eq. \ref{softX}) means that the factor $l^2/pbr^2$ in this
equation must be a constant. Since $r \simeq (15/2)\dot m$ (Eq. \ref{eq:resph}), consistency with Eq. (\ref{softX}) 
requires
\begin{equation}
    b \simeq \frac{73}{\dot m^2},
\label{eq:beam}    
\end{equation}
where we have assumed that the slowly--varying factors $l \sim p \sim 1$. (Tight beaming means that we
observe the source along the disc axis, so we view the central disc plane orthogonally, making $p \simeq 1$.)

The physical origin of the simple form (\ref{eq:beam}) is straightforward. The beaming factor is the total solid
angle of the two open cones around the disc axis, i.e
\begin{equation}
    b = (1-\cos\theta)
\end{equation}
where $\theta$ is the opening half--angle. Near the spherization radius, where 
the wind outflow is maximal (see Section 3.3),
the flow geometry scales with $\dot m$. But all flows, regardless of $\dot m$, reduce locally to the 
same Eddington inflow near the accretor, so 
we expect $\theta \sim  {\rm constant}\times\dot m^{-1}$. For sufficiently 
large $\dot m$ we have $\theta \ll 1$ and so
$b \sim \theta^2/2 \propto \dot m^{-2}$. 

We see that the simple form 
(\ref{eq:beam}) formally applies only for 
$\dot m \gtrsim 8$, since, at such rates, $b <1$. Evidently it is legitimate to adopt a form of 
$b$ interpolating smoothly between 
$b = 1$ for $\dot m = 0$, and the form (\ref{eq:beam}) for $\dot m \gtrsim \sqrt{73} 
\simeq 8$. For example, \citet{Hameury1120} use
\begin{equation}
    b \simeq \frac{73}{73 + \dot m^2}.
    \label{eq:beam2}
\end{equation}
Importantly, although we have derived this relation by considering only the soft X--ray emission
of ULXs, the fact that it is purely geometrical and involves only electron scattering means that it must 
apply to all forms of radiation from ULXs.

The true emission from the central accretion disc 
is $L \sim \Ledd(1 + \ln\dot m)$, so the apparent 
luminosity is
\begin{equation}
    L_{\rm sph} = \frac{\Ledd}{b} \sim M\dot m^2 \sim \frac{\dot M^2}{M},
    \label{eq:lsph}
\end{equation}
where $\dot M$ is the mass supply rate, since $\dot m = \dot M/\dot M_{\rm Edd} \propto \dot M/M^2$.
So for a given mass supply rate, accretors of lower mass have {\it higher} apparent luminosities, 
as they are more super--Eddington, and the consequent increase in beaming outweighs the lower Eddington 
luminosity. 
This means that neutron stars are favoured as ULXs over black hole accretors in binaries where the mass transfer rate is insensitive to the accretor mass.
This condition holds for mass transfer driven by the evolutionary expansion of the donor star, (see 
Sections \ref{sec:evolulx} and \ref{sec:pop}), and also for cases where the increased mass transfer is a transient effect caused by a disc instability,
e.g. in Be--X--ray binaries, and in soft X--ray transients, where the disc is subject to the thermal--viscous instability (cf \citealt{Hameury1120}).

The beaming formula (\ref{eq:beam}) gives a simple explanation of the observed soft X--ray correlation 
(\ref{softX}). We can easily see that other explanations are problematic. 

If we assume there is no beaming,
i.e. $b = 1$, then (\ref{LT}) requires
\begin{equation}
    r = 62\left(\frac{l}{p^{1/2}}\right).
    \label{rl}
\end{equation}
The true accretion luminosity is $L = l\Ledd$, and so (\ref{LT}) would require an accreting mass
\begin{equation}
   M_{\rm accretor} = \frac{500\msun}{T_{\rm 0.1\, keV}^{4} }\left(\frac{1}{l}\right)
   \label{mimbh}
\end{equation}
for each object. 

Without beaming, the typical size, $r$, of the soft X--ray emitting region cannot self--consistently
be much larger than $O(1)$ for an accretion disc spectrum. 
Then since $p \sim 1$, (\ref{rl}) requires $l \sim 1/62$, we find 
$M_{\rm accretor} \gtrsim 3\times 10^4\msun$ from (\ref{mimbh}), i.e. a significantly massive IMBH. 

A magnetar model would
require a neutron star mass, so (\ref{mimbh}) requires $l \gtrsim 350$. But then from (\ref{rl}) there must be an implausibly large soft X--ray
region $r \sim 1.6\times 10^4$, which is $\sim 500$ neutron star 
radii\footnote{We will see in Section (\ref{sec:feedback}) that a large photosphere is possible in ULXs producing strong outflows, i.e. if 
a strong accretion--disc wind produces beaming.}. 
In addition, the implied luminosity is
$L \sim 5\times 10^{40}\,\ergs $, beyond the domain of validity for magnetar models for ULXs (Canuto et al., 1971).

In the context of the model described above, the observed correlation
(\ref{softX}) provides strong evidence in favour of accretion--disc wind beaming (and indeed, yields sensible results: \citealt{King2003}). We also note that the innermost (hottest) regions will naturally be the most collimated and therefore beamed, whilst one would expect radiation produced from around $r_{\rm sph}$ to be more isotropic. The final luminosity we observe may therefore be a somewhat complicated function and will naturally be inclination dependent. 
However, regardless of the final functional form of the beaming, it is inevitable that where the wind is optically thick and subtends a large scale-height (as seen in every MHD simulation of super-critical accretion), collimation and beaming must result.

\subsection{Models of pulsing ULXs (PULXs)}
\label{sec:pulx}
The discovery (beginning in 2014) of a small group of ULXs showing coherent pulsing has had a significant effect on the study of ULXs in general. First, it subjected existing models of ULXs to a stringent new test, {   as the mass of the accretor must be close to $1.4\msun$}. As remarked above, it removed the main motivation for the IMBH model, and we will see in Section \ref{sec:klk17} below, that it fitted naturally and self--consistently into the the disc--wind beaming picture, even though this was not developed with PULXs in mind.

But the removal of one candidate model (IMBH) did not simplify the discussion, as the discovery of PULXs naturally stimulated suggestions that magnetic effects on the accretion flow might be significant.

\subsubsection{The KLK17 model of PULXs}
\label{sec:klk17}

The discovery of PULXs has shown that ULX apparent luminosities can be highly super-Eddington. 
The luminosities of the classic (non Be-X) PULXs  range
from $\sim 20$ to $\sim 385$ times the Eddington value for a neutron star (see Table \ref{tb:combined}). For 
normal (pulsar-like) magnetic fields $\sim 10^{11} - 10^{13}$G, such values
exclude the possibility that they correspond to the actual luminosities since this would require implausibly 
high accretion rates (see Eqs. \ref{eq:lss} and \ref{eq:ladv}).
We will show in Sect \ref{sec:nomagnet} that the hypothesis that PULXs contain magnetars is extremely unlikely. Then the only 
physically reasonable option left is that the luminosity
of magnetic ULXs is beamed according to Eq.(\ref{eq:lsph}). Given this, \citet{King1605} proposed a 
beamed-emission model for the first detected PULX, M82 ULX-2. This model was successfully used
by \citet{King1702} (hereafter KLK17) to explain the properties of the three first--discovered PULXs, and later 
\citet{King1905} applied this model to the growing number of observed PULXs, including the Be-PULXs. 
\citet{King2003} explained how one obtains significant pulsed X-ray luminosity fraction in beamed radiation (see Section 3.7.5).
{   It is assumed throughout that for neutron star magnetic fields of the standard X-ray binary strength (i.e. $B \lesssim 10^{13} $~G) the surface emission is locally isotropic. 
}

\citet{Kluzniak1503} noted that PULXs are sharply distinct from other X--ray pulsars, not simply in their luminosities (by definition $L>10^{39}\rm erg\,s^{-1}$) but also in
their very large spinup rates ($\dot\nu > 10^{-10}\rm s^{-2}$ -- up to two orders of magnitude larger than normal XRPs). Between superoutbursts, transient Be-PULXs are normal {\sl spinning-down} X-ray pulsars, but when their luminosity reaches $\sim 10^{39}\,\ergs$
they become rapidly {\sl spinning-up} sources with $\dot\nu \gtrsim 10^{-10}\rm s^{-2}$ (see Table \ref{tb:combined}), demonstrating that such high spin-up rates are a generic property of PULXs\footnote{
One might insist that it is a coincidence that Be-X systems show the same spin-up as other PULXs when they reach the same luminosities, but here one rather expects that ``coincidences mean you're on the right path''\citep{Vbooy}.}.
\begin{figure}[h!]
\begin{center}
\includegraphics[width=0.8\columnwidth]{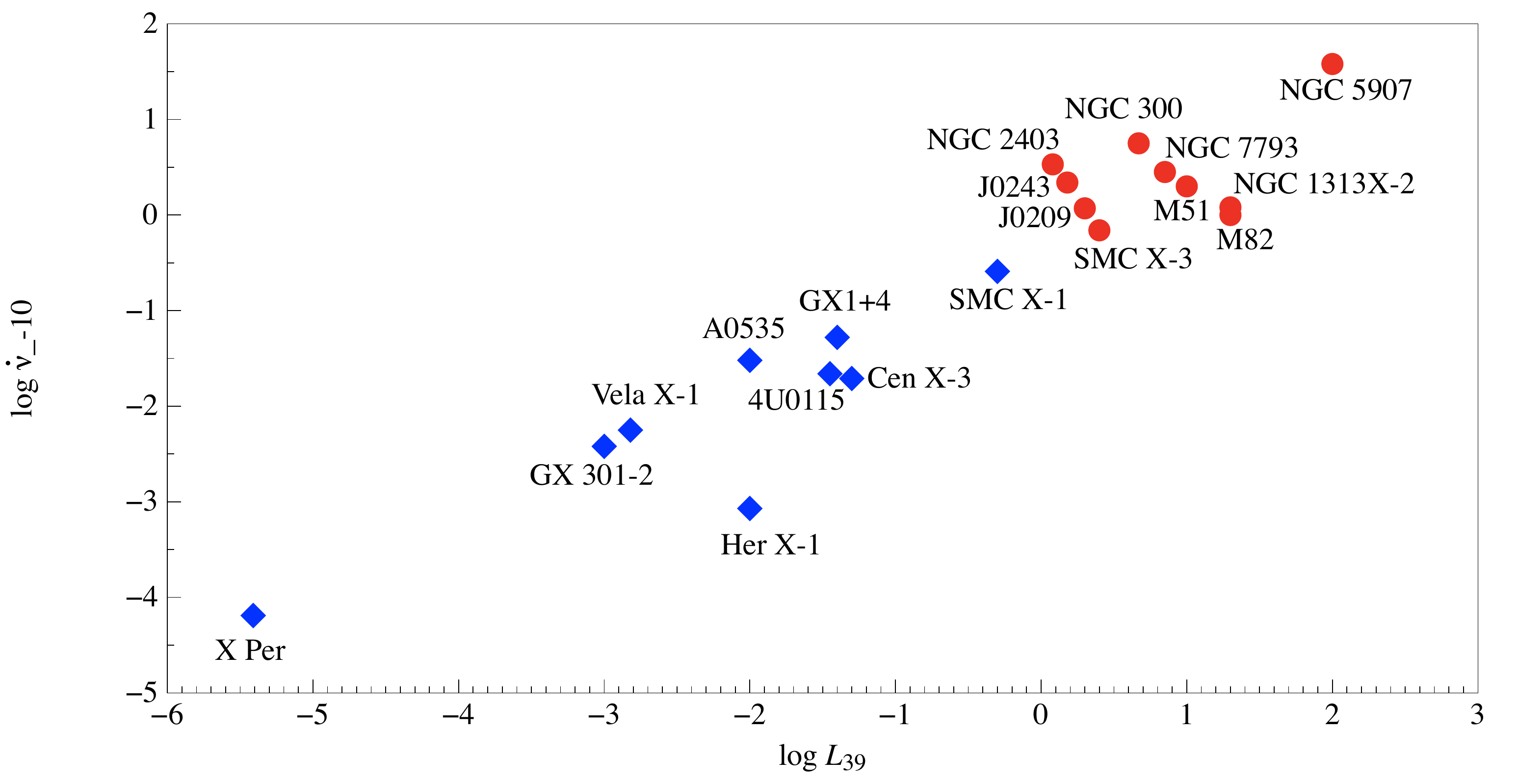} 
\caption{The $L_{39}$ -- $\dot \nu_{-10}$ diagram for XRPs and PULXs. Red dots: the ten PULXs with known spin-up rates (values from KLK17). Blue diamonds:
selected (for comparison) X-ray pulsars (see Table \ref{tab:XRP} - in the Appendix)}
\label{fig:dotnul}
\end{center}
\end{figure}
\FloatBarrier

The luminosities and spin-up rates  of PULXs are tightly correlated 
(Fig. \ref{fig:dotnul}). This correlation strongly implies that the dominating 
torque in the system is provided by accretion, as assumed in KLK17 and \citet{Vasilopoulos1218}, since
\begin{equation}
\dot\nu = \frac{\dot J(R_M)}{2\pi I} = \frac{\dot M (GMR_M)^{1/2}}{2\pi I} \propto \dot M^{6/7}
\label{eq:dotnudef}
\end{equation}
where $R_{\rm M} \propto \dot M^{-2/7}$ (e.g. Frank et al., 2002) 
is the magnetospheric radius and $I$ the neutron star's moment of inertia.

If we define PULXs by the twin properties of luminosities larger than $10^{39}\,\ergs$ \textsl{and} spin-up rates larger than $\sim 10^{-10}\rm s^{-2}$, these two quantities should form the basis
for models of these sources. At the very least, any cogent model must be able to reproduce these values.
KLK17 chose the first option, using the observed quantities $L_X$, $\dot \nu$ as input for their PULX model. Although there are no direct mass measurements
for neutron stars in PULXs, the likely mass range is limited. KLK17 make no assumptions about the magnetic field strengths, and these are outputs of the model. The resulting values lie in the typical range for standard pulsing X--ray binaries (see Table \ref{tab:ulx3b}), and compare favourably with the dipole magnetic field strengths measured from CRSFs (one of them for a non-pulsing neutron star ULX - see also section 2.2.3).

For the mass inflow, KLK17 use the
super--critical (`windy') solution of \citet{Shakura73} described in Section \ref{sec:ss73}. In this picture, the super--Eddington part of the mass supply below $R_{\rm sph}$  is eventually expelled as
a wind. This simultaneously provides the beaming, making PULXs appear super--Eddington 
and explains the strong outflows observed from PULXs.
The neutron star gains mass only at its effective Eddington rate (Eq. \ref{eq:lss}).

The spherization radius defined in Eq. (\ref{eq:resph}) can be written as
\begin{equation}
R_{\rm sph}\simeq 2.3 \times 10^6 \dot m_0 m_1\, \rm cm,
\label{eq:rsph}
\end{equation}
where $\dot M_0=\dot m_0\, \dot M_{\rm Edd}$  is the accretion rate at $R_{\rm sph}$, assumed equal to the mass transfer rate. For $R > R_{\rm sph}$ the disc is assumed to be a standard SS73
accretion disc.
From Eq. (\ref{eq:mdotr}) we have
\begin{equation}
\dot M(R) \simeq \dot m_0 \dot M_{\rm Edd}\frac{R}{R_{\rm sph}}.
\label{eq:mdotr}
\end{equation}
for ${R} < {R_{\rm sph}}$. 

KLK17 assume that the \citet{Shakura73} `windy' model describes the accretion flow between $R_{\rm sph}$ and the magnetospheric radius $R_{\rm M}$ defined 
by the equation \citep[][]{Frank0201} 
\begin{equation}
R_{\rm M} = 1.2 \times 10^8 q\, \dot m^{-2/7} m_1^{-3/7} \mu_{30}^{4/7}\, \rm cm,
\label{eq:rm}
\end{equation}
where $q\sim 1$ is a factor taking into account the geometry of the
accretion flow at the magnetosphere. Then from Eq. (\ref{eq:mdotr})
\begin{equation}
\dot M(R_{\rm M}) \simeq {\dot M_0}\frac{R_{\rm M}}{R_{\rm sph}}.
\label{eq:mr}
\end{equation}
Applying the model to PULXs with measured spin-up gives ${R_{\rm M}} < {R_{\rm sph}}$ in all cases. This implies that in these objects the accretion flow inside the magnetosphere is highly super--Eddington. Since it is hypersonic but
forced to follow the magnetic fieldlines it is highly dissipative, and one expects it to generate an outflow similar to 
that of \citet{Shakura73}, limiting the local luminosity to its Eddington value.

The total luminosity is then given by Eq. (\ref{eq:lss})
\begin{equation}
L \simeq L_{\rm Edd}\left[1 + \ln \dot m_0\right].
\label{eq:Llog}
\end{equation}

The luminosity 
from both parts of the super--Eddington outflow is assumed to be beamed. Outside the magnetosphere the beaming factor is taken to be
\begin{equation}
b \simeq \frac{73}{\dot m^2},
\label{eq:b2}
\end{equation}
as in Eq. (\ref{eq:beam}).
\begin{table*}
\centering
\caption{KLK17 model: derived properties of neutron star ULXs.}
\vskip5pt
{
\setlength{\tabcolsep}{1pt}
\label{tab:ulx3b}
{\small
\hfill{}
\begin{tabular}{ ||l|||c||c||c||c||c||c||c||c||c||} 
 \hline\hline
 Name &  $\dot m_0$ \ \ \ &$ b$ \ \ &  ${\bm B}\, q^{7/4}m_1^{-1/2}I_{45}^{-3/2}R^3_6$ [G]& ${\bm R_{\rm sph}}m_1^{-1}$ [cm] &  ${\bm R_M} m_1^{-1/3}I_{45}^{-2/3}$ [cm] & ${\bm P_{\rm eq}}q^{-7/6}m_1^{1/3}$ [s] &  ${\bm t_{\rm eq}}$ [yr]$^1$\\
 \hline\hline
 M82 ULX2 &   46  & 0.03 & $4.0\times 10^{10}$ & $1.1\times 10^8$ & $1.0\times 10^7$  &  0.02 & 15600 \\
 \hline
 NGC 7793 P13  &  25 &0.12  &  $1.1\times 10^{11}$ & $5.8\times 10^7$ & $1.6\times 10^7$ &  0.09 & 1386 \\
 \hline
 NGC5907 ULX1 & 95 &0.01 & $9.4\times 10^{12}$ & $2.2\times 10^8$ & $1.1\times 10^8 $  & 1.86 &  0 \\
 \hline
 NGC300 ULX1 & 24 & 0.13  & $5.3\times 10^{11}$$^{\heartsuit}$ & $5.5\times 10^7$ &  $3.2 \times 10^7$ & 0.19& 297 \\
 \hline
 M51 ULX7$^a$ &28 & 0.09 & $1.9\times 10^{11}$  &  $6.4\times 10^7$ & $2.0\times 10^7$ &0.08 & 1337  \\
 \hline
 M51 ULX7$^b$ &28 &0.09  & $6.9\times 10^{9}$  &  $6.4\times 10^7$ & $4.6\times 10^6$ &0.01 & $\sim 10^5$  \\
 \hline
 NGC 1313 X-2 & 46  & 0.03 & $5.3 \times 10^{10}$ & $1.1 \times 10^8$ & $1.8 \times 10^6$ & 0.03& 8641 \\
 \hline\hline
 SMC X-3$^{\rm Be}$   & 18 &0.23 & $2.3\times 10^{10}$ & $4.1 \times 10^7$ &  $7.1 \times 10^6$& 0.006&76621 \\
 \hline
 NGC 2403 ULX$^{\rm Be}$ & 13 & 0.43 & $2.5\times 10^{11}$ & $3.0\times 10^7$ & $2.3 \times 10^7$ & 0.16& 578 \\
 \hline
 Swift J0243.6+6124$^{\rm Be}$  & 14  &0.37 & $1.3\times 10^{11}$  &  $3.2\times 10^7$ & $1.7\times 10^7$   & 0.07 & 2047 \\ 
 \hline
 RXJ0209.6-7427$^{\rm Be}$ & 17 & 0.25 & $5.3 \times 10^{10}$ & $3.2 \times 10^{7}$ & $1.8\times 10^6$ &0.03 & 8665\\
 \hline
 NGC 1313 ULX$^{\rm Be;c}$  & 15 & 0.32 & ? & $ 3.5 \times 10^7$& ? &? & ?\\
 \hline\hline
 M51 ULX8& 17 & 0.25 & $\sim 3\times  10^{11}$$^{\clubsuit}$ & $3.2\times 10^7$ &  $2.7 \times 10^7$$^{\spadesuit}$  & non-pulsing & \\
\hline\hline
\end{tabular}}
}
\hfill{}
\vskip 0.2truecm 
%
\begin{itemize}
\item[]
{\footnotesize
--  Systems with $^{\rm Be}$ superscript have Be--star companions.\\
{$^1$ - calculated using the value of  $P_{\rm eq}q^{-7/6}m_1^{1/3}$ from the previous column.}\\
$^{\heartsuit}$--  consistent with observations \citet{Walton1804}.\\
$^{\clubsuit}$-- from observations  \citep{Brightman1804,Middleton2021_SS433}.\\
$^{\spadesuit}$ -- for a $\sim 10^{12}$\,G magnetic field.\\
$^a$ -- for $\dot \nu = 2.8\times 10^{-10}$;
$^b$ -- for $\dot \nu = 3.1\times 10^{-11}$ \citep{Vasilopoulos1910}.\\
$^c$ -- unknown $\dot \nu$.
}
\end{itemize}
\end{table*}

For accretion rates such that  radiation is geometrically beamed as described by Eqs. (\ref{eq:Llog}) and  (\ref{eq:b2}), one can deduce $\dot m_0$ from the observed X-ray luminosity $L=L_X$ by combining these two equations into
\begin{equation}
L_{X} \simeq 2.0 \times 10^{36}\dot m^2_0\left[1 + \ln \dot m_0\right]m_1 \ergs.
\end{equation}
Having $\dot m_0$ one obtains $R_{\rm sph}$ from Eq. (\ref{eq:rsph}). 

The second observed quantity, the spinup, follows from Eqs. \ref{eq:dotnudef} and \ref{eq:rm} as 
\begin{equation}
\dot\nu=3.3 \times 10^{-11} q^{1/2}\dot m^{6/7}m_1^{6/7}\mu_{30}^{2/7}I_{45}^{-1}\rm s^{-2}.
\label{eq:dotnumag}
\end{equation}
This gives the values of  the magnetic moment $\mu = BR^3$ (where $B$ is the field and $R$ the neutron--star radius) which in turn allows one to calculate the values of $R_M$ and $\dot m(R_M)$ from  Eqs (\ref{eq:rm}) and  (\ref{eq:mr}):
\begin{equation}
\mu_{30} = 0.04\,  q^{-7/4}\dot \nu^{3/2}_{-10} m_1^{1/2} I_{45}^{3/2}\, \rm G\,cm^{3}
\label{eq:mures}
\end{equation}
\begin{equation}
R_{\rm M} = 1.0 \times 10^7  \dot \nu^{2/3}_{-10} m_1^{1/3} I_{45}^{2/3}\, \rm cm
\label{eq:rmres}
\end{equation}
\begin{equation}
\dot m(R_{\rm M}) = 9.9\,  \dot \nu^{2/3}_{-10} m_1^{-2/3} I_{45}^{2/3},
\label{eq:mdotres}
\end{equation}
where $10^{-10}\dot \nu_{-10}=\dot \nu$.

The results are presented in Table \ref{tab:ulx3b}, with the values of the equilibrium period $P_{\rm eq} = 0.23 q^{7/6}m_1^{-1/3} \mu_{30}^{2/3}$\,s,
and the lower limit on the time to reach equilibrium at the present spin-up rate
\begin{equation}
t_{\rm eq}\equiv \frac{1}{\dot \nu}\left(\frac{1}{P_{\rm eq}} - \frac{1}{P }\right).
\end{equation}
(The  corotation radius
$R_{\rm co}\equiv \left(GMP_s^2/4\pi^2\right)^{1/3}$ is always larger than the magnetospheric and spherization radii. Thus only possibly NGC 5907 ULX1 is close to its equilibrium spin, and none of the observed PULXs is a propeller.)

For NGC300 ULX-1, the KLK17 model predicts a neutron--star magnetic 
field of $B > 5 \times 10^{12}$G (since $q < 1$ and $m_1 > 1$, the entries provided in the 4th column of Table \ref{tab:ulx3b}
are a factor of few lower than the predicted magnetic field values).
A magnetic field of very similar strength is deduced by \citet{Walton1804b} from the  CRSF inferred to be present in the X-ray 
spectrum of this PULX.  According to the KLK17 model, all seven PULXs have magnetic fields in the range $10^{11}-10^{13}$G, i.e. always in the
standard pulsing X--ray binary range and below the 
value defining magnetars.

The predicted values of the beaming factor $b$ are in the range $\sim 0.03$ -- $0.5$, except for the very luminous source in NGC 5709,
for which $b\approx 0.01$. For NGC 300 ULX-1, the model gives $b\approx 0.13$: \citet{Binder1808} use the KLK17 model and obtain $b\approx 0.25$,
because they deduce $\dot m_0$ from the average rather than the maximum luminosity. The beaming factor $b\approx 0.12$ for P13 in NGC7793 is consistent with observations of the X-ray irradiation of the neutron-star companion
\citep{Motch2018}. According to the KLK17 model, the luminosity of most of the known PULXs is only mildly beamed. {   This is worth stressing since
critics of the KLK17 model often claim that there are problems with `strong beaming'.}

We see that for all seven PULXs with known spin-up rates $\dot \nu$,
one has $R_{\rm M}\lesssim R_{\rm sph}$, which is probably the 
condition for observing pulses at all if the mass transfer is super--Eddington. The small difference between $R_{\rm sph}$ and $R_{\rm M}$
means that the flow is strongly super–Eddington on reaching $R_{\rm M}$. Most of this cannot land on the neutron star (let alone its polecaps only) and so must be ejected.

It is worth stressing that nothing in the assumptions of the model guarantees its self--consistency, i.e. that $R_{\rm sph} > R_{\rm M}$. But in every case this is satisfied by the deduced values. {   Further, exactly the same equations describe both the Be--star magnetic accretors and those PULXs which do not have Be--star companions: the first group cannot of course be modelled at all by assuming very strong dipole magnetic fields.}

{   The spin-up time-scales $t_{\rm eq}$ are probably much shorter than the lifetimes of the individual PULXs 
(see Table for NGC 1313 X-2)
so it seems very unlikely that we observe these systems during
their only approach to spin equilibrium. Instead, although most of them are probably rather far from their $P_{\rm eq}$  they have nevertheless the time to alternate spin-up and spin-down phases. Evidently, we can only see these systems during spin-up phases
(so that $\dot \nu$ has its maximum value) because centrifugal repulsion
during spin-down presumably reduces the accretion rate and so the luminosity. In addition the magnetic fields of the
PULXs appear to be significantly lower than is usual
for a new-born neutron star ($10^{12} - 10^{13}$G) which is consistent
with hypothesis that accretion of even a relatively small mass severely reduces the surface fields
of neutron stars – this is central to the concept of pulsar recycling,
which is implicated in the production of millisecond pulsars \citep[see][and references therein]{Bhatacharya0191}.
Things are even more complicated by the fact that the vast majority of ULXs do not pulse, despite
containing (magnetized) neutron stars, which suggests that alignment of spin and
central disc axes is rapidly suppressing (perhaps temporarily?) the pulsations, which are observed to be transient anyway. As expected from the KLK17 model, 
Be-star systems are prominent among the PULXs, since accretion is relatively weak and transient.}

Parametrizing  $R_{\rm M}=f R_{\rm sph}$, with $f\approx 0.3 - 0.9$, 
then from Eq. (\ref{eq:mr}), $\dot m(R_{\rm M})= f\dot m_0$ (KLK17) and from Eqs 
(\ref{eq:dotnumag}) and (\ref{eq:mdotres}) we find
\begin{equation}
\dot \nu = 7.8 \times 10^{-10}\, f^{7/6} q^{7/6} \mu^{2/3}_{30} m_1^{-1/7} I_{45}
^{-1},
\label{eq:nueq}
\end{equation}
in agreement with the observed spin-up values of PULXs (cf Fig \ref{fig:dotnul}).

{   It has been claimed \citep[e.g.][]{Mushtukov2011} that the KLK17 picture suffers from fundamental weaknesses since `strong beaming' would be incompatible
with the high observed pulsed fraction of the ULX radiation. As we noted (two paragraphs above), the beaming is not `strong'. Further, in Sect. (3.7.5) 
using the results of \citet{King2003}, we will show that the KLK17 model is fully compatible with the observed X--ray pulse fraction
(which varies in time and frequency), and also explains the very small number of observed PULX.}

\subsubsection{The G\'urpide et al. PULX models}

\citet{Gurpide9221} suggest an alternative
scenario in which the hardest ULXs are powered by strongly magnetised neutron
stars, so that the high-energy emission is dominated by the hard direct emission from the accretion column. It is assumed that one can explain softer sources as weakly magnetised neutron stars or black holes, in which the presence of outflows naturally explains their softer
spectra through Compton down-scattering and their spectral transitions. Outflows are also invoked to explain the dilution of the pulsed emission in
sources containing neutron stars. According to these authors NGC 7793 P13 and NGC 1313 X-2 would be strongly magnetized with $R_M > R_{\rm sph}$ in contradiction with the prediction of the KLK17 model. However, they find that NGC 5907 ULX-1 appears harder when dimmer. This is difficult to reconcile with their scenario and they have to conclude that for this source $R_M \approx R_{\rm sph}$, in agreement with the KLK17 model prediction.
However, \citet{Gurpide9221} do not take into account the possibility that the flow geometry is as described in Sect. \ref{sec:sinus}. In this case, under appropriate conditions one sees  the ``magnetic column" directly when $R_M > R_{\rm sph}$. In any case, one would expect that the flux--hardness relation would also depend on the intrinsic properties of the emitter and not on the source geometry only. The model geometry of \citet{King2003} is used by \citet{Pintore0321} with additional ingredients to interpret the spectral variation during the flares of NGC 4559 X7. 

Although M51 ULX-8 is not a PULX, we know that it is a magnetic system.
Assuming that its X-ray radiation is beamed, one can obtain its
$\dot m_0$ and so both $R_{\rm M}$ and $R_{\rm sph}$. 
As in PULXs this gives $R_{\rm M} \lesssim R_{\rm sph}$  (see Table \ref{tab:ulx3b}).
\citet{King1905} (see also Sect. \ref{sec:beamp}) speculate that M51 ULX-8 might
be a PULX, and suggest that pulsations may be seen in future observations of this system. \citet{Brightman1804} point out that rather long
XMM-Newton observations would be needed to detect $\sim1$s pulsations if the pulsed fraction is $\lesssim 45\%$, as in most
PULXs. \citep[Interestingly the exception is NGC 300 ULX1, in which ][inferred the presence of a cyclotron feature]{Walton1804b}.

Although simple, the KLK17 model correctly reproduces the main observed properties of PULXs and has passed
this test with each newly discovered PULX, in particular with the Be-PULX systems (see Table \ref{tab:ulx3b}). In the KLK17 picture, PULXs are ``normal'' (non-magnetar)
X-ray pulsars caught in a high accretion--rate episode of their binary evolution (non--Be systems: see Sections 3.11, 3.12) or orbital history (Be systems: see Section 3.15). 

\subsubsection{The \citet{Erkut0820} PULX models}

\citet{Erkut0820} (hereafter ETEA20) base their models of PULX on the well--known
 approach to the disc--magnetosphere interaction best known from the paper of
 \citet{Ghosh0879}. They assume that the luminosity in the magnetosphere
 is at most critical, using, as the relation between
the observed X-ray luminosity and the accretion rate:
\begin{equation}
\label{eq:lxerkut}
    L_X = \frac{\eta}{b}\dot M_{\star} c^2,
\end{equation}
where $\dot M_{\star}$ is the accretion rate at the neutron-star surface and $b$ is a beaming factor, and assumes $\dot M_{\star}=\dot  M_{\rm in}$, where
$\dot M_{\rm in}$ is the accretion rate at the innermost disc radius $R_{\rm in}$. (This is separated from the magnetosphere
by a boundary region of width $\Delta R\equiv \delta R\cdot b$.)

In other words
\begin{equation}
 L_{\mathrm{X}} \leqslant L_{\mathrm{c}} \simeq\left[1+311\left(\frac{B}{B_{\mathrm{c}}}\right)^{4 / 3}\right] L_{\mathrm{E}},
\end{equation}
where $B_{\mathrm{c}} \equiv m_{\mathrm{e}}^{2} c^{3} / \hbar e=4.4 \times 10^{13} \mathrm{G}$ {(see Eq. \ref{eq:sigmaB})}.

The beaming considered by ETEA20 is assumed {\sl magnetic} {  (i.e. not due to an outflow} in origin, with beaming factor
\begin{equation}
 b_{\rm M}=\frac{A_{\mathrm{c}}}{\gamma A},   
\end{equation}
where $A_{\mathrm{c}}$ is the polar cap area, $A = 4\pi R^2_{\star}$
is the total surface
area of the neutron star, and $\gamma < 1$ ``is a normalization constant
corresponding to the maximum fractional area of the polar cap'' \citep{Erkut0820}.
Since they are physically different, and act in different parts of the accretion flow in PULXs, direct comparison of the values
of ETEA20's $b_{\rm M}$ with those of KLK17's outflow--generated $b$ is not of much interest. 

The accretion flow between $R_{\rm in}$ and $R_{\rm sph}$ (defined as in Eq. (\ref{eq:resph}), with a factor with 27/2, instead of 15) is allowed
to be super--Eddington and is assumed to be described by the
`windy' \citet{Shakura73} solution (Eq. \ref{eq:mdotr}).

The inner radius of the disk is determined by the balance between magnetic and material stresses
\begin{equation}
\label{eq:mageq}
    \frac{d}{d R}\left(\dot{M} R^{2} \Omega\right)=-R^{2} B_{\phi}^{+} B_{z},
\end{equation}
where $\Omega$ is the angular velocity of the innermost disc matter
within the boundary region, $B_z$ is the poloidal magnetic field,
$B^+_\phi=\gamma_\phi B_z$ is the toroidal magnetic field at the surface of the
disc, and $\gamma_\phi$ is the azimuthal pitch that can be expressed as $\gamma_\phi \simeq \omega_{\star} -1$, where
$\omega_{\star}$ is the so-called ``fastness parameter''
\begin{equation}
    \omega_{*} \equiv \frac{\Omega_{*}}{\Omega_{\mathrm{K}}\left(R_{\mathrm{in}}\right)}=\left(\frac{R_{\mathrm{in}}}{R_{\mathrm{co}}}\right)^{3 / 2}.
\end{equation}
At spin equilibrium, the fastness parameter is assumed to be equal to a (model-dependent) critical value $\omega_{\star}=\omega_c$.
ETEA20 use $\omega_c=0.75$.

Integrating Eq. (\ref{eq:mageq}), using Eq.(\ref{eq:lxerkut}) and defining {   poloidal field} $B_z \simeq - \mu /R^3$ one obtains an expression for the inner disc radius
\begin{equation}
\label{eq:erin}
    R_{\text {in }}=\left(\frac{ \mu^{2} \sqrt{G M_{\star}}\delta}{b_{\rm M} R_{\star} L_{{X}}}\right)^{2/7}.
\end{equation}
Since one can write for a strong enough magnetic field
\begin{equation}
    \frac{A_{\mathrm{c}}}{A} \simeq \frac{R_{\star}}{4 R_{\mathrm{in}}} \cos ^{2} \alpha,
\end{equation}
where $\alpha$ is the angle between the rotation and magnetic axes \citep[][the star's spin and the disc rotation axes are assumed to be aligned]{Frank0201}, one gets
$b=\left(\cos ^{2} \alpha / 4 \gamma\right) R_{\star} / R_{\text {in }}$.

Using $\mu = B_z R_{\star}^3$  
one can then write
\begin{equation}
\label{eq:bm}
    b_{\rm M}=\left(\frac{\cos ^{7} \alpha L_{{X}}}{32 \gamma^{7 / 2} \sqrt{G M_{\star}} B^{2} R_{\star}^{3 / 2} \delta}\right)^{2 / 5},
\end{equation}
for the beaming factor and express the fastness parameter as
\begin{equation}
\label{eq:efast}
    \omega_{\star}=\left(\frac{\gamma \sqrt{G M_{\star}} B^{2} R_{\star}^{4} \delta}{\cos ^{2} \alpha L_{{X}}}\right)^{3 / 5} \frac{2 \pi}{P \sqrt{G M_{\star}}},
\end{equation}
where $P=2\pi/\Omega_{\star}$.
In addition to observed quantities, $L_X$ (ETEA20 use the observed flux $F_X$, but since, with one exception, PULXs are extragalactic, this precaution does not seem to be necessary), and $P$, Eqs/ \ref{eq:bm} and \ref{eq:efast} contain unobservable (at least in PULXs) quantities $\alpha$ and $\gamma$.
They can be eliminated by the use of the torque equation that ETEA20 write under the form
\begin{equation}
\label{eq:etorque}
    2 \pi I\, \dot{\nu} =\omega_{\star}^{1 / 3} n_{0} \sqrt{G M_{\star} R_{\mathrm{co}}} \dot{M}_{\star},
\end{equation}
where $n_0$ is a constant of order unity corresponding to a dimensionless torque.
Given the neutron--star mass (ETEA20 try $M_{\star} =2\Msun$ for M51 ULX-7), moment of inertia and radius,    one then obtains two equations for the four quantities defining the physics of the system: $B$, $b_{\rm M}$, $\omega_{\star}$ and $\delta$ as a function of the observables $L_X$, $P$ and $\dot P$ (or $\dot \nu$):
\begin{equation}
    B=\frac{\omega_{\star}}{R_{\star}^{3}} \sqrt{\frac{2 G M_{\star} I|\dot{P}|}{\pi n_{0} \delta}},
\end{equation}
and
\begin{equation}
    \omega_{\star}={16 \pi^4 G M_{\star}}\left(\frac{I|\dot{P}|}{n_{0} L_{{X}} P^{7 / 3} R_{*} b_{\rm M}}\right)^{3}.
\end{equation}
These two equations allow us only for delimiting the range of magnetic-field strength and fastness-parameter values compatible
with observation. The boundary region parameter is assumed to lie in the range $0.01 \leq \delta \leq 0.3$. The determined range of
physical parameters can be made narrow by assuming that $L_X=L_c$, or $L_X < L_c$ and the fastness parameter can be set equal to the equilibrium value
$\omega_{\star}=\omega_c=0.75$.

In their paper, ETEA20 consider several alternative scenarios: (i) the PULXs are away from spin equilibrium; an efficient standard
spin-up torque is used to account for the observed spin-up rates, (ii) the PULXs are so close to spin equilibrium that the
fastness parameter is given by its critical value, (iii) the X-ray luminosities of the PULXs can be well represented by the
maximum critical luminosity, (iv) the conditions described in (i) and (iii) both apply, (v) the conditions described in (ii) and
(iii) both apply, and (vi) the X-ray luminosities of the PULXs are subcritical and either the condition described in (i) or the
condition described in (ii) applies.
They found that the narrowest ranges for $B$, $b_{\rm M}$, and $\omega_{\star}$ are found when they use
the critical luminosity condition along with either the observed
spin-up rates or the spin-equilibrium condition.  Scenario (iv),
based on the observed spin-up rates at critical luminosity, is the only one that works
well for all the PULXs, implying $B$ in the $ 10^{11} - 10^{13}$ G range. The results for scenario (iv) are presented in Table \ref{tab:etea} \citep[Table 5 in ][]{Erkut0820}.
\begin{table*}[ht]
\caption{ETEA20 models of PULXs [scenario (iv)]}
\vskip 5.0pt
\centering
{
\setlength{\tabcolsep}{3pt}
{
\hfill{}
  \begin{tabular}{|l||c|c|c|c|c|}
    \hline
    Source & $L_X\left(\rm erg\, s^{-1} \right)$ & $\dot v \left(\rm  s^{-2}\right)$ & $b_{\rm M}$  & $\omega_*$   & $B\left({\rm G}\right)$   \\
    \hline \hline
    M82 X-2  & $6.5 \times 10^{39}$&  $1.1 \times 10^{-10} $    &0.16--0.25 & 0.060--0.22 & 2.0--2.9 $\times 10^{12}$\\
    ULX NGC 7793 P13 & $9.4 \times 10^{39}$ &  $1.2 \times 10^{-11} $     &0.021--0.027 & 0.065--0.14 & 1.2--3.2 $\times 10^{11}$ \\
    ULX NGC 5907 & $1.5 \times 10^{41}$   &   $3.9 \times 10^{-9} $     &0.19--0.30 & 0.16--0.61 & 2.7--3.8 $\times 10^{13}$\\
    NGC 300 ULX1 & $4.3 \times 10^{39}$&  $5.6 \times 10^{-10} $ &  $\sim 1$ & 0.015 & 6.8 $\times 10^{12}$\\
    M51 ULX-7 & $7.0 \times 10^{39}$ &  $3.1 \times 10^{-11} $ & 0.052--0.0 &0.019--0.063 & 6.2--10 $\times 10^{12}$\\
    Swift J0243.6+6124 & $1.7 \times 10^{39}$  &   $2.4 \times 10^{-10} $ &  $\sim 1$ &0.041& 3.5$\times 10^{12}$ \\
\hline\hline
  \end{tabular}}
  }
  \hfill{}
  \vskip 0.2truecm
  \begin{itemize}
\item[] {\small {   Note.} In this scenario the magnetic field, beaming fraction, and fastness parameter for Swift J0243.6+6124
are only marginally obtained at $b_{\rm M}\sim1$ for $M_{\star}=1.2\msun$ and
$R_{\star}=10$ km, and no common solution is found for NGC~1313~X-2 (even a mass as low as $0.9 \Msun$ does not help). For all other sources $M_{\star}=1.4\msun$ and
$R_{\star}=10$ km are used.}
\end{itemize}
  \label{tab:etea}
\end{table*}
The ETEA20 models do not allow for a prediction of the pulsed fraction of the emitted radiation, nor the pulse shapes, and the authors refer to the ``accretion curtain'' of \citet{Mushtukov1705} as a possible source of the pulsed luminosity.

The conclusions of the \citet{Erkut0820} paper: PULXs have sub--critical magnetic fields and their luminosity is subject to medium beaming are the same as those obtained by \citet{King1605,King1702,King1905,King2003} but they differ in their physical motivations. First, the beaming mechanisms are different in the two approaches to the problem. Second, while ETEA20 require the accretion flow to be always at most critical and assume that the accretion rate is constant inside the magnetosphere, KLK17 allow the accretion flow to be super-Eddington and losing mass also when following the magnetic field-lines. It is perhaps worth noting that these two very different
models both exclude magnetar--strength fields.
\begin{figure}[h!]
    \centering
    \includegraphics[width=0.4\columnwidth]{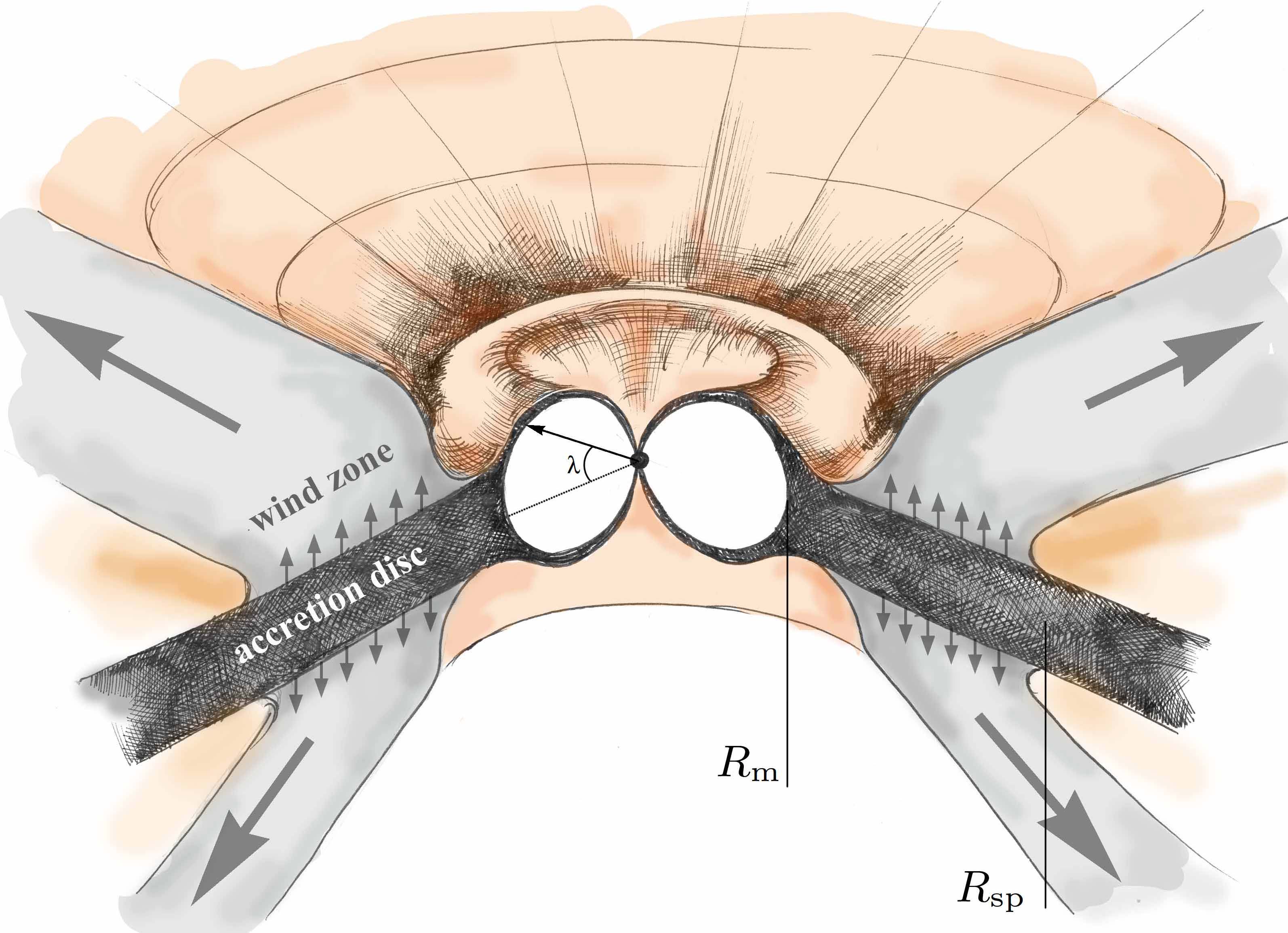}
    \caption{Schematic picture of a PULX according to \citet{Mushtukov1705,Mushtukov1903}. $R_{\rm m}$ and $R_{\rm sp}$ are respectively the magnetospheric and  spherization radius. $\lambda$ is the angle between the disc plane and the magnetic axis. Accreting matter not lost in the wind is accreted onto the central object forming an envelope which is optically thick at accretion rates  typical of  PULXs. In the KLK17 model, the configuration is similar, but $R_{\rm m} \lesssim R_{\rm sp}$ and the wind does not end up at the magnetospheric radius. 
    (see Fig. \ref{fig:alignement}). {    Note that the Figure of Mushtukov et al., (2017, 2019) shown here assumes that the neutron star spin is aligned with the disc axis, although this is not explicitly stated. We argue in subsection \ref{sec:sinus} that this assumption is in general not justified. In addition the configuration shown appears to be axisymmetric, i.e. has the magnetic axis parallel to the spin (and disc) axes. This would of course preclude pulsing, so there must in reality be some deviation from axisymmetry.}
    \label{fig:mushtuenv}}
\end{figure}
\FloatBarrier

\subsubsection{The origin of ``sinusoidal'' pulses in PULXs}
\label{sec:sinus}

X--ray pulse light curves for PULXs are generally described as being `sinusoidal', i.e. without obvious eclipses, and continuously modulated (see Sect. \ref{sec:puls}).  
In other words, the X-ray light curves of PULXs are never flat and never zero.

\citet{Mushtukov1705,Mushtukov1903} argued that this was understandable
if, perhaps 
because of the large optical depth, the photospheric surface emitting the X-rays occupied a large fraction of the magnetosphere.
Then this region would neither be permanently 
in view, nor periodically occulted. Accordingly,
\citet{Mushtukov1705,Mushtukov1903} proposed that the X--rays are emitted
by an optically thick envelope defined by the accretion flow over the
neutron--star magnetosphere. This is consistent with the 
accretion along fieldlines being a bending hypersonic flow, which must  therefore shock as 
suggested by \citet{King1905} (see also Sect.\ref{sec:klk17}). The spin modulation occurs because the magnetic axis is inclined to the spin axis.

\citet{Mushtukov1705,Mushtukov1903} do not attempt to reproduce the observed PULX lightcurves and do not explain why magnetized systems such as ULX-8 in M51 do not pulse. The simplest explanation would be that the spin and magnetic axes are aligned, but then one would have to explain why the few pulsing systems are different in this respect from the probable majority of NULXs.

\subsubsection{Beamed pulsed X-ray radiation}
\label{sec:beamp}

An argument often made against the beaming model of PULXs is the belief that the observed high ($\lesssim 45\%)$ pulsed fraction of the X-ray emission cannot appear in radiation passing through a funnel because reflections from the funnel walls would destroy the signal coherence.

\begin{figure}[h!]
\begin{center}
\includegraphics[width=0.6\columnwidth]{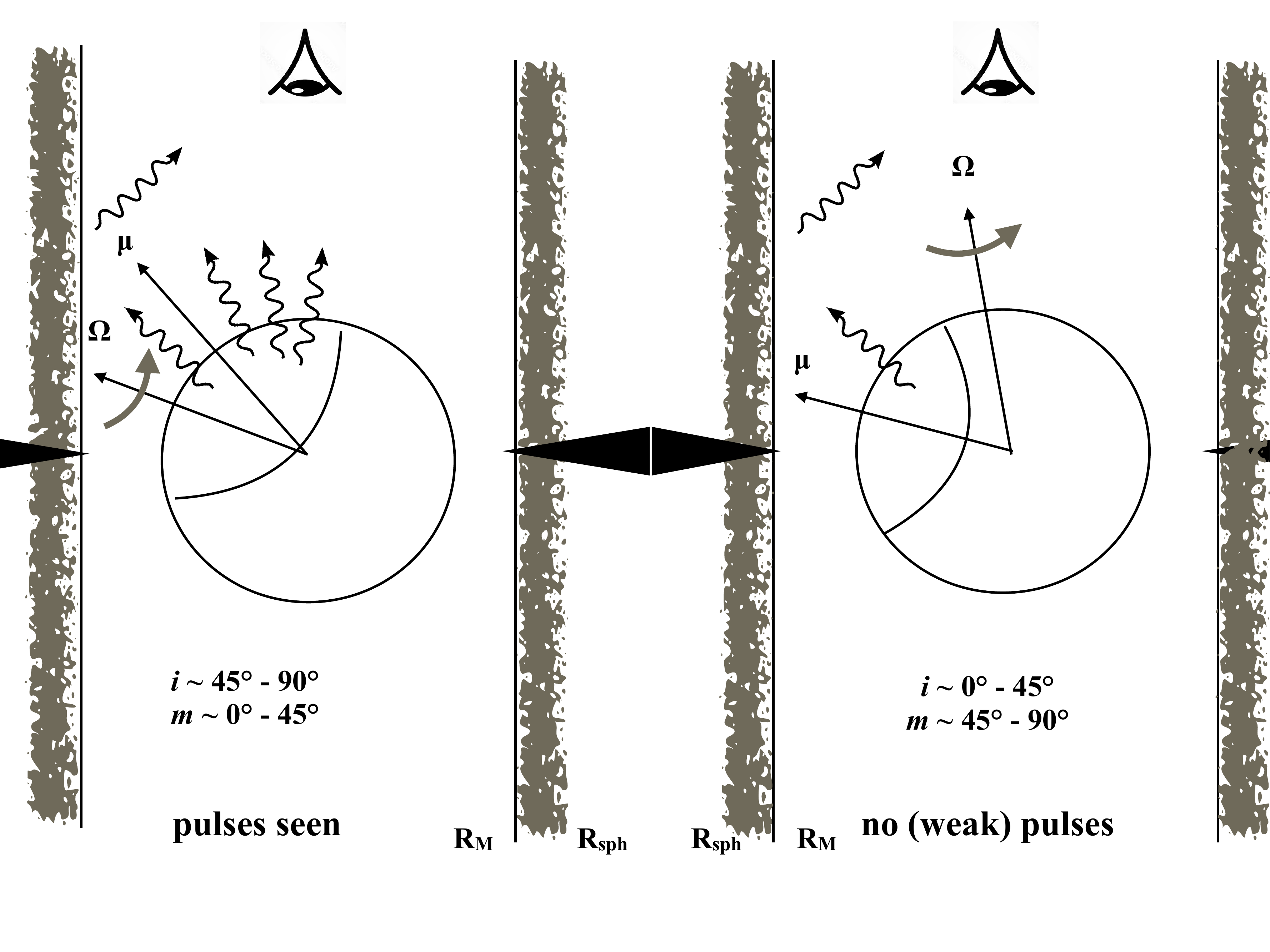}
\caption{Effect of spin orientation on pulsing. Left: the neutron--star spin
and the accretion disc beaming axes are strongly misaligned, so that a 
significant part of the pulsed emission is periodically occulted by the neutron--star body at certain spin phases.  This gives a large pulse fraction. Right: the neutron--star spin 
and central disc axes are assumed to be substantially
aligned {    (this special condition is assumed without any explicit statement in papers asserting that beaming cannot produce pulsed emission)}. Much
of the primary X--ray emission is scattered by the walls
of the beaming `funnel' before escaping. In this kind of configuration the pulse fraction is
reduced or suppressed. \citep{King2003}.}
\label{fig:alignement}
\end{center}
\end{figure}

We note first that 
although many (if not most)
ULXs probably contain magnetised neutron stars, only a small fraction show pulses. Second, \citet{King2003} point out that the argument above assumes {    (without any explicit statement)} that the spin axis of the neutron star is aligned with the accretion disc axis, and so with the beam axis\footnote{This unstated assumption has a similar status to the implicit belief that ULX nebulae must be Str\"omgren spheres powered by the direct radiation of the ULX, rather than wind bubbles made by its powerful near--spherical wind -- see Sections \ref{sec:small} and \ref{sec:feedback}. In both cases the effect is to `rule out' beaming.}.
In fact observations of high--mass X--ray binaries show that the neutron star spin is generally not aligned with the disc axis, probably because of the asymmetric nature of the supernova giving birth to the neutron 
star. 
\\
The beamed emission {    will appear sinusoidal, (i.e. the light curve has no constant intervals) 
if three conditions hold simultaneously:
\\
(a) there is significant misalignment of spin and beam (disc) axes  (cf Figure \ref{fig:alignement}), and 
\\
(b) the emitting region is not essentially symmetrical around the spin axis, and
\\
(c) its linear scale is comparable to the neutron star radius, so that a significant fraction of its area, but not the whole of it, is occluded by this star
as it spins}
\\

Two processes can align the spin and disc (beam) axes. One
is warping of the central disc into the neutron--star 
spin plane through differential torques (both magnetic and precessional), 
and the other is direct 
accretion of angular momentum (characterized by $R_M$) from an 
unwarped disc lying in the binary orbital plane. 
The latter seems more 
likely, as any interruption
in the accretion flow means that disc warping has to `start again'. 
Neither process will be be very efficient
in the Be--star PULXs, where accretion rate is less than $\lesssim 10 \dot M_{\rm Edd}$
 and confined to short transient episodes.
 
The outcomes of the two spin orientations  are very different. 

\begin{itemize}

\item If there is strong misalignment of the spin and
beaming axes, a significant part of the pulsed emission can escape without
scattering, giving a large pulse fraction (Fig. \ref{fig:alignement}, left). This must be 
maximal at the
highest X--ray energies, as scattering makes the X--rays both softer and 
less pulsed. This 
correlation of pulse fraction and X--ray energy is well known for observed 
PULXs \citep{Kaaret1708}. 

\item If instead there is substantial
alignment of the spin and central disc axes,
much of the primary X--ray emission is scattered by the walls
of the beaming `funnel' before escaping (Fig. \ref{fig:alignement}, right).
Since the light--travel time across
the funnel is usually
comparable with the pulse duration, the pulse amplitude can be
severely reduced. {This would result in an inability to detect many highly collimated PULXs through pulse searches (\citealt{Mushtukov1903}); as accumulate and improve in quality, this picture will be tested more
thoroughly.} The pulse fraction can of course also be reduced or 
entirely removed if enough matter accretes to weaken the neutron--star 
magnetic field. This presumably happened in the case of Cyg X--2,
which is a survivor of a phase of strongly super--Eddington accretion 
\citep{King9910}. Here the neutron star is not noticeably magnetic,
and has probably gained a few tenths of $\msun$ during the 
super--Eddington phase.

\end{itemize}

These outcomes appear to be consistent with observations of ULXs. Very
few ULXs show pulsing. But if most ULXs are the
direct descendants of high--mass X--ray binaries (HMXBs, 
once the companion
star fills its Roche lobe), or Be--star HMXBs,
the vast (unpulsed) majority must contain neutron stars.
First, almost all HMXBs contain neutron stars rather than black
holes \citep{King1605,King1702}. Second, 
neutron--star systems are more
super--Eddington and  more beamed than black--hole systems 
with the same mass transfer rate, and 
therefore have {\it higher} apparent ULX luminosities (see Section 3.6).
In addition, there is at least one ULX  whose 
spectrum shows
a CRSF corresponding to a pulsar--strength magnetic field \citep{Middleton2021_SS433},
but which does not show pulsing, strongly suggesting that is it an 
aligned accretor. 

The vast majority of ULXs do not have detected pulses, despite
containing neutron stars. Unless this is caused by a lack of data for ULX pulse 
searches 
it implies 
that alignment of spin and central disc axes is rapid, and possibly that
field suppression through accretion occurs also. As expected in this picture, Be--star
systems are prominent among the PULXs since accretion rates are relatively low 
and the episodes of increased luminosity are transient. 

\subsection{Numerical simulation of super-Eddington accretion flows}
\label{sec:numer}

Most of the numerical simulations of super-Eddington accretion on to compact bodies concern black holes. The purely absorbing inner boundary conditions make the problem simpler than for neutron stars, which have hard surfaces and, in the most interesting case, strong and rotating magnetic fields. We first review the simulations of black--hole super-Eddington accretion.

\subsubsection{Numerical simulations of black-hole accretion}
\label{sec:bhnum}

Analytical or quasi-analytical solutions of super--Eddington accretion on to black holes represent two extreme possibilities: either the local emission is kept close to the Eddington value by blowing out the ``excess" matter \citep{Shakura73}, or by advecting the ``excess'' energy towards the black hole, where it disappears from the Universe \citep{Abramowicz8809}. In both cases the total luminosity radiated by the accretion flow is $1 + \ln \dot m_0$ times the Eddington luminosity, where $\dot m_0$ corresponds either to a constant accretion rate (slim disc) or is equal to the rate at which matter is brought to the accretion flow, the accretion rate decreasing to keep the local flux at the Eddington value (wind/outflow solution). Although the accretion rate reaching the black hole can differ between the two cases by several orders of magnitude, in neither case can the emitted luminosity  be larger than $\sim 10\Ledd$\footnote{For ``hyper-Eddington'' accretion on to an intermediate-mass or super-massive black hole with $\dot m = 5000$, the luminosity is still $\sim 30\,\Ledd$ \citep{Sakurai1016}.}. Then for a ULX with $L_X > 10\,\Ledd$, this apparent luminosity cannot be equal to the physical luminosity, whatever the accretion flow model: the observed super-Eddington luminosity must be beamed. Both extreme models imply beaming depending on the accretion rate.

For slim discs, \citep{Wiktorowicz1709} fitted the relative semi-height of the scattering photosphere calculated analytically and numerically by \citet{Lasota1603} by the formula:
\begin{equation}
\frac{H_{\rm phot}}{R}\approx \frac{1.6}{1+\frac{4}{\dot{m}}},   
\end{equation}
and from 
\begin{equation}
\left(\frac{H_{\rm phot}}{R}\right)^{-1}=\tan \frac{\theta}{2},    
\end{equation}
obtained the beaming factor 
\begin{equation}
    b=1-\cos \frac{\theta}{2}.
\end{equation}
This beaming becomes nearly constant for $\dot m \gtrsim 100$ which gives the maximum beaming for $b\approx 0.15$. 
For the windy \citet{Shakura73} model with beaming given by Eq. (\ref{eq:beam}), $b=0.15$ corresponds to $\dot m=22$.
One can expect that in reality the solution describing super-Eddington accretion on to a black hole may have both
outflows and advection, but the relative proportions can only emerge from numerical calculations and comparison with
observations. Unfortunately both these methods encounter serious difficulties. 

Before we discuss numerical solutions describing
super-Eddington mass supply to black holes, we should note that, without estimates of the black-hole masses in ULXs, these models remain untested, since we do not know the true Eddington luminosities. We now know that the masses of black holes descending from stellar evolution can differ
by an order of magnitude \citep{Abbott1020}, so this is a real handicap. Currently there is only one upper limit on the compact-object mass in a ULX \citep{Motch1410}, and here the accretor was later found to be a neutron star \citep{Furst1611}.

In discussing numerical simulations of super-Eddington accretion flows, we emphasize that these
cannot so far represent the 
conditions expected in ULXs, i.e. a disc of at least $\gtrsim 10^5\,R_g$ in size, geometrically thin for $R \gtrsim 15 \dot m R_g$. This is not a surprising result: MHD codes attempting to model the disc viscosity from a first--principles treatment of the magnetorotational instability currently find values of the dimensionless viscosity parameter $\alpha$ significantly smaller
($\lesssim 0.01$) than observational estimates from time variations, which strongly suggest $\alpha \gtrsim 0.3$ 
\citep{Tetarenko0218}. 
The reason for this discrepancy is likely to be numerical diffusivity, as already recognized in the solar and MHD--fluids literature. One should add, however, that the viscosity parameter can be reliably determined in the outer disc's regions only (see, e.g., \citealt{Kotko1209}).

Most, if not all, numerical simulations of such flows start with a torus, with an assumed distribution of angular momentum that after relaxation forms an accretion flow. In most cases the ``circularization radius" of a torus--generated flow (the Keplerian orbit corresponding to the torus angular momentum) is kept rather small ($\sim 30 - 50\, R_g$). This is because the simulations aim to get quasi-stationary solutions, but this requires run times significantly longer than the viscous time of the accretion flow.

Since the viscous time at radius $R$ is 
\begin{equation}
t_{\mathrm{vis}} \sim \alpha^{-1} \left(\frac{H}{R}\right)^{-2}     \left(\frac{R^3}{GM}\right)^{1/2},
\end{equation}
one can write
\begin{equation}
t_{\mathrm{vis}} \sim 1.1 \times 10^{5} \left(\frac{R}{100 R_{g}}\right)^{3 / 2}\left(\frac{\alpha}{0.1}\right)^{-1}\left(\frac{H / R}{0.5}\right)^{-2} \, t_g,   
\end{equation}
where the dynamical time $t_g\equiv R_g/c$, is  often used to express the duration of numerical calculations ($t_g=5\times 10^{-5}$ s, for a $10\Msun$ black hole). Typical 3D GRRMHD (general-relativistic radiation magneto-hydrodynamic) calculations by \citet{Sadowski0316} require runs of $20 000\, t_g$, so it is not surprising that their quasi-stationary solutions require rather small discs, with $R \lesssim 40 R_g$.

The pioneering simulations of super-Eddington accretion by \citet{Eggum0788} were able to represent only 0.6 seconds of physical evolution of the system, while the improved scheme of \citet{Okuda0402} increased this duration to 1.6 seconds, both shorter than the corresponding viscous time. \citet{Ohsuga0507}, using 2D radiation hydrodynamic (RHD) simulations were the first to obtain a quasi-steady structure of
a supercritical accretion flow onto a black hole. It took another 10 years for other groups to join the Japanese team of Ken Ohsuga in trying to model supercritical accretion onto compact bodies. \citet{Sadowski1511,Sadowski0316} used a general-relativistic RMHD code first in two, then in three dimensions, while \citet{Jiang1412} performed 3D simulations in the framework of the so--called pseudo--Newtonian potential of \citet{Paczynsky0880}\footnote{Since the Paczy\'nski-Wiita ``potential" mimics the general--relativistic description of massive particle orbits in the black--hole's equatorial plane, it should actually be called ``pseudo--relativistic", but it is too late to change the widely accepted nomenclature. (The name ``Paczyński'' was misspelled in the original publication, hence the spelling used in the reference list.)}. Except for \citet{Takahashi0716}, who used the general-relativistic description of the black-hole accretor, other simulations of the Ohsuga group \citep{Ohsuga0507,Ohsuga1107,Hashizume0815,Kitaki1218,Kitaki0121} used the pseudo-Newtonian description of particle motion in the gravitational field of the accreting body. 
\begin{table*}[ht]
\centering
\caption{    Numerical Models of super-Eddington black hole accretion \citep[Adapted from][]{Kitaki0121}}
\vskip 5pt
{
\label{tab:numermod}
\centering
{
\hfill{}
  \begin{tabular}{|l|l|l|l|l|l|l|l|l|}
    \hline\hline
    Paper & Method & Compton & $R_{\rm out}$ \  &  $R_{\rm K}$  & $R_{\rm qss}$ & $R_{\rm trap}$ & $\dot{M}_{\rm BH}$ & $\dot{M}_{\rm outflow}$ \\
     &             & [Yes/No] & $[R_g]$ & $[R_{g}]$ & $[R_{g}]$ & $[R_{g}]$ & [$\dot m$] & $[\dot m$] \\
    \hline
    Kitaki$+$21 & 2D-RHD       & Yes  & $6000$ & $4860$& $\sim 1200$  & $\sim 540$   & $\sim 18$ &$\sim 2.4$\\
    \hline
    Ohsuga$+$05    & 2D-RHD       & No  & $1000$  & $200$ & $\sim60$   & $\sim400$  & $\sim260$ &\\
    Ohsuga$+$11    & 2D-RMHD      & No  & $105$  & $40$  & $\sim10$   & $\sim150$  & $\sim100$ &\\
    Jiang$+$14     & 3D-RMHD      & No  & $100$   & $50$  & $\sim 40$   & $\sim660$  & $\sim 22$ &$\sim 40$\\
    S\c{a}dowski$+$15 & 2D-GR-RMHD & Yes & $5000$ & $42$  & $\sim 70$   & $\sim 1280$  & $\sim42$ &$\sim700$\\
    S\c{a}dowski$+$16 & 3D-GR-RMHD & Yes & $1000$  & $40$  & $\sim20$   & $\sim520$  & $\sim18$ &$\sim52$\\
    Hashizume$+$15 & 2D-RHD       & No  & $10000$ & $200$  & $\sim200$ & $\sim460$  & $\sim15$ &$\sim50$\\
    Takahashi$+$16 & 3D-GR-RMHD    & No  & $250$  & $34$& $\sim20$   & $\sim600$  & $\sim20$ &\\
    Kitaki$+$18    & 2D-RHD       & Yes & $6000$ & $600$ & $\sim400$  & $\sim840$  & $\sim28$ &$\sim30$\\
    Jiang$+19^{*} $    & 3D-RMHD      & Yes & $1600$  & $80$  & $\sim30$   & $\sim760$  & $\sim25$ &\\
    \hline\hline
  \end{tabular}}
  }
  \hfill{}
  \vskip 0.2truecm
  \begin{itemize}
      \item[] {\footnotesize
    $R_{\rm out}$ -- radius at the outer boundary,
    $R_{\rm K}$ -- initial Keplerian radius (``circularization radius"), 
    $R_{\rm qss}$ -- radius, inside which a quasi--steady state is established,
    $R_{\rm trap}$ -- photon-trapping radius derived based on equation \ref{eq:trapp} (with right hand side multiplied by $\sim$1.5),
    $\dot{M}_{\rm BH}$ is the accretion rate onto the black hole,
    $\dot{M}_{\rm outflow}$ is the outflow rate at around $R_{\rm out}$.
    It is indicated whether the Compton scattering effect is taken into account or not.\\
    $^{*}$ In these simulations the black-hole mass is $5\times 10^8\Msun$ }
  \end{itemize}
  \label{table1}
\end{table*}
All these simulations produce rather similar results: strong outflows and regions of the inflow dominated by advection. They differ mostly
in details, except for \citet{Jiang1412} who observe an effect not seen by the other teams: vertical advection of radiation caused by magnetic buoyancy 
that transports energy faster than photon diffusion.
This allows a significant fraction of the photons to escape from the surface of the disk before being advected into the
black hole; vertical ``advection" reduces horizontal advection. The reason for this discrepancy between simulation results is not understood.
The main suspect is the treatment of radiation in the code. While \citet{Jiang1412} solve the full radiative diffusion equations, other authors use approximations, such as flux limited diffusion (FLD) method or the M1 closure. This might explain differences between the radiative energy distribution and radiation collimation observed between the two sets of simulations. On the other hand, in \citet{Jiang1412} the inner edge of the simulation box is located outside
the horizon which, as showed by \citet{Mckinney0102} can lead to reflection of energy that would have normally been advected
into the black hole.

However, as pointed out and discussed in some detail by \citet{Kitaki0121}, all the simulations above suffer from one serious drawback that puts into question their relevance to direct modelling of ULXs: their circularization radii, as well as the outer radii of the quasi-steady disc region are always {\sl inside} the trapping and spherization radii (see Table \ref{tab:numermod}). Therefore although all numerical models predict wind plus advection dominated flow configurations it remains to be seen where such features occur, and what their properties are in real systems with, say,  $R_{\rm out} > 10^4 R_{g}$ and $R_{\rm sph} > 750 R_{g}$ (corresponding to $R_{\rm out}> 3\times 10^{10}$ cm for a 10$\Msun$ black hole and $\dot m > 10$, respectively).

\citet{Kitaki0121} have made a substantial contribution towards reaching this objective (see first line of Table~\ref{tab:numermod}). They performed simulations with an outer radius of 6000$R_g$ and circularization radius 4860$R_g$. Their calculations achieved a quasi-stationary state inside $R_{\rm qqs} \approx 1200 R_g$, a radius 3 to 120 times larger than in previous simulations of super-Eddington accretion flows.
\begin{figure}
    \centering
    \includegraphics[width=0.7\textwidth]{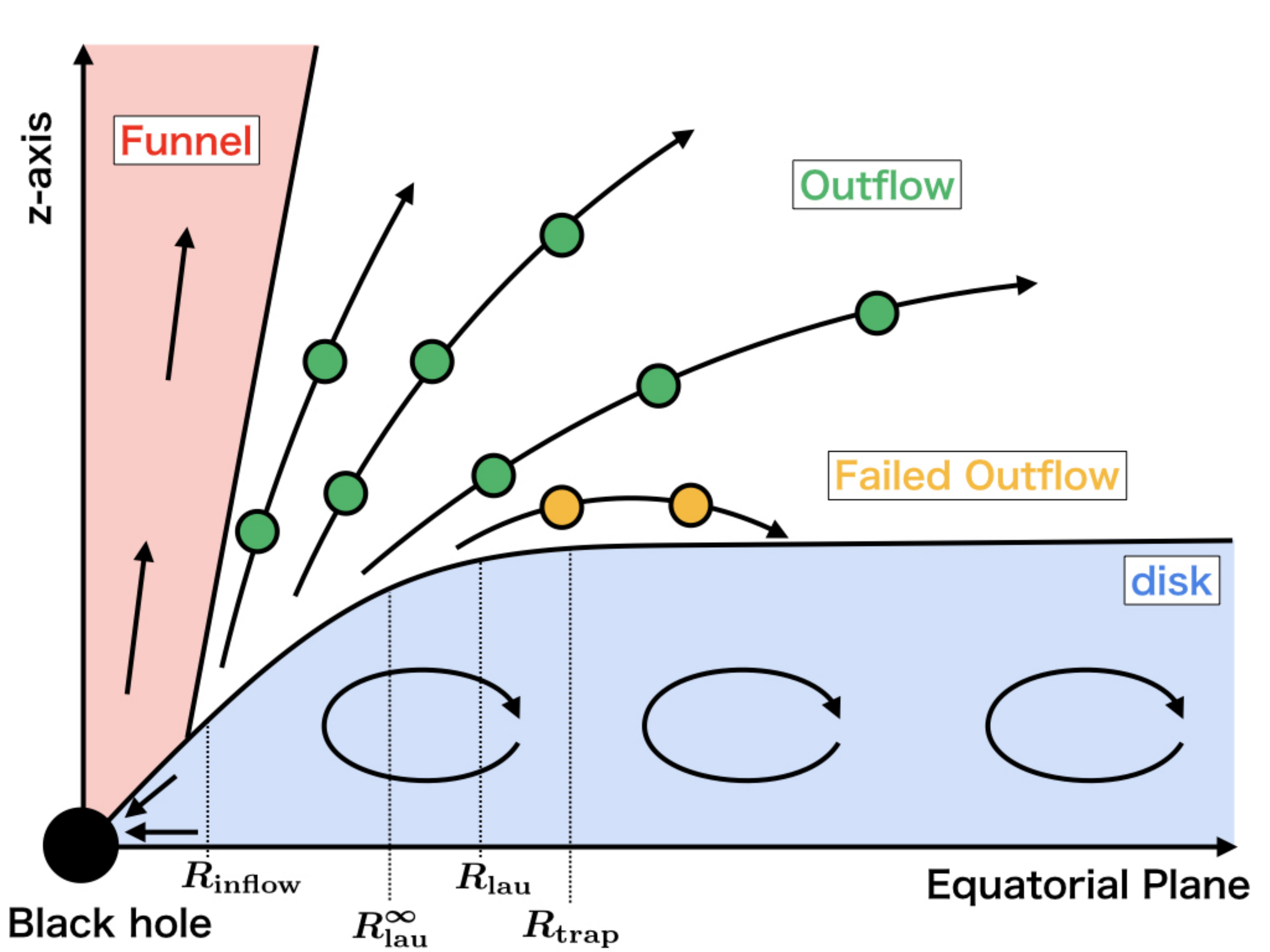}
    \caption{Schematic view of the structure of the super-Eddington accretion
flow and associated outflow based on our numerical results. The black arrows
indicate the gas motion. \citep{Kitaki0121}}
    \label{fig:kitaki}
\end{figure}
\citet{Kitaki0121} used a 2D R-RH code, with a viscosity parameter $\alpha=0.1$. The black hole mass is $10\Msun$ and a Paczyński-Wiita ``potential'' is used to describe massive--particle motions. Matter is continuously added at $R_{\rm out}$ at a rate $\dot m=70$, with specific angular momentum $3.12\times 10^{18}$cm$^2$s$^{-1}$. All matter arriving at $R_{\rm in}=4R_{g}$ is absorbed.

The structure obtained by \citet{Kitaki0121} is schematically shown in Figure \ref{fig:kitaki}. The accretion rate onto the black hole is $\dot m_{\rm BH} = 18$.
The flow forms a well--delimited disc structure with a surface defined as the loci
at which the radiation force balances the gravity in the radial direction. Near the black hole and up to $R\sim R_{\rm trap}$ the disc height $H$ is proportional to
$H$ ($H/R \sim 1$), above this radius it is constant $H \sim 22 - 34 \dot m_{\rm BH} R_g$. The trapping radius $R_{\rm trap}\sim 50 \dot m_{\rm BH} R_g$. This means, as expected from simple models 
(Section \ref{sec:sediscs}), that for $R\lesssim R_{\rm trap}$ the flow is advection dominated, while at larger radii it behaves like a radiation--pressure dominated, Shakura-Sunyaev--type accretion disc. Such discs should be thermally unstable, but this  does not seem to occur in \citet{Kitaki0121} simulations despite the calculation time amply allowing the instability to develop. The reasons for this remain unclear \citep[see, however,][]{Jiang1113} but are worth investigating.

Closer to the black hole, at $R_{\rm lau} \sim 26 \dot m_{\rm BH} R_g$, an outflow is produced, but the expelled gas is not able to reach infinity and falls back on to the disc. Only at $R^{\infty}_{\rm lau} \sim 15 \dot m_{\rm BH} R_g$ is a real outflow (wind) blown out. The formula for $R^{\infty}_{\rm lau}$ has the same form as that for $R_{\rm sph}$ (Eq. \ref{eq:resph}) but with $\dot m_{\rm BH}$ instead of $\dot m$ which in Eq. (\ref{eq:resph}) corresponds to the rate at which matter is fed to the outer radius (in the simulations under discussion, $\dot m=70$, while $\dot m_{\rm BH}=18$). We discuss this difference later in this section. Finally, in the regions closest to the black hole, below $R_{\rm inflow} \sim 4.4 \dot m_{\rm BH} R_g$, most of the gas flows in at a rate $\dot m=18$, with some weak outflow $\dot m^{\rm out}_{\rm outfl}\sim 1.3$.

The \citet{Kitaki0121} solution is fully consistent with what one expects from an accretion--launched wind. The wind speed is  $\varv \sim 0.1 c$, and $L_{\rm mech} \approx 0.05 - 0.08 L_{\rm rad}$, with $L_{\rm rad} = L_X \approx 2  - 3 L_{\rm Edd}$, so
$L_{\rm mech} \approx 0.1 - 0.24L_{\rm Edd}$. The launch radius of the wind is $R_{\rm lau} \sim{\rm \,few}\, 100 R_g$, in agreement with the wind 
speed $\varv \sim (2GM/R_{\rm lau})^{1/2}$. This gives a wind momentum rate  $\dot M_{\rm out} \varv  = (2/\varv)\, L_{\rm mech}  \sim 20L_{\rm mech}/c \sim L_{\rm rad}/c$ as one would expect
when the mass supply is not hyper-Eddington (here it is  $\sim \dot m=18$). These values are similar to analytic estimates \citep[][]{King1003} for momentum--driven winds. 

This solution suggests that for black-hole accretors, below $R\sim 1000 R_g$, most of the mildly ($\dot m \sim {\rm few}\, \times 10 $)
super-Eddington mass accretion is simply swallowed, as predicted by slim-disc models. The outflow to accretion ratio $\dot m_{\rm outflow}/\dot m_{\rm BH}-0.14$ is much lower than in previous simulations which \citet{Kitaki0121} attribute to the sizes of the simulated flows, theirs being much larger than in the preceding calculations, thus avoiding having the whole simulated domain in the region where the flow is puffed--up from the start. But we note that although $R_{\rm qss}\sim 1200R_g$ is 20 to 30 times larger than the corresponding radius in R-GR-MHD calculations, it is still ``only'' $1.8 \times 10^9$cm, well inside the circularization radius. The net accretion flow for $R \lesssim R_{\rm qss}$ is roughly constant ($\dot m \sim 18$) while the mass inflow rate decreases from $\dot m \sim 30$ at $R \approx R_{\rm qss}$ to $\dot m=18$ for $R \sim 40 R_g$ ($\dot m \sim R^{-0.2}$), but it is not clear what happens above 1200$R_g$, where the simulated configuration is not relaxed.

For example, for $R > 2012 R_g$, the gas pressure dominates over radiation pressure and the disc should be geometrically thin (see Eq. (\ref{eq:rradpres}) with $\dot m=70$). This should reduce the outflows, thus making the whole configuration potentially inconsistent. 
It seems that we are still some distance from having a complete description of the whole picture of super-Eddington accretion onto black holes. 

{Indeed, recently, \citet{Hu0322} simulated super-Eddington accretion onto a $10^3\Msun$ black hole. The physical outer radius of their simulation domain is $R=1500 R_g$, four times shorter then in \citet{Kitaki0121}. In their case, however, the steady-state radius $R_{\rm qss}=1500 R_g$, slightly longer than in \citet{Kitaki1218}. The flows they study settle down to a quasi-steady state in millions of the orbital timescale, which in their case corresponds to the viscous time at $\sim 1000 R_g$ (they take $\alpha=0.01$).
In their case $\dot{M}_{\text {in }} \propto r^{p} \text { with } p \sim 0.5-0.7$ and contrary to the \citet{Kitaki0121} only a small fraction of the inflowing matter feeds the central black hole. Indeed, their solutions can be represented as
\begin{equation}
\begin{aligned}
\dot{M}_{\mathrm{BH}} &=17.1 \dot{M}_{\mathrm{Edd}}\left(\frac{\dot{m}_{0}}{300}\right)^{0.5} \\
\dot{P}_{\mathrm{out}} &=0.51 c \dot{M}_{\mathrm{Edd}}\left(\frac{\dot{m}_{0}}{300}\right)^{0.5} \\
\beta &=17.5\left(\frac{\dot{m}_{0}}{300}\right)^{0.5}-1
\end{aligned}
\end{equation}
and the wind velocity is
\begin{equation}
\left\langle \varv_{\text {wind }}\right\rangle=1.7 \times 10^{-3} c\left(\frac{\dot{m}_{0}}{300}\right)^{-0.5},
\end{equation}
where $\beta =\dot M_{\rm out}/\dot M_{\rm in}$ is the wind loading fraction. It is only $\sim 0.13$ in \citet{Kitaki0121} but $\sim 20$ in the model of \citet{Hu0322}, so that even in the case of a black hole, advection does not play a significant role when accretion rate are very strongly super-Eddington.
Therefore the main difference between the two numerical models is that in \citet{Kitaki0121} most of the outflowing gas fails to escape from the outer boundary
and falls back onto the disc, while in \citet{Hu0322} most matter is lost in the wind. \citet{Hu0322} attribute the large differences between the two approaches to several factors, numerical and physical. First, \citet{Kitaki0121} impose in their computational domain equatorial mirror-symmetry of the accretion 
flow across the equatorial plane. It seems that this imposed symmetry tends to suppress global convective motion that crosses the equator and is known to produce outflows coherently through the equator rather than the polar regions \citep[see][]{Li0413}.
The second reason would be that \citet{Kitaki0121} assume a viscous parameter of $\alpha=0.1$  \citep[][assume $\alpha=0.01$]{Hu0322} which is
apparently known to cease convective/turbulent motion of advection dominated flows and thus to reduce the outflow rate \citep[see, e.g.][]{Inayoshi0513}. These differences do not necessarily favour the model with stronger outflows. For example, it seems that high values of the viscosity parameter $\alpha > 0.1$ are appropriate to the description of high--accretion rate flows \citep{Tetarenko0218}. Only a self-consistent calculation with a MRI--determined (anomal) viscosity
could decide if numerical models can reproduce the mechanical power observed in some ULXs.
}
{    In the most recent paper of the Ohsuga group \citep{Yoshioka0922} it is concluded that the fraction of the failed outflow decreases with the decrease of $\dot M_{\rm in}$. They leave the comparison with the \citet{Hu0322} simulation for a future paper.
What's more important, their simulations confirm the effect of radiation beaming at high accretion rates since they found that ``the higher $\dot M_{\rm in}$ is, the more vertically inflated becomes the disk surface, which makes radiation fields more confined in the region around the rotation axis".}

An interesting result of most numerical simulations is that they do not see any evidence of the photon bubbles proposed as a solution of the super-Eddington luminosity problem by \citet{Begelman0204}. When such bubbles do appear, they play no significant role in super-Eddington accretion flows \citep[see, e.g.,][]{Jiang1412}.

We are even further away from understanding super-critical accretion through numerical simulation when the accretor is a rotating, magnetized neutron star, as discussed in the next section.

\subsubsection{Numerical simulations relevant to PULXs}
\label{sec:nsnum}

The difficulty of simulating super--Eddington accretion on to rotating, magnetized neutron stars is probably best illustrated by the fact that until now only four papers have dealt with the problem: \citet{Takahashi0817,Takahashi0118,Abarca1809,Abarca0321}. The maximum dipole field strength considered is $\sim 10^{10}$G (in two cases $B=0$ G), and none of the simulations describes a rotating accretor with a hard surface.

All attempts to describe super--Eddington accretion on to neutron stars use GRRMHD codes, similar to, or inspired by, the codes used in the description of black--hole accretion, described above. There are, however, two further difficulties that must be overcome when such codes are applied to magnetized neutron stars.
First, it is not easy to describe in the same code both  magnetically--dominated flows with a force--free character and the ``normal'' properties expected from the (GRR)MHD equations. This difficulty has been partially removed by \citet{Parfrey1712}.

The second difficulty is in fixing the inner boundary conditions, i.e. the physical properties of the accreting matter at the the neutron star surface. In contrast to black--holes, whose surface is always fully absorbing independent of what -- and how much -- is falling onto it, the properties of matter arriving
at the neutron--star surface depend on 
the magnetic field strength and the accretion rate \citep[see][and references therein]{Abarca0321}. As in most black--hole simulations, limits on the available computational time restrict the range of radii for which the solutions relax to quasi--stationarity; in all simulations $R_{\rm QSS} \lesssim 50 R_g$, while the spherization radius is $R_{\rm sph} \lesssim 500R_g$ and the computational domain extends up to $1000 R_g.$

\citet{Takahashi0118} and \citet{Abarca1809} consider the case of super--Eddington ($\dot m \sim 10$) accretion onto a non-magnetic and non-rotating neutron star with a reflective surface, and compare it with the case of accretion onto a black hole. They both observe more powerful outflows in the neutron--star case and \citet{Takahashi0118} find that for a  neutron--star accretor, the accretion flow inside $R \lesssim 10 R_g$ (in their case $R_{\rm QSS} \sim 10R_g$) corresponds to the SS73 windy solution (Sect. \ref{sec:ss73} ), whereas in the black-hole case, the flow is advection dominated\footnote{This clear physical difference between
accretion flows onto compact objects which have, or do not have, surfaces, makes the advection-dominated solutions of \citet{Chashkina0619} rather unrealistic.}.
\begin{figure}
    \centering
    \includegraphics[width=0.6\columnwidth]{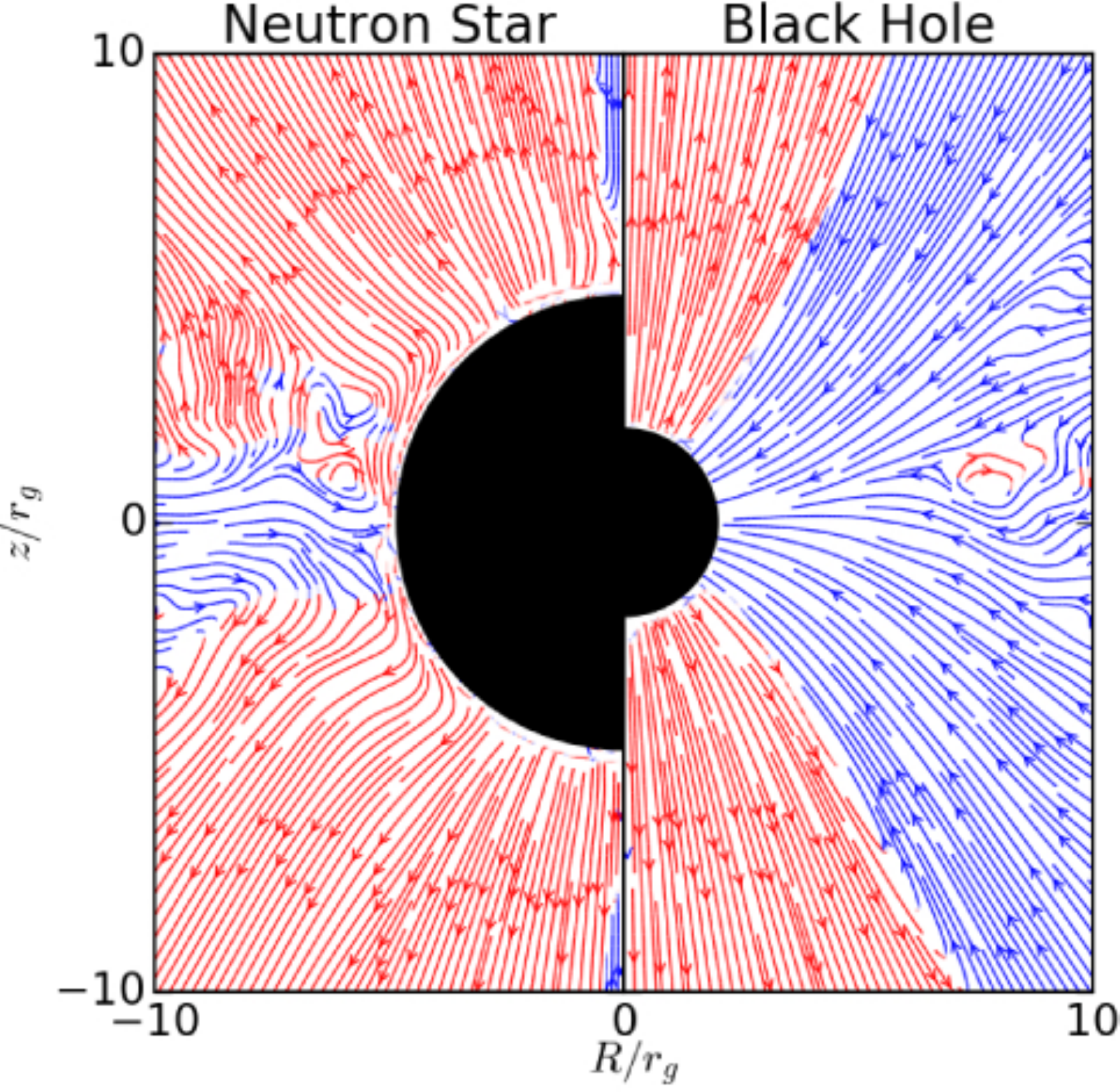}
    \caption{Stream lines around a ($1.4\Msun$) neutron star (left) and a ($10\msun$) black hole (right). Red (blue) lines indicate that the
radial velocity is in the positive (negative) direction. \citep[From][]{Takahashi0118}.}
    \label{fig:takah}
\end{figure}
The strong wind blown out from the vicinity of a super--Eddington accreting neutron star is very optically thick to electron scattering 
``which would lead to the obscuring of any NS pulsations observed in corresponding
ultraluminous X-ray sources'' \citep{Abarca1809}. Since the model of a non--rotating and unmagnetized neutron star would produce no pulses in any case, this is not a big drawback but stresses that the presence of optically--thick outflows from super--Eddington accreting neutron stars cannot be neglected when describing their X--ray emission.

 \citet{Abarca0321} performed a 2D axisymmetric radiative GRMHD
simulation of accretion onto a neutron star with a
$2 \times 10^{10}$ G dipolar magnetic field. The combination of
the hard surface and confinement of the gas into an accretion column near the star,
allows the flow to release radiative energy at a rate of several times the Eddington limit.
They compared their results to the KLK17 model and found a larger beaming intensity,
but they believe that post-processing would show a less intensely
beamed distribution of radiation at infinity. They also note
that they do not model the same system
as considered in KLK17 (e.g. they have super-Eddinton accretion rates in the accretion column). In their case the distance between the Alfvén
radius and spherization radius is large, and their star is non-rotating. This paper reveals the difficulties current
numerical simulations still face in modelling real PULXs.

In all simulations, the estimated minimum values of the inverse of the beaming parameter $b$ (maximum beaming) obtained from various models 
range from $1/b=1/30 - 1/20 \approx 0.033 - 0.05$, which compares quite well 
with the range of values $b = 0.009- 0.6$ from Table \ref{tab:ulx3b} in Section \ref{sec:klk17}. i.e. with the results of the KLK17 model.

The simulations of super--Eddington accretion onto neutron stars thus provide strong support for the assumptions of the KLK17 PULX model, i.e. that the accretion flow inside the spherization radius is well described by the SS73 windy model, and that the X-ray luminosity is beamed.  Although these simulations assume weakly magnetized and non--rotating neutron stars, their basic results: weak advection, unlike in the black--hole case, and  therefore strong optically thick outflows, probably hold for stronger fields and faster rotation of the accretor.

\subsubsection{Numerical simulations of PULX accretion columns}
\label{sec:colsim}

Numerical models of super-Eddington accretion onto a strongly magnetized neutron star are given by \citet[][hereafter K16]{Kawashima1610} and \citet[][hereafter K20]{Kawashima2002}.
The simulations describe the last part of the accretion flow when the infalling plasma forms a column above the neutron star surface.
In both papers, the accretion column is represented by a truncated cone with a base (``polecap'') occupying a fraction 0.07 of the stellar surface. Since the method used is radiation hydrodynamics, the value of the magnetic field does not appear explicitly, but in K16 it is estimated to be $\sim 10^{10}$G, while in K20 the magnetic field strength is $\gtrsim 10^{12}$G. In the first case, the plasma is allowed to move transversely, in the second, the motion of matter is restricted to the radial directions. In K16 the initial uniform density in the column gives a time-averaged accretion rate of $\dot m=50$; K20 uses the same initial set-up but also a lower-density model, giving  $\dot m=3$. In both types of model, a radiative shock forms above the neutron-star surface (fixed at 10 km) at a height  $\sim 3$ km for the "weak magnetic field" and at $\sim 10$km in the case where the accreting matter motion is restricted to the radial direction.
\begin{figure}
    \centering
    \includegraphics[width=0.4\columnwidth]{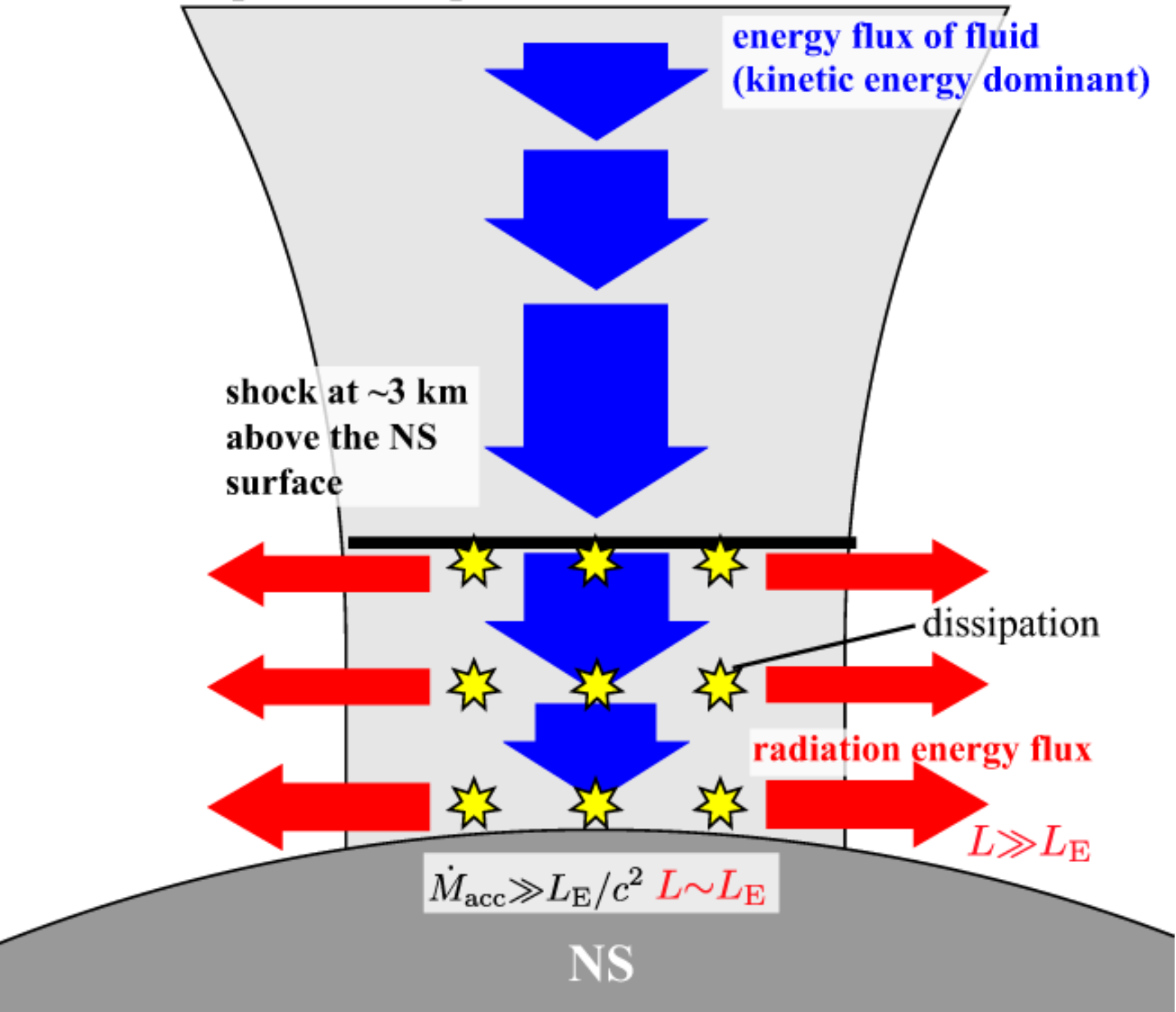}
    \caption{Schematic picture explaining energy flow from gas (potential
energy) to outgoing radiation within a supercritical accretion column.
The blue arrows represent the energy flow carried by gas: their length
and width are drawn in proportion to the kinetic energy flux and the
mass accretion rates, respectively, whereas the red arrows represent
energy flow carried by radiation and their widths are drawn in proportion
to the radiation energy flux \citep{Kawashima1610}}.
    \label{fig:kawashima}
\end{figure}
In the ``weak-field" simulations, practically all the radiation is released below the shock and radiated sideways. The apparent luminosity is highly anisotropic; it can reach $\sim  50\Ledd$ sideways, but is close to Eddington vertically along the accretion column. In contrast to the ``weak-field" case, where the radiation field below the shock is homogenized by circular motions of radiation bubbles, the ``strong-field" simulations show that matter is accreted along the side walls of ``hollow cones" while matter inside the cones is blown away by radiation. For $\dot m=3$, only one hollow cone forms, while for very high accretion rates ($\dot m=50$), the column contains three such structures. Also in this case the apparent luminosity is highly anisotropic, radiation still managing to get out mainly sideways, but radiated from a larger surface than in the ``weak-field" case. The sideways luminosity can reach $30\Ledd$.

Using the K16 model, \citet{Inoue0420} calculated the light-curves produced by such an accretion column, taking into account general-relativistic effects on light propagation near the neutron-star surface. The pulses they obtain are quasi-sinusoidal. They obtain a high pulsed fraction of the emitted light, from 5\% up to 50\%, depending on the magnetic and observational inclination angles. However, a narrow magnetic column for a highly super-Eddington accretion rate, and field of $\sim 10^{10}$G does not seem to give the best representation of a real PULX {    since high accretion rates and low magnetic fields should instead favour wide accretion columns.}

The papers K16 and K20 are important contributions to the understanding of supercritical accretion in strong magnetic fields, but it is unlikely that they correspond to the accretion structure of real PULXs. The accretion flow in these systems probably does not form a column extending to $2\times 10^{8}$\, cm ($\sim 200$ neutron--star radii) above the neutron star surface. Since PULXs are binary systems, matter transferred from the neutron-star's companion forms a disc extending down to a magnetosphere where the pressure of the accreting matter is comparable to the magnetic pressure. The size of the magnetosphere will determine the angular size of the column, which  will vary with magnetic--field intensity, differing by two orders of magnitude. For the parameters of K16, the magnetospheric radius is close to the neutron star surface at $\sim 3 \times 10^{6}$ cm (see Eq. \ref{eq:rm}) and for the K20 set-up $R_{\rm M} < 10^{8}$cm. What is most important, however, is that accretion within the magnetosphere is most likely highly dissipative, since the flow along field-lines requires a highly supersonic flow to bend and shock, while in K16 and K20, the first shock occurs very near the neutron-star surface, because in the simulations (different from the schematic view of Fig. \ref{fig:kawashima}) the field-lines are straight. By necessity, the K16 and K20 simulate only the last part of the PULX accretion flow and they assume that it arrives at the magnetosphere at a super-Eddington rate, which is rather unlikely in reality \citep[][]{Takahashi0118}.

\subsection{PULXs as magnetars in binary systems}
\label{sec:nomagnet}

The discovery that ULX M82 X–2,  with an apparent luminosity
$L \simeq 1.8\times 10^{40}\, {\rm erg\, s}^{-1}$,  is a pulsar
with a coherent periodicity P = 1.37 s \citep{Bachetti1410}, raised the possibility that its
apparent super--Eddington luminosity might be intrinsic \citep[cf][]{Tong1511, Eksi1503,DallOsso1512,Mushtukov1512}. If the 
strength of its magnetic field is magnetar-like ($B \sim 10^{14}\rm G$), the reduction in electron scattering opacity
for some directions and polarizations can be quite substantial.

At first sight, this idea is similar to the one motivating \citet{Paczynski0792} to explain
the luminosity $L \sim 10^4 \Ledd$ observed in the soft-gamma repeater (SGR), GB 790305 (``the 5 March event''),
by the presence of a magnetic field $B \approx 3\times 10^{14} \rm G$. In a very strong magnetic field, the Thomson
and Compton scattering cross-sections are strongly reduced for photons with energies $E_\gamma$ much lower than the
energy corresponding to cyclotron frequency $E_{\rm cyc} = 11.6 B_{12}\,\rm keV$, where $B_{12}10^{12}\,\rm G=B$ \citep{Canuto0571}. From \citet{Herold0579} one has
\begin{align}
\label{eq:sigmaB}
\frac{\sigma_1}{\sigma_T}\approx &  \sin^2\theta +\left(\frac{E_\gamma}{E_{\rm cyc}}\right)^2\cos^2 \theta \\
\frac{\sigma_2}{\sigma_T}\approx &  \left(\frac{E_\gamma}{E_{\rm cyc}}\right)^2, \, \rm for \, \frac{E_\gamma}{E_{\rm cyc}}\ll 1,
\end{align}
where indices 1 and 2 correspond to the two linear photon polarisations, and $\theta$ is the angle between the directions of the magnetic field
and light propagation. \citet{Paczynski0792} gives an approximate relation between the ``new'' critical luminosity and the Eddington luminosity,
defined through the Rosseland mean opacity:
\begin{equation}
\label{eq:Lcrit}
\frac{L_{\rm crit}}{\Ledd} \approx 2 B_{12}^{4/3}\left(\frac{g}{2\times 10^{14}\rm cm\,s^{-2}}\right)^{-1/3},
\end{equation}
where $g=GM/R^2$, when $L_{\rm crit} \gg \Ledd$. This critical luminosity is of course only approximate because it is not properly integrated over angles, but estimates of the Rosseland mean
opacities vary with angle only by a factor two \citep{Paczynski0792}.

From Eq. (\ref{eq:Lcrit}), for a $1.4\Msun$ neutron star, a sub-critical luminosity $L \lesssim 0.1 L_{\rm crit} $ implies 
\begin{equation}
L \lesssim 2 \times 10^{40} \left(\frac{B}{10^{14}\, \rm G}\right)^{4/3} \,\ergs,  
\label{maglimit}
\end{equation}
so that explaining PULXs with $L > 10^{40}\,\ergs$ assuming isotropic radiation, requires magnetic field strengths that would be considered extreme, even among magnetars. 

In PULX models by \citet{Tong1511,DallOsso1512,Eksi1503,Chashkina0917}, the fields are in the range $\sim 3 \times 10^{13} - 6\times 10^{16}$ G (the last value considered by the authors to be ``rather unphysical''), so that the critical luminosities corresponding to the first three PULXs discovered put them safely into the sub-critical luminosity regime, even with a magnetic moment derived by magnetic torque considerations.

But there is a fundamental difference between SGRs and PULXs: SGRs are powered by magnetic energy \citep{Duncan0692}, while the luminosity of
PULXs comes from accretion. For SGRs the magnetic field allowing $L\approx 10^4\,\Ledd$ is consistent with pulsar slowdown and
the required magnetic energy. There is no known alternative to magnetar-strength fields in SGRs and anomalous X-ray pulsars, and there is so far no observational
evidence against the presence of fields with these strengths.

This is not true of the magnetar hypothesis for PULXs.
First, no magnetar is observed to be a member of a binary system. {\sl All} $\sim 30$ known
magnetars (or ``candidate" magnetars) are isolated neutron stars \citep{Olausen0514}. Conversely,
all neutron stars in X--ray binaries are observed to have magnetic field strengths $< 10^{13}$G, below the magnetar range \citep[see, e.g.,][]{Revni16}.

The super-strong magnetic field in magnetars is thought to be formed in a process which destroys (or merges) the binary, of which its progenitor was a member \citep{Popov0116}. 
This would explain their absence in binary systems. Magnetars are supposed to form through neutron-star mergers \citep[e.g.,][]{Giaco0713,Piro0817} or  in type Ibc supernova outbursts in 
massive binaries. The first mechanism obviously produces a single object. In the second, the superluminous supernovae disrupt the  
binary and produce an isolated magnetar, as demonstrated by the magnetar CXOU J1647-45 in the young massive cluster,
Westerlund 1 \citep{Clark0514}. 

{We note that ``most of theoretical considerations do not favour even existence, not speaking about active decay, of magnetar-scale
fields in neutron stars older than $10^6$ yrs'' \citep{Popov0122}.}

These considerations make the presence of rapidly rotating, strongly accreting magnetars in binary systems highly unlikely.
{Their presence in ULXs would evidently require a cosmic conspiracy.}

\subsection{ULX populations}
\label{sec:ulxpop1}

The beaming formula (\ref{eq:beam}) allows one to study the populations of disc--wind beamed ULXs in galaxies. 
If the host galaxies have space density $n_g\,{\rm Mpc}^{-3}$,
and each host contains $N$ ULXs with randomly oriented beams, one would need to search through $\sim 1/Nb$ galaxies, occupying a space volume
$\sim 1/n_gNb$, to be sure of being in the beam of one ULX. 
Then the nearest ULX must be at a distance
\begin{equation}
    D_{\rm min}\sim\left(\frac{3}{4\pi n_gNb}\right)^{1/3} \sim 0.7(n_gN)^{-1/3}\dot m_1^{2/3}\, {\rm Mpc}
    \label{Dmin}
\end{equation}ń
where $\dot m_1 = \dot m/10$, and has apparent (isotropic) luminosity
\begin{equation}
    L = 2.2\times 10^{39}m_*\dot m_1^2\, \ergs
\end{equation}
where $m_* = m/(10\msun)$. The scalings $m_*, \dot m_1$ are appropriate for black holes or mass $\sim 10\msun$, so by the argument following Eq. (\ref{eq:lsph})
the luminosity normalization here would be $2.2\times 10^{40}\, \ergs$ for neutron star ULXs. These relations can be used to show that the ULX
luminosity function of the Local Group is consistent with the idea that beamed ULXs are the evolutionary stage following the standard high--mass X--ray binary phase \citep[cf][]{Mainieri1005}.

\subsubsection{Pseudoblazars?}
The beaming described by (\ref{eq:beam}) potentially allows one to see suitably--oriented ULXs out to very large distances. The extreme 
object SS433 has $\dot m \sim 300 - 10^4$ \citep[][see Section \ref{sec:ss433}
]{King0002,Begelman0607}.  
With $\dot m = 10^4\dot m_4$ the nearest object of this kind which we would observe within the beam is at a distance
\begin{equation}
     D_{\rm min}\sim\left(\frac{3}{4\pi n_gNb}\right)^{1/3} \sim 660N^{-1/3}\dot m_4^{2/3}\, {\rm Mpc}
\end{equation}
where we have taken $n_g \sim 0.02\, {\rm Mpc^{-3}}$ as appropriate for $L^*$ galaxies. 
 The apparent isotropic luminosity of such an object would be
\begin{equation}
 L_{\rm sph} =2.2\times 10^{45}m_*\dot m_4^2\, \ergs.
\end{equation}
This distance and apparent luminosity are typical for active galactic nuclei (AGN), so without further information
one could not easily pick out a bright ULX like this from a sample of observed AGN. But unlike a genuine AGN, 
there is no reason to suppose that a given ULX lies in the nucleus of the host galaxy. There are possible candidates for such systems: the
BL Lac system PKS 1413+135 \citep{Perlman1102} has distance $D \simeq 1\, {\rm Gpc}$ and isotropic luminosity $\sim 10^{44}\, \ergs$ , but lies 
at $13 \pm 4$~mas from the centre of the host galaxy, and HLX-1
could be a system of this type (cf Section \ref{sec:hlx}).
Of course it is difficult to exclude the possibility that this system may lie at the nucleus
of a fainter line--of--sight galaxy.

\subsection{ULXs and binary stellar evolution}
\label{sec:evolulx}

Sections \ref{sec:windbeam}, \ref{sec:pulx} and \ref{sec:numer} above have shown how the phenomena characterizing ULXs arise, in many cases, from strongly super--Eddington mass transfer in binaries containing compact objects (white dwarfs, neutron stars or black holes) which have
reached the end of their stellar evolution. We should now
ask what causes these very high mass transfer rates. To do this we need to consider the evolution of binary systems 
where one of the stars is compact (a black hole, neutron star, 
or white dwarf) and the companion star transfers mass to it, powering accretion. The rate of mass transfer is determined
by the evolution, both of the companion star, and of the binary orbit, as mass transfer in general changes this.  
Dissipation in the mass transfer process generally
makes the binary orbit 
circularize.

Kepler's laws give the orbital angular momentum of a
circular binary system consisting of stars of mass $M_1, M_2$ as
\begin{equation}
    J = M_1M_2\left(\frac{Ga}{M}\right)^{1/2},
    \label{J}
\end{equation}
where $M = M_1 + M_2$ is the total binary mass, and $a$ is the orbital separation. We take star 1 as the compact
object, and star 2 as an extended star. In a high--mass X--ray binary (HMXB), star 2 is hot, and loses mass
in a wind. The X--rays result when some of this wind is captured by the compact star (a neutron star or 
black hole). But as the extended star 2 approaches the end of core hydrogen--burning it begins to expand. 
Eventually its radius $R_2$ fills its Roche lobe 
\begin{equation}
    R_L = f(q)a
    \label{roche}
\end{equation}
(where $q = M_2/M_1$ is the binary mass ratio, and $f(q)<1$ is a slowly--varying function of $q$). The star begins to 
transfer mass through the inner Lagrange point, the saddle point of the combined gravitational--centrifugal, Roche 
pseudopotential between the two stars. It is generally a good approximation to assume that all the mass lost
by star 2 is accreted by the compact star 1, and that the orbital angular momentum $J$ is conserved in this
process. Then $\dot J = \dot M = 0, \dot M_1 + \dot M_2 = \dot M =0$, and
logarithmic differentiation of Eq. (\ref{J}) gives
\begin{equation}
    \frac{\dot a}{a} = 2\frac{(-\dot M_2)}{M_2}\left(1 - \frac{M_2}{M_1}\right)
    \label{mtp}
\end{equation}
Since $-\dot M_2 >0$, we see that if the extended, mass--losing star is more massive than the compact star, 
the effect of mass transfer is to shrink the binary separation ($\dot a < 0$). As $f(q)$ varies only
weakly with mass ratio $q$, $R_L$ also shrinks, tending to increase the rate of mass transfer. The result
is that mass transfer from the more massive to the less massive star is self--sustaining, at least for a time. 

But there is a limit to this process. The outermost stellar gas -- above the saddle point of
the Roche potential -- is lost on a dynamical timescale, and gas from below flows towards the 
saddle point on the same timescale, trying to refill the Roche lobe. But this gas is not now in thermal equilbrium
because its rapid rise means that it retains the same specific entropy it had when deeper in the star. But
in
stars with a radiative envelope (typically, blue stars with effective temperatures $T_{\rm eff} \gtrsim 10^4\, {\rm K}$), 
the equilibrium entropy increases outwards, so the new gas has too little entropy
for its new position. This means that it is denser and cooler than the gas it replaces, making the radius of this part of the
star smaller, and reducing the mass transfer rate. To reach thermal equilibrium, 
this gas must absorb some heat from the stellar radiation field, on a thermal timescale. In doing so
it expands, returning this part of the star to the equilibrium radius for its new 
(slightly smaller) mass\footnote{If the star instead has a convective 
envelope, and so constant specific entropy, mass transfer can reach very high (near--dynamical) values, and the system 
develops a deep common envelope. This may lead to a merger of the two stars. \citet{King0799} show that this is
very unlikely to happen with a radiative envelope, even for transfer rates as high as $\sim 10^{-4}\msun\, {\rm yr}^{-1}$.}.

The argument above shows that mass transfer from a more massive radiative companion
cannot happen on a timescale shorter than the star's thermal
timescale,  $t_{\rm th} \sim GM_2^2/R_2L_2$ (where
$R_2, L_2$ are the star's radius and luminosity respectively).
But this is usually much shorter than the nuclear timescale 
in stars of mass a few $\msun$, and we find values $-\dot M_2 \sim M_2/t_{\rm th}$ as large as 
$\sim 10^{-7} - 10^{-4}\msun\, {\rm yr}^{-1}$, depending on the stellar mass $M_2$. 

Rates like this are far higher than the value $\dot{M}_{\rm Edd}$ defined above, which would give the Eddington luminosity if it
could be accreted {and so it is very likely that the super--Eddington excess is likely to be expelled}. 

As we mentioned in the Introduction and Section \ref{sec:windbeam}, the neutron star in the long--period ($P = 9.84$~d) low--mass X--ray 
binary Cyg X-2, appears to have
expelled most of the mass transferred to it in this way. The distinctive feature of this system is the unusual state
of the companion star. Combining spectroscopic information, photometry of the ellipsoidal variation of the companion, 
and the geometry of the Roche potential, \citet{King9910} showed
that this star had a low mass $M_2 \simeq 0.5 - 0.7\Msun$, and yet a large radius
($R_2\simeq 7\Rsun$) and very high luminosity ($L_2 \simeq 150\Lsun$). This is just what one expects at the end of 
an episode of 
thermal--timescale mass loss: the hot central region of what was initially a much more massive 
star has been exposed, and has had no time to cool. The evolution followed by Cyg X-2 is called early 
massive Case B, in the terminology of \citet{Kippenhahn0167}.

The most important feature of this evolution is that the neutron star accretor has evidently gained almost none of the $\sim 3\msun$
transferred to it by the companion: its current mass is only $1.78\pm 0.3\msun$. This expulsion cannot have
come from common--envelope evolution -- the orbital binding energy released in shrinking the binary to the separation given by
the current period of almost 10 days is far too small to expel any matter. This leaves only accretion energy -- the accretion disc around the neutron star
must have expelled most of the transferred mass. This is just the process discussed by \citet{Shakura73}, and summarized in Section \ref{sec:ss73}.
As we discuss in Sections \ref{sec:numer} and \ref{sec:feedback}, it leads to the production of a quasi--spherical wind with typical outflow
velocity $v \sim 0.1c$, with open funnels along the disc axes which collimate a large fraction of the accretion energy released as
radiation. In this picture, ULXs are the evolutionary stage following the wind--fed HMXB phase, once the massive companion star
fills its Roche lobe. This seems particularly likely for the pulsing
PULXs, which contain neutron stars very similar to those in HMXBs. 

Although thermal--timescale mass transfer is very probably the cause of ULX behaviour in many cases, it is not the only way to make a ULX. 
Evidently any process that produces a significantly super--Eddington mass
transfer rate can do this. We have already mentioned that some Be -- X--ray binary systems become 
ULXs from time to time, because of dynamical effects on the Be--star disc. Three other possible ways involve mass transfer
on the nuclear timescale of a massive donor \citep{Rappaport0501},  transient outbursts involving the
thermal--viscous disc instability \citep[][see Sect. \ref{sec:trans}]{King0209,Hameury1120}, and gravitational--wave driven mass transfer from a white dwarf onto a neutron star \citep{King1105}
-- a so--called ultracompact binary (UCB). UCBs are often found in globular clusters \citep[see][and references therein]{Dage0321}, where they are formed by dynamical capture, so these sources offer
a possible explanation for ULXs sometimes claimed in elliptical galaxies. We will discuss other evolutionary paths, especially for weak ($L_X \simeq 3 \times 10^{39}\ergs$)
ULXs  in Section \ref{sec:pop}.

\subsection{`Super--Eddington Accretion'}
These possibilities illustrate that ULXs are likely to be binaries containing a compact accretor in an epoch of strongly super--Eddington mass
transfer. It is worth emphasizing that the commonly used phrase `super--Eddington accretion' is highly ambiguous, and a frequent cause of confusion. As the example of Cyg
X--2 shows, {    the accretors 
in ULXs do not gain mass at significantly super--Eddington rates:} it is the mass {\it supply rate} which is super--Eddington,  not the accretion. Accordingly, the phrase {\it `super--Eddington mass supply'}
is strongly preferable.

\subsection{Population syntheses}
\label{sec:pop}

Population synthesis of ULXs was first performed by \citet{Rappaport0501} who showed that the population
of ULXs in spiral galaxies can be explained by short phases of
high mass--transfer (with nuclear or thermal timescales) in black--hole plus evolved--star binaries. However, they neglected the pre--supernova evolution phase in their calculations and did not take into account other types of binaries, such as systems containing neutron--star accretors, or evolved donors. Their study was expanded by \citet{Madhusudhan0812} who completed it by predicting the observational properties of the ULX population. \citet{Linden1012} used the StarTrack population synthesis code \citep{Belczynski0206b,Belczynski0801} to show that the bulk of ULXs can be interpreted as the high-luminosity
tail of high-mass X-ray binaries. Neutron--star accretors were included in population studies of ULXs only after the discovery of PULXs.

\citet{Wiktorowicz1509} performed a proof--of--concept study where they showed that any ULX, including the most luminous ones, and neutron--star systems, can be a short-lived phase in the life of a binary star. The detection of double compact object mergers \citep{Abbott1602} triggered investigations of the connection between binary compact objects and ULXs \citep[e.g.,][]{Finke1712,Marchant1708,Klencki1811}.

A problem encountered when performing ULX population synthesis is the unfortunate choice of the threshold ULX luminosity at $10^{39}\ergs$, because it fails to separate ``standard'' X-ray binaries (XRBs) from ``generic'' ULXs. At least five transient Galactic XRBs reach luminosities $>10^{39}\ergs$ \citep{Tetarenko1602}. \citet{Middleton1503} suggested that one should take $3 \times 10^{39}\ergs$ as the minimum luminosity defining ULXs whose nature is ``contentious''. The idea was to ensure that if the accretor was a black hole (BH), the apparent luminosity would be super-Eddington. But we now know that stellar evolution can produce black holes with masses $\gtrsim 30\Msun$ \citep{Belczynski1005, Abbott1602}, up to $50 \Msun$ under favorable conditions (see, e.g., \citealt{Belczynski2012}; the fact that the maximum value of black--hole masses observed in X-ray binaries is $\sim 20\Msun$ follows from the limited range of metallicities scanned by X-ray observations \citealt{Olejak2006}). So only sources with luminosities $\gtrsim6\times10^{39}\ergs$ qualify as
(apparently) super-Eddington for a BH resulting from stellar evolution. BH\footnote{The limit can be even higher for He-rich accretion, and for the most massive BHs \citep[e.g.][]{Belczynski2012}}. The current outdated but default definition of ULXs as $L \gtrsim 10^{39}\ergs$ tends to coerce population syntheses to use it, at the cost of producing ``excesses'' of sources with luminosities which in most cases are actually {\it sub}--Eddington. 

Existing population syntheses do not include Be-X systems, and until recently they considered only Roche--lobe overflowing (RLOF) companions of the compact accretor, so excluding wind--fed systems. The second omission was corrected by \citet{Wiktorowicz0221}, and wind--accreting ULXs will be discussed below, but there is not much hope that the difficulties inherent in the first (Be-X) will be overcome any time soon. Although current models of Be stars allow simple modelling of their discs at radii far from the stellar surface, they also contain hidden assumptions that are not physically plausible \citep{Nixon1220}. If the model for Be--star discs proposed by the latter authors (the disc material originating from small-scale magnetic flaring events
on the stellar surface) is correct, it will be problematic to incorporate it in a population--synthesis code.

\citet{Wiktorowicz1709,Wiktorowicz1904} performed comprehensive simulations of ULX populations by using the {\tt StarTrack} population
synthesis code \citep{Belczynski0206b,Belczynski0801} with significant updates
described in \citet{Dominik1211,Wiktorowicz1409}.
They simulated the evolution of $2\times 10^7$ binary systems for every model and
scaled the results to a Milky-Way equivalent galaxy
\citep[$M_{\rm MWEG}={6}\times 10^{10}\msun$;][]{Licquia1506}. Two star formation cases are considered: at a constant rate of
$6.0\msy$ for $10$Gyr, and $600\msy$ in
bursts of star--formation lasting $100\myr$). They consider three metallicity prescriptions: solar 
(\Zsun), $10\%$ of solar (\Zsun/10), and $1\%$ of solar (\Zsun/100). The fiducial accretion model for ULX evolution in this study assumes the SS73 windy solution and beaming described by Eq. (\ref{eq:beam}), with a saturation limit at $\dot m = 150$.
For higher mass accretion rates, the beaming is assumed to be constant
and equal to $b\approx\ttt{3.2}{-3}\,(\theta\approx9^\circ)$. For comparison, seven other models with different accretion and/or beaming configurations are
considered. In general, the results do not depend strongly on the assumptions about the accretion mode or the beaning model, except for the case where beaming is assumed to be constant ($b=0.1$) which, by lowering the ULX luminosity threshold, results in neutron--star ULX always dominating the population at late times.

It is known from observations, that ULXs are often associated with low metallicity
environments \citep[e.g.][]{Pakull0202,Soria0501,Luangtip1501,Mapelli1010}
Massive black holes apparently form more easily in low--metallicity environments
\citep[e.g.,][]{Zampieri0912,Mapelli0905,Belczynski1005}, and 
RLOF onto a compact object is therefore more frequent than in high metallicity cases \citep{Linden1012}. 

\citet{Wiktorowicz1709} found that metallicity has a strong impact on the number of ULXs, but
only in star--forming regions. They show that the number of
black--hole ULXs increases significantly for \zsun/10 in comparison with \zsun. But
the number of neutron--star ULXs is virtually unaffected by metallicity. Interestingly, for models
with the lowest investigated metallicity ($\zsun/100$) they find fewer ULXs than
for $Z=\zsun/10$. This suggests that the relation between the number of ULXs and metallicity
is not monotonic \citep[see][]{Prestwich1306}. The reasons for this inversion are currently not understood.

The general conclusions of the ULX population synthesis by \citet{Wiktorowicz1709} are:

\begin{itemize}
    \item ULX with NS accretors dominate the post-burst ULX populations,
    and 
    {    those with}
    constant stellar formation (duration $>1\gyr$) in high-$Z$ environments,
    consistent with current understanding of binary evolution.
    Neutron--star ULXs are present in significant numbers ($\gtrsim10\%$) also during the
    star--formation bursts and in lower-$Z$ ULX populations.
    
    \item ULXs appear in a very specific sequence after the start of the star formation
        ($t$ denotes the time since the beginning of  star formation): 
        \begin{enumerate}
            \item \makebox[3cm][l]{$t\approx4\dash40\myr$} BH\dash MS
                ($5.6\dash11\msun$),
            \item \makebox[3cm][l]{$t\approx6\dash800\myr$}   NS\dash MS
                ($0.9\dash1.5\msun$),
            \item \makebox[3cm][l]{$t\approx430\dash1100\myr$} NS\dash HG
                ($0.6\dash1.0\msun$),
            \item \makebox[3cm][l]{$t\approx540\dash4400\myr$}  NS\dash RG
                ($\sim1.0\msun$);
        \end{enumerate}
    (BH--black hole, NS -- neutron star, MS -- main sequence, HG -- H--rich Hertzsprung gap, RG -- red giant.)

    \item {Neutron--star ULXs may reach luminosities as high as those of
    black--hole ULXs} ($\lxmax> 10^{41}\ergs$).

    \item { The most luminous ULXs ($\lx\gtrsim 10^{41}\ergs $) contain HG donors
    (black--hole ULXs; $M_2\approx 1.2\dash3.7\msun$) or evolved helium stars (neutron--star ULXs;
    $M_2\approx 1.7\dash2.6\msun$)}, which overfill their Roche lobe and transfer mass
    on a thermal timescale.
    They form typically within $15\dash75\myr$ after the
    ZAMS.
\end{itemize}
\begin{table*}
    \caption{    Number of ULXs per MWEG $100\myr$ after the start of stellar formation}
    \vskip 5pt
    {
    \centering
\begin{tabular}{llcccc}
\hline\hline
    Metallicity &  &  $>10^{39}\ergs$ & $>$ \nergs{3}{39} & $>10^{40}\ergs$ & $>10^{41}\ergs$ \\
    \hline
               & \nulx   &  407  &  65  &  14  &  0.5 \\
    \zsun    & \nbhulx &  372  &  40  &  4.7  &  0.38 \\
                & \nnsulx &  35  &  25  &  9.4  &  0.13 \\
                \hline
                & \nulx   &  7281  &  1940  &  200  &  12 \\
    \zsun/10    & \nbhulx &  7200  &  1900  &  180  &  11 \\
                & \nnsulx &  81  &  40  &  12  &  0.43\\
                \hline
                & \nulx   &  5000  &  2400  &  390  &  15 \\
    \zsun/100   & \nbhulx &  4900  &  2400  &  380  &  15 \\
                & \nnsulx &  16  &  12  &  9.0  &  0.075 \\
                \hline
    \end{tabular}}\\
{\citep[Adapted from][]{Wiktorowicz1709}. `MWEG' = `Milky Way Equivalent Galaxy'.}
\label{tab:num_sfb_lxmin}
\end{table*}

Finally, as seen in Table \ref{tab:num_sfb_lxmin},  the ULX population is in all cases strongly dominated by sources with $L_X < 3 \times 10^{39}\ergs$.

The results of \citet{Wiktorowicz1709} describe the evolution of the {\it intrinsic} ULX population, but since at least these sources are supposed to be beamed, the observed ULX population is different. \citet{Wiktorowicz1904} use the results of the population synthesis 
described above to derive the properties of the observed ULX population. They use the fiducial model defined in \citet{Wiktorowicz1709} but remove the beaming saturation--condition because
extremely beamed sources are not only hard to observe, but also extremely rare and short-lived so have no influence on the final results.

\citet{Mondal2001} discuss the full range of possible ways of making a ULX. This paper
also estimates that at low redshift, about 50 per cent of merging BH--BH progenitor binaries evolved through a ULX phase.
They find that observed ULXs with black--hole accretors typically emit isotropically ($b=1$) and undergo nuclear--timescale
mass transfer, whereas those with neutron--star accretors are predominantly beamed (typically $b=0.7-0.2$) and in most cases the mass transfer occurs
on a thermal timescale. They also show that beaming depends on the stellar environment; very young (burst)
populations (age $< 10$ Myr), dominated by black-hole ULXs, are significantly beamed, while black--hole ULXs in older stellar populations are usually isotropic emitters. In contrast, the majority of neutron-star ULXs are always beamed, irrespective of the stellar environment.
The ratio of neutron--star ULXs to black--hole ULXs is higher in the total sample
than in the observed sample. For continuous star formation, black--hole ULXs typically outnumber the NS ULXs in the observed
sample. Black--hole ULXs also outnumber neutron--star ULXs in the observed sample for burst stellar formation at early times, but after the star formation burst, neutron star ULXs tend to dominate the observed population. Here the observed neutron--star ULXs represent only 20\% of the
total neutron--star ULX population, and many are expected to be obscured  \citep[in the absence of precession, which may act to bring some into view, see,][]{Dauser1704,Middleton_2018_Lense_Thirring}.

\citet{Heida1910} pointed out that the growing number of ULXs with (tentatively identified) red--supergiant donors (for which RLOF is dynamically unstable) motivates the study of wind-driven mass--transfer, previously neglected in the context of ULXs. Including wind accreting X-ray binary models into {\tt StarTrack}, \citet{Wiktorowicz0221} showed that wind-fed ULXs can constitute a significant fraction of all ULXs, and in some environments may even be the majority.  Bondi-Hoyle-Lyttleton  accretion is a standard form of accretion in {\tt StarTrack} \citep{Belczynski0801,Belczynski2004} and they additionally adapt a wind RLOF (WRLOF) scheme from \citet{Mohamed0709}, already used by \citet{Ilkiewicz1906}, to analyse the relation between SNIa and wide symbiotic binaries. With somewhat optimistic assumptions for wind--accretion efficiency, they suggest that wind-fed ULXs should not be neglected. The wind--fed ULX sample they found contains a significant fraction of RSG companions, supporting the suggestion that the apparent superposition of some ULXs and RSGs results from coexistence as a binary. Although some of these systems will evolve into double compact objects, none of the ULX systems with a RSG donor is a viable progenitor of double compact object mergers with $t_{\rm merge}<10$Gyr, because they have large orbital separations.

It is important to note that models of wind accretion and emission suffer from serious uncertainties. This problem cannot be solved only by extensive numerical simulations, but also requires systematic quantitative observational data.
\citet{Wiktorowicz1709,Wiktorowicz1904} and \citet{Wiktorowicz0221} do not consider the origin or evolution of neutron--star magnetic fields, considering that this problem contains too many unknowns to be worth including in population synthesis.

\citet{Kuranov2020} do not share this view, and perform population synthesis of binaries containing magnetized neutron stars only. 
They assume solar abundances and neglect magnetic--field evolution. They consider two accretion--flow models: the standard \citet{Shakura73} model adapted to magnetized accretors by \citet{Chashkina0917} and for the supercritical accretion they use the \citet{Chashkina0619} ``advective'' model (for the suitability of advection models to describe accretion onto neutron stars see Sect. \ref{sec:nsnum}). The results are presented as a $P_s - L_x$ figure on which the observed values of 10 PULXs are superposed. Although Be/X stars are not present in the modelled population, the authors use also four Be/ULX data points. When the data points fall into the regions where the calculations predict the existence of PULXs, they consider this as a success of their model. Which is the case for all accretion modes, only NGC 5097 ULX-1 shows, as usual, some resistance. 
They conclude that that standard evolution of close--binary systems and accretion
onto magnetised neutron--stars can quantitatively explain  the observed properties of PULX
i.e. their X-ray luminosities, spin periods, orbital periods and masses of visual components,
without assuming beamed X-ray emission. In a model galaxy with star formation rate
3–5 $\msun\,{\rm yr}^{-1}$ ``there can exist" several PULXs. If PULXs are not beamed, it is not clear why they are invisible in the Milky Way. 
(The single PULX observed in the Galaxy is in a Be-X binary -- because of their uncertain evolutionary status, such systems are not taken into account in population syntheses.)
\citet{Kuranov2020} assert that 
discovery of powerful winds from PULXs with
$L_X \sim 10^{41}\ergs $ may be a signature of super-Eddington accretion onto magnetised neutron stars. If so, this raises two questions.
First, in the absence of beaming why would such sources not be observable as bright X-ray sources? Second, with such powerful outflows, how one can avoid beaming? Finally, although these authors say that, in their models, the PULXs represent ``a subset of neutron stars at the stage of disc accretion in close binary systems with luminosity $L_X > 10^{39}\ergs $'', they take no account of the other distinctive parameter that singles out the PULXs: their extremely high spin frequency derivatives $\dot \nu$ (see Fig. \ref{fig:dotnul}).

\subsection{Winds and Feedback from ULXs}
\label{sec:feedback}

We have seen (in Section \ref{sec:windbeam}) that the characteristic feature of the disc--wind beaming picture of ULXs is that their accretion discs eject the super--Eddington part
of the mass transferred to them. If $-\dot M \gg \dot M_{\rm Edd}$ this is of course most of the transferred mass. In all cases we 
expect a quasispherical accretion disc wind, with most of the emitted radiation escaping through open funnels along the disc axis. The
dynamics of the wind are controlled by radiation pressure, and therefore by the electron--scattering optical depth 
distribution of the 
outflow. For modest Eddington factors $\dot m = \dot M/\Mdoted \gtrsim 1$ we expect that the photons produced by the conversion of accretion energy will 
find the open funnels around the rotational axis of the disc and so escape 
after only a small number of scatterings. The front--back symmetry of 
non--relativistic electron scattering means that, on average, a photon gives up all of its momentum (but not energy) in each scattering 
event. Then we expect that the accretion disc wind should have an outgoing momentum of order that in the original radiation field, i.e.
just the Eddington momentum
\begin{equation}
    \dot M v \simeq \frac{\Ledd}{c},
    \label{momwind}
\end{equation}
so that the outflow velocity is
\begin{equation}
    v \simeq \frac{\eta}{\dot m}c \sim 0.1c,
    \label{momwindv}
\end{equation}
where $\eta = \Ledd/\Mdoted c^2 \simeq 0.1$ 
is the radiative efficiency of accretion \citep{King1003}. 
Winds with the properties (\ref{momwind}, \ref{momwindv}) are seen in many AGN \citep[see][for a review]{King0815}, where the Eddington factor is unlikely to be large. 

But in ULXs, the condition $\dot m \gg 1$ is a distinct possibility, and here we 
expect considerably more scatterings before a photon finds the open funnels around the disc axis and escapes. Instead we expect
that the radiation field may put a significant fraction of its energy into the wind, i.e.
\begin{equation}
    \frac{1}{2}\dot M {v'}^2 \simeq l'\Ledd, 
\end{equation}
with $l' \sim 1$. This now leads to an outflow velocity 
\begin{equation}
    v' \simeq \left(\frac{2l'\eta}{\dot m}\right)^{1/2}c,
\end{equation}
\citep{King0816}.
Despite the different physics involved, the two types of winds have surprisingly similar velocities,
i.e. 
\begin{equation}
   v,\, v' \simeq 0.05c - 0.2c 
   \label{vwind}
\end{equation}
for values $10< \dot m < 100$. Observations show the presence of winds with velocities of this order in 
{    most ($\sim 70 \% $)} ULXs (Sect. \ref{sec:HEres}).

An observer not located very close to the central accretion disc axis (i.e. not in the narrow beam defining the ULX)
would observe the photosphere of the quasi-spherical wind. \citet{King0816} show that this has radius
\begin{equation}
    R_{\rm ph} \simeq 10^4\times l'^{-1/2}\dot M_{w10}^{3/2}{\eta_{0.1}}^{-3/2}M_{10}\,{\rm km}
\label{rphot}
\end{equation}
and effective temperature
\begin{equation}
    T_{\rm eff} = 1\times 10^6l'^{1/4}\dot M_{w10}^{-3/4}\eta_{0.1}^{3/4}M_{10}^{-1/4}\,{\rm K}
    \label{Teff}
\end{equation}
where $\dot M_{w10}$ is the wind mass outflow rate in units of $10\dot m$, $\eta_{0.1}$ is the 
accretor efficiency in units of $0.1$, and $M_{10}$ is the accretor mass in units of $10\msun$. 

Objects almost precisely like this  -- blackbody emission at temperature $\sim 100$\, eV from photospheres of order (\ref{rphot})
are observed, in the form of the ultraluminous supersoft sources \citep[ULSS,][]{Kong0603}. Evidently these are ULXs
viewed `from the side'. The well--know extreme Galactic source SS433 is probably a member of this class, as we discuss below.

ULX/ULS winds are quasispherical, and
hypersonic with respect to the interstellar medium (ISM) surrounding the ULX, and so must
be slowed in a strong reverse shock against it. A weaker forward shock propagates into the surrounding ISM
ahead of the contact discontinuity between the ULX wind and the ISM, sweeping this gas outwards in a shell.
As in the similar problem for supermassive black holes \citep[see][for a review]{King0815}
the nature of ULX feedback on the interstellar gas depends on whether the shock slowing the ULX wind is cooled by radiation losses
within a flow timescale or not. In the first case, where the shock is efficiently cooled, most of the energy of the wind is lost to
radiation, and the wind transmits only its momentum to the ISM. If instead shock cooling is inefficient,
the outflow gives all of its energy adiabatically to this gas. These two cases are often called momentum--driven
and energy--driven 
respectively\footnote{
In SMBH accretion, feedback switches from momentum-- to energy--driven once the 
SMBH mass reaches the $M - \sigma$ value $M \propto \sigma^4$. This fixes the limit to SMBH mass growth, since the 
ISM gas ultimately fuelling its growth is expelled at this point, in a galaxy--wide high--speed outflow. See King \& Pounds (2015)}.

For ULXs, we will see that feedback is 
in practice always energy--driven, which in turn means that ULXs generally have a significant effect on their 
surroundings. The temperature of the reverse shock slowing the ULX wind is 
$T_s \sim m_{\rm H}u^2/k \sim 10^{10} - 10^{11}$~K, where $u \sim v, v'$, 
and $k$ is Boltzmann's constant.
Since this is far higher than the $T \lesssim 10^8$~K radiation field of the ULX, inverse Compton scattering
off the ULX radiation field is the main cooling mechanism, as in the SMBH case. (The thermal bremsstrahlung
timescale is far longer.)
King (2003) shows that the ratio of Compton cooling time and flow timescale $R/v$ is
\begin{equation}
    \frac{t_c}{t_{\rm flow}} \simeq \frac{2}{3}\left(\frac{m_e}{m_p}\right)^2\frac{R}{R_g}\frac{c}{v}
\end{equation}
at distance $R$ from the ULX. From (\ref{vwind}) this defines a cooling radius
\begin{equation}
    R_c = 5\times 10^5R_g \sim 5\times 10^{10}\,{\rm cm},
\end{equation}
which is generally smaller than the ULX binary's orbit. Shock cooling is never important for ULX outflows, 
and all ULX feedback is energy--driven.

We can easily see the effect of this feedback. If the interstellar gas has uniform density $\rho$ and the
shock exerts pressure $P$ on it, momentum conservation at the radius $R$ of the interface between the ULX wind and the ISM gives
\begin{equation}
    4\pi R^2P = \frac{\rm d}{{\rm d}t}\left(\frac{4\pi}{3}R^3\rho \dot R\right)
\end{equation}
so that
\begin{equation}
    \frac{P}{\rho} = \frac{1}{3}R\ddot R + \dot R^2.
    \label{P}
\end{equation}
Since energy is transmitted adiabatically from the wind we have
\begin{equation*}
    \frac{{\rm d}}{{\rm d}t}\left[2\pi R^3P\right] = \Ledd - P\frac{{\rm d}}{{\rm d}t}\left(\frac{4\pi R^3}{3}\right)
\end{equation*}
Eliminating $P$ gives the equation of motion for the shock as
\begin{equation}
    \frac{\Ledd}{2\pi\rho} = \frac{1}{3}R^4\dddot R + \frac{8}{3}R^3\dot R\ddot R + 5R^2\dot R^3
\end{equation}
All solutions of this equation tend to the attractor 
\begin{equation}
    R = \left(\frac{125\Ledd}{101\pi\rho}\right)^{1/5}t^{3/5}, 
    \label{nebularradius}
\end{equation}
(cf Eq. \ref{neb1}) or numerically
\begin{equation}
    R \simeq 50\left(\frac{L_{39}}{\rho_{-25}}\right)^{1/5}t_5^{3/5}\,{\rm pc},
\end{equation}
where $L_{39} = \Ledd/10^{39}\ergs$, 
$t_5 = t/10^5\, {\rm yr}$, 
and  
$\rho_{-25} = \rho/10^{-25}\,{\rm g}\,{\rm cm^{-3}}$. The conditions correspond to a ULX with a lifetime
of order the thermal timescale of a massive companion star, driving a wind into an ISM of number density 
$\sim 0.1\,{\rm cm}^{-3}$ H atoms per cm$^3$.
The shell of swept--up ISM gas expands at a speed
\begin{equation}
\dot R =300\left(\frac{L_{39}}{\rho_{-25}}\right)^{1/5}t_5^{-2/5}\,{\rm km\, s^{-1}},
\label{bubble}
\end{equation}
so the forward shock into the ISM is much weaker than the reverse shock slowing the ULX wind (cf Eq. \ref{vwind}).
For longer thermal timescales or more powerful wind mechanical luminosities, $L_{39} \sim 10$, and the
predicted nebulae can have sizes $\sim 500$~pc. Observations (Sec. \ref{sec:envin}) show that many ULXs are associated
with wind--blown nebulae of order 50 - 500~pc in size. These are significantly larger than supernova remnants ($\sim 3$~pc).

\subsubsection{SS433}
\label{sec:ss433}

The well--known extreme Galactic binary system SS433 fits closely into this picture.
\citet{King0002} show that it probably has a mass transfer rate implying $\dot m_{w10} \sim 10^3$, and $t_5 \sim 1$. The
mechanical luminosity is then. $\sim 10^{39}\ergs$.
For a $10\msun$ black hole accretor, Eq. (\ref{bubble}) implies a spherical nebula (excluding the `ears' formed by the precessing jets) comparable with the observed $\sim 30\, {\rm pc}$
size. \citet{King0816} note that this implies a quasispherical wind speed close to the observed velocity 
width $\sim 1500\,{\rm km\,s}^{-1}$ of the so-called stationary H alpha line. 
Despite its very high mass transfer rate, SS433 is a remarkably weak X--ray source -- \citet{Watson0986} estimate $L_X \simeq 10^{36}\ergs$
and a mechanical luminosity $\simeq 10^{39}\ergs$, and it is likely that the softer ($<$ 10 keV) X--rays come only from the prominent jets the system produces (whilst the harder X-ray emission may originate from scattering inside the wind-cone, \citealt{Middleton2021_SS433}). 
In a detailed study, \citet{Begelman0607} show that SS433 has a viewing angle appropriate to a ULS, but is not
observable as one: from (\ref{Teff}) we have
$kT_{\rm eff} \sim 10\approx,{\rm eV}$, which is too soft to be detectable as this system is heavily reddened ($A_V \simeq 7$). 
SS433 is distinctive because of the huge periodic redshift and blueshifts of the H--alpha lines emitted by its precessing jets.
These evidently result from its extreme Eddington factor, and precess (and are baryon--loaded and slowed to $0.26c$) probably because they are 
deflected by a thick precessing central disc. As a mark of this, the precession rate of the jets does show power at the binary orbital period \citep{Begelman0607}.

\subsection{Be--star ULXs}
\label{sec:bemodel}

The luminosity of ULXs is generally variable, but two classes of ultraluminous X-ray sources are transient. One of these consists of systems appearing in the transient X-ray sky only as ultraluminous sources. We discuss these in the next section. 

The other class is a small set of systems which normally
appear as high--mass X--ray binaries, but occasionally become ULXs. This class has only recently been recognised, but actually includes
A0538-66, the first system observed to have an apparently super--Eddington X--ray luminosity \citep{White0478}, and also the 
Milky Way's only known ULX, Swift J0243.6+6124, first detected on 3 October 2017 during a giant X-ray outburst \citep[][see Table \ref{tb:combined}]{Kennea2017}.
In all these systems, a neutron star, often with a noticeable magnetic field, accretes from a much more massive Be (or Oe) star in a wide eccentric orbit.

Be stars are unusual in having a large disc of material surrounding their rotational equator. The origin of the disc is unclear, but is evidently
related to the star's rotation, and possibly mediated by small--scale surface magnetic fields \citep[][for a recent discussion]{Nixon1220}.
Be X--ray binaries generally produce most of their X--ray emission when the neutron star is near pericentre and interacts with the circumstellar
disc. The X--rays therefore generally indicate the orbital period.

But it is now known that, on a much longer timescale, this interaction is much stronger, and the system becomes a ULX. The origin of this behaviour
is probably von Zeipel--Lidov--Kozai (ZLK) oscillations of the Be star disc \citep{Martin0914}.
Highly misaligned test particle orbits around one component of a binary cyclically exchange inclination for eccentricity
\citep{VZ1910,Kozai1162,Lidov1062}. This process conserves the component of the angular momentum orthogonal to the binary orbital plane, i.e.
\begin{equation}
 (1 - e_p^2)^{1/2}\cos i_p = {\rm constant}.   
\end{equation}
Here $i_p$ is the inclination of the particle orbit to the binary orbit and $e_p$ is the test--particle eccentricity. A test particle initially on a circular strongly misaligned orbit 
oscillates to closer alignment. Then $|\cos i_p|$ increases, requiring larger $e_p$. The oscillations occur only if the initial inclination $i_{p0}$ inclination 
satisfies
\begin{equation}
   \cos^2 i_{p0} < \frac{3}{5}, 
\end{equation}
so $39^{\rm o}< i_{p0} < 141^{\rm o}$, and the maximum eccentricity that an initially circular test orbit can reach is
\begin{equation}
    e_{\rm max} = \left(1 - \frac{5}{3}\cos^2 i_{p0}\right)^{1/2}.
\end{equation}

\citet{Martin0914} show that this test--particle behaviour extends to full hydrodynamical discs, which undergo global KL cycles of the disc inclination and eccentricity.
The spins of Be stars are known in general to be strongly misaligned  from the orbits of their neutron star companions in Be -- X--ray binaries, probably as a result of the asymmetric 
supernova explosion producing the neutron star. At maximum disc eccentricity, it is likely that the neutron star accretes much more gas from the Be--star disc, and this may turn it into a ULX. Global KL cycles typically have periods an order of magnitude longer than the binary orbit, so the transition to the ULX state is relatively infrequent.

\subsection{Modelling transient ULXs}
\label{sec:trans}

Accretion discs in low-mass X-ray binaries are subject to a thermal--viscous instability when the mass transfer rate from the companion falls below a critical value depending mainly on the size of the disc \citep[orbital period;][]{Coriat0911}. The instability is similar to that in cataclysmic
variables leading to dwarf nova outbursts \citep{Lasota0106,Hameury0920}, but X-ray self--irradiation of the disc plays a crucial role both in the instability criterion \citep{vanParadijs9606} and the outburst development 
\citep{King9801,Dubus9902,Dubus0107}. 

The thermal--viscous disc instability is naturally associated with low mass transfer rates, but since the critical accretion rate increases strongly with the disc size,
for large enough binary systems, even super--Eddington mass transfer can correspond to an unstable disc \citep{Lasota1503,Hameury1120}. So for sufficiently large orbital periods, accreting systems  have large unstable discs and can produce outbursts bright enough to explain some ULXs \citep{King0002}.
Although as discussed in Section \ref{sec:opt} it is very difficult to determine the binary parameters of these distant systems, it is clear that many of them have giant or supergiant companion stars (Section \ref{sec:ir}) and have orbital periods longer than 2 days. Several ULXs are classified as transient (see e.g. \citealt{Brightman0620}). Usually their light curves are only sparsely covered, but the recent well sampled observations of a transient ULX source in the galaxy M 51 by \citet{Brightman0620}  provides a light curve that can be compared with model predictions. Some long--period, sub-Eddington X-ray transients show long-lasting outbursts (tens of years), so some apparently steady ULXs could actually be transients observed during long outbursts.

The critical rate below which a disc is unstable is well approximated by:
\begin{equation}
    \dot{M}_{\rm crit}^+ \approx 2.4 \times 10^{19} M_1^{-0.4}f_{\rm irr}^{-0.5} \left( \frac{R_{\rm out}}{10^{12} \; \rm cm} \right)^{2.1} \; \rm g \, s^{-1},
\label{eq:mdotcrit}
\end{equation}
where $R_{\rm out}$ is the outer disc radius and $f_{\rm irr} = {\eta_{\rm t} \mathcal{C}}/{(5 \times 10^{-4})}$ and $\mathcal{C}$ is defined through 
\begin{equation}
\sigma T_{\rm irr}^4 = \mathcal{C} \frac{\eta_{\rm t} \dot{M} c^2}{4 \pi r^2}.
\label{eq:tirr}
\end{equation}
$\mathcal{C}$ contains all the physics of the irradiation process, and $\eta_{\rm t}$. \citet{Dubus0107} found that the light curves of low-mass X-ray transients are reasonably well reproduced, and  \citet{Coriat1208} found that the corresponding stability criterion provides the observed division of sources into steady and outbursting, if one uses a constant $\eta_{\rm t} \mathcal{C}$ of order $10^{-3}$. The radiative accretion efficiency $\eta_t$ is defined as the ratio of the true luminosity (without the beaming factor) to $c^2$, and is therefore:
\begin{align}
                \eta_{\rm t}  & =  0.1  \;  \ \ \ \ \  \hskip 1.8cm  {\rm if} \; \dot m < 1,       \nonumber \\
                    \,        & = 0.1 (1+\ln \dot m) / \dot m  \ \ \ \ \ \;  {\rm if} \; \dot m \geq 1.
\label{eq:eta}
\end{align}
From Eq. (\ref{eq:mdotcrit})  it follows that neutron-star ULXs may be unstable for super--Eddington mass transfer rates at orbital periods $\gtrsim 10$ days (slightly shorter for black-hole ULXs). But 
if the front fails to reach the outer disc edge, the peak outburst luminosity is given by
Eq. (\ref{eq:mdotcrit}) with $R_{\rm out}$ replaced by $R_{\rm tr,max}\leq R_{\rm out}$, where
$R_{\rm tr,max}$ is the maximum radius reached by the instability--triggering heating front. In this case, the peak luminosity can be up to 50 times larger \citep{Hameury1120}. This means that wide binaries in which the mass transfer is sub-Eddington might become transient ULXs at, and close to, the maximum outburst luminosity
if their discs are unstable.

\citet{Hameury1120} derived analytical formulae describing typical X-ray transient light curves with great accuracy. They apply also to transient ULXs. 
When the heating front brings the whole disc to a hot state ($R_{\rm tr,max}\approx R_{\rm out}$), the $\dot M(t)$ relation during the decay from outburst maximum is very well described by \citep{Ritter0101}\footnote{The corresponding formula in the original publication contains a misprint.} 
\begin{equation}
\dot{M}=\dot{M}_{\rm max} \left[ 1+ \frac{t}{t_0} \right]^{-10/3},  
\label{eq:varmdot}
\end{equation}
where the integration constant $t_0$ is given by: 
\begin{equation}
    t_0 = 3.19 \alpha_{\rm 0.2}^{-4/5} M_1^{1/4} R_{12}^{5/4} \dot{M}_{\rm max,19}^{-3/10} \; \rm yr,
\end{equation}
with $\dot{M}_{\rm max,19} = \dot{M}_{\rm max}/10^{19}$~g\,s$^{-1}$. Here $t_0$ is {\sl not} the characteristic timescale of  disc evolution, which is instead
\begin{equation}
    \tau = \frac{M_d}{\dot{M}} = 0.81 \psi f^{-0.3} M_1^{0.37} f_{\rm irr}^{0.15} R_{12}^{0.62} \alpha_{0.2}^{-0.8} \; \rm yr, 
\label{eq:tau}
\end{equation}
where $M_d$ is the disc mass and $f=\dot{M}_{\rm max}/\dot{M}_{\rm crit}^+(r_{\rm out}) > 1$ \citep[for the derivation see][]{Hameury1120}.
One can then relate the maximum outburst accretion rate to the characteristic decay time.
\begin{equation}
    \dot{M}_{\rm max} = 2.0 \times 10^{19} \alpha_{0.2}^{2.71} f^{2.02} M_1^{-1.65} \left( \frac{\tau}{1 \; \rm yr}\right)^{3.39} f_{\rm irr}^{-1.01}  \; \rm g \, s^{-1},
\label{eq:mdot-tau}
\end{equation}
Then when  both $\dot{M}_{\rm max}$ and $\tau$ are known from observations, this relation determines $f$, and hence $\dot{M}_{\rm crit}^+$ and the size of the accretion disc. 

This phase of the outburst corresponds to a viscous decay, during which X--ray irradiation prevents the formation and propagation of a cooling front. This ends when the decreasing accretion rate falls  to the critical value $\dot{M}_{\rm crit}^+$ and the front starts propagating into the disc. 

If the heating front is unable to reach the outer regions -- which happens quite often in huge discs -- the viscous decay phase is missing and, from the start, the decay-from-maximum $\dot M(t)$ relation corresponds to the cooling--front propagation, and has the form
\begin{equation}
    \dot{M} = 1.31 \times 10^{17} \alpha_{0.2}^{2.71} M_1^{-1.65} f_{\rm irr}^{-1} \left[{\xi}\frac{(t_{0}^{\prime}-t)}{1 \; \rm yr} \right]^{3.39} \; \rm g \, s^{-1},
    \label{eq:decay}
\end{equation}
where $t_0^{\prime}$ is a constant that is determined by the condition that, when the cooling front starts, $\dot{M}$ is equal to $\dot{M}_{\rm crit}^+$ at the maximum transition radius. Then $t_0^\prime$ can  be written as:
\begin{equation}
    t_0^{\prime}=4.64 \; \xi^{-1} M_1^{0.37} f_{\rm irr}^{0.15} \alpha_{0.2}^{-0.8} r_{12}^{0.62} \; \rm yr,
    \label{eq:t0}
\end{equation}
and $\xi$ is a constant to be calibrated by numerical simulations; it is found to be in the range $2.47 < \xi \lesssim 8.0$. Here $t_0^{\prime}$ is equal to the {\sl duration} of this phase of the outburst, and can be directly compared to observations. As before, there exists a relation between $\dot{M}_{\rm max}$ and ${t_0^\prime}$:
\begin{equation}
    \dot{M}_{\rm max} = 3.1 \times 10^{19} \alpha_{0.2}^{2.71} M_1^{-1.65} \left(\frac{\xi}{5}\frac{t_0^\prime}{1 \; \rm yr}\right)^{3.39} f_{\rm irr}^{-1.01}  \; \rm g \, s^{-1},
\label{eq:mdot-tp0}
\end{equation}
which is essentially the same as in Eq. (\ref{eq:mdot-tau}), with $f=1$.

It is important to realise that, for super-Eddington accretion rates, $\dot M(t)$ does not directly represent the observed $L_X(t)$ light curve, as the relation between the luminosity and accretion rate is no longer linear. Then when considering an outburst whose apparent luminosity is larger than $\Ledd$ \citet{Hameury1120} use the following relations
\begin{align}
    L_{\rm x} & =  (1+\ln \dot m) \left[1+\frac{\dot m^2}{\tilde b} \right] L_{\rm Edd} \; \ \ \ \ \ \  {\rm if} \; \dot m \geq 1  \nonumber \\
              & =  \dot m \, L_{\rm Edd}  \; \hskip 3cm {\rm if} \; \dot m < 1,
\label{eq:lx}
\end{align}
where $b=(1+\dot m^2/\tilde b)^{-1}$ is a beaming term; the larger the beaming parameter $\tilde b$, the larger $b$ and hence the smaller the beaming effect. Eq. (\ref{eq:beam}) corresponds to $b=\tilde b/\dot m^2$ with $\tilde b=73$. Since the beaming factor $\tilde b /\dot m^2$ applies only for $\dot m > \sqrt{\tilde b}$, in time-dependent calculations, one replaces it by $(1+\dot m^2/\tilde b)^{-1}$ in order to get a smooth transition with the case where beaming is negligible. 

Further, when $\dot m \gg 1$ so that the apparent luminosity scales approximately as $\dot m^2$, the decay time of the luminosity $\tau_{\rm L}$ is one half of the decay time of the accretion rate, and Eq. (\ref{eq:mdot-tau}) has to be written as:
\begin{equation}
    \dot{m}_{\rm max} = 132\, \alpha_{0.2}^{2.71} f^{2.02} M_1^{-2.65}
      \left( \frac{ \tau_{\rm L}}{1 \; \rm yr}\right)^{3.39} f_{\rm irr}^{-1.01}.
\label{eq:mdot-tau_edd}
\end{equation}

If not all of the disc reaches the  hot state, the total outburst duration is related to the maximum mass accretion rate via Eq. (\ref{eq:mdot-tp0}), and we must have:
\begin{equation}
    \dot{m}_{\rm max} = 24.0\, \alpha_{0.2}^{2.71} M_1^{-2.65}
      \left(\frac{\xi}{5}\frac{t_0^\prime}{1 \; \rm yr}\right)^{3.39}
      f_{\rm irr}^{-1.01},
\label{eq:mdot-tp0_edd}
\end{equation}
where $t_0^\prime$ now refers to the outburst duration.
\begin{figure}[h!]
\includegraphics[width=0.8\columnwidth]{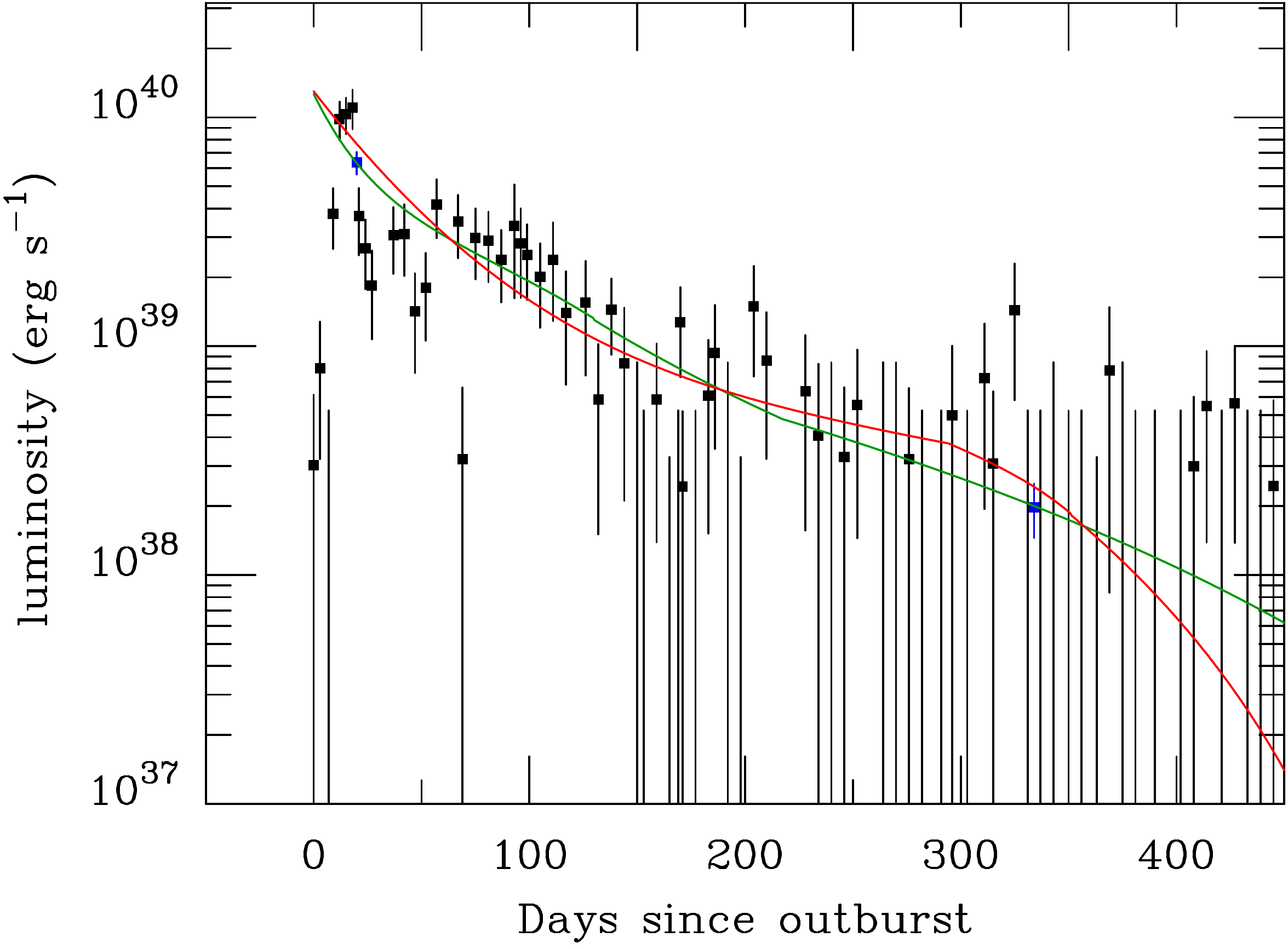}
\caption{Observed flux from M51 XT-1 as compared with model predictions for a 1.4~M$_\odot$ neutron star (red curve) and a 10~M$_\odot$ black hole (green curve). Black points correspond to {\sl SWIFT} data and blue points to {\sl Chandra} or {\sl XMM-Newton} observations \citep[courtesy M. Brightman, see also][]{Brightman0620}.}
\label{fig:m51_xt}
\end{figure}

\subsubsection{M51 XT-1}

M51 XT-1 is currently the only ULX transient source whose light curve is suitable for comparison with the prediction of the disc--instability model. 
Figure \ref{fig:m51_xt} shows how observational data of M51 XT-1 \citep{Brightman0620} compare with the predictions of the disc--instability model when the light curve comes from Eqs. \ref{eq:varmdot} and \ref{eq:decay}, combined with Eq. (\ref{eq:lx}) with the beaming parameter $\tilde b = 73$. Two cases are shown. A 1.4 M$_\odot$ accreting neutron star with a disc size $4.8 \times 10^{11}$~cm; and a 10 M$_\odot$ accreting black hole, with a disc extending to $4.9 \times 10^{11}$~cm. As can be seen, the agreement is reasonably good, and as acceptable as the original fit with a power law with index $-5/3$ \citep{Brightman0620}. These data can also be fitted with a 10~M$_\odot$ accreting black hole, with a disc extending to $4.9 \times 10^{11}$ cm.

In the neutron star case, the maximum accretion rate is $6 \times 10^{19}$~g\,s$^{-1}$, while in the black hole case it is $1.5 \times 10^{20}$~g\,s$^{-1}$. Hence the emission from a black-hole system would not be beamed, but for a neutron-star accretor the beaming factor would be $b=0.06$, as expected for $\dot m =33$. The mass transfer rate can be very roughly estimated to be of  order  1 -- 2\% of the maximum accretion rate, i.e. $1 - 3 \times 10^{18}$~g\,s$^{-1}$.

\citet{Hameury1120} showed that taking $\tilde b=200$ instead of $\tilde b=73$ in Eq. (\ref{eq:lx})
also provides an acceptable fit, but $\tilde b=20$ is excluded by the {\sl Chandra} and {\sl XMM-Newton} points, thus suggesting only moderate beaming.

\subsubsection{HLX-1 in ESO 243-49}
\label{sec:hlx2}

HLX-1 in the galaxy ESO 243-49 \citep{Farrell0907} attracted a lot of attention as the best IMBH candidate among ULXs. It is the brightest ULX candidate known; its luminosity is variable and can exceed $10^{42}$~erg~s$^{-1}$ at maximum. Its association with the galaxy ESO 243-49 at a distance of 95 Mpc is fairly well established \citep{Wiersema1010,Soria1301} so there is little doubt about its luminosity \citep[see, however,][]{Lasota1503}. \citet{Greene0820} conclude that
a self-consistent scenario is that HLX-1 is a central massive black hole on the low--mass end of the mass distribution of nucleated $10^9 -10^{10}\msun$ galaxies. Its small host galaxy was accreted and tidally stripped by ESO 243-49, leaving only the bare nuclear star cluster near the black hole (e.g., \citealt{Mapelli1311}). Star formation, perhaps encouraged by the merger, could have increased the capture rate for  the hole to acquire a companion star on which it is currently feeding. Then by the definition adopted in this review HLX-1 is not an IMBH candidate, but was the central MBH of a small galaxy, which (unusually) has been accreted and disrupted by a larger one.

In line with this picture, HLX-1's historical light curve (see e.g. \citealt{Godet1410}) initially showed a promising series of almost annual outbursts, only to lose this quasi--periodicity later on. Subsequent to its discovery, a number of so--called QPE (`quasiperiodic eruption') sources have recently been found from the centres of small galaxies (e.g. \citealt{Miniutti1909,Miniutti2001,Arcodia2104}). These can be modelled as tidal disruption near--misses (\citealt{King_GSN}): instead of being fully disrupted by filling its tidal radius near the MBH (a tidal disruption event or TDE) an infalling star fills its tidal lobe only at the pericentre of a very eccentric orbit about the MBH. In the small galaxy GSN 069 the star is probably a low--mass white dwarf (\citealt{King_GSN}),
and the predicted CNO anomaly was later found (\citealt{Sheng2110}). It appears that this model works well for all the currently recognised QPE systems \citep{King0622}.
These systems all have very short lifetimes as the orbits shrink under gravitational radiation, which is what drives the mass transfer.

Modelling HLX-1 instead as a transient IMBH source undergoing disc outbursts shows that the outburst timescales are incompatible with a central mass of $\gtrsim 10^4\, \msun$, as would be required to explain the peak luminosity of $10^{41}\,{\rm erg\, s^{-1}}$ (\citealt{Hameury1120} and references therein).

\subsection{Beaming or not?}

The arguments in favour of luminosity beaming in PULXs presented above are simple and fundamental. The main one is that the observed X-ray luminosities of magnetic ULXs  lie between 6 and 500 $\Ledd$, with the classic (non BeX) systems having $L_X > 20 L_{\rm Edd}$. Even if mass is supplied at a super--Eddington rate, the luminosity is only $L\approx \Ledd (1 + \ln \dot m)$, so even $\dot m \approx 1000\, (> 10^{-5} \Msun{\rm yr}^{-1})$ produces only $\sim 8\Ledd$. 

Explaining these huge excess luminosities by appealing to super--strong magnetic fields appears unpromising.
We have argued above that the presence of magnetars in PULXs is extremely unlikely. Among other problems, it requires a cosmic conspiracy confining magnetars to just these binary systems. Be--PULXs spend most of the time as standard pulsed X--ray sources, and clearly do not have magnetar--strength fields. Assertions that the ``apparent luminosity [of PULXs] is close to the actual one'' \citep{Mushtukov2011} are rather surprising, especially when it is not explained how this is supposed to be achieved.

\citet{Mushtukov2011} refer to a table in \citet{King2003} to claim that these authors require $a=1/b\sim 20$ for M82 ULX-2, NGC 7793 P13 and NGC 300 ULX1 and $\sim100$ for NGC 5. But Table \ref{tab:ulx3b} shows that the values of $1/b$ for the first three sources are respectively 17, 6, 6, to which one can add 10 for ULX-7 in M51. 

\subsection{Measuring mass transfer via the orbital period derivative?}

The most important parameter in understanding ULXs, as in any accreting system, is the instantaneous mass transfer rate $-\dot M_2$ in the host binary system. Unfortunately, precisely because we do not yet have a clear picture of how exactly 
the mass transfer rate specifies the luminosity of ULXs, it is difficult to measure $-\dot M_2$ directly from observation. 
But if we assume conservative mass transfer (i.e. conserving both total mass and angular momentum) we see from Section (\ref{sec:evolulx}) that all quantities such as the companion mass $M_2$, the binary separation $a$,
and by Kepler's law, the binary period $P \propto a^{3/2}$,
evolve on the mass transfer timescale $M_2/(-\dot M_2)$ -- see Eq. (\ref{mtp}). This makes it tempting
to use the much more easily measured orbital period derivative $\dot P$ to estimate $-\dot M_2$. The recent papers by \citet{Bachetti2112,Bachetti2208} try to use  this method for the ULX M82 X-2. 

As we noted above, the problem of measuring $-\dot M_2$ is common to all interacting binary systems, not simply ULXs. There is a long history of efforts to find $-\dot M_2$ by using $\dot P$ and assuming conservative evolution,
going back at least to early studies of cataclysmic variables (where a low--mass main--sequence star transfers mass to a white dwarf in a very circular orbit). But all these attempts give uniformly disappointing results -- not least because even the {\it sign} of $\dot P$ often flatly contradicts that expected from conservative evolution. The underlying reason for these difficulties is that relations like (\ref{mtp}) hold {\it only} if the quantities $\dot P, -\dot M_2$ are
{\it averages over a timescale long compared with the time $t_H$ for the Roche lobe to move one scaleheight within the donor star}. 
This requirement arises because stars have smooth density distributions rather than sharp edges. On timescales shorter than $t_H$ the instantaneous value of $\dot P$ is always in practice dominated by large but short--term effects, with varying signs often in conflict with the long--term rate. This point has been made repeatedly in the literature, going back at least to \citet{Pringle7503}; see also e.g. \citet{Ritter8808} and \citealt{Frank0201} (Section 4.4). Pringle (1975) concluded that ``... an accurate estimate of the rate of mass transfer cannot be deduced from a change of binary period".

In response to \citet{Bachetti2112}, \citet{King2112} estimated $t_H \sim 1000\, {\rm yr}$ for M82 X-2. Then observations of $\dot P$ over humanly accessible timescales cannot give information about $-\dot M_2$. 
\citet{King2112} further pointed out that direct application of the method of \citet{Bachetti2112} to three well–studied X–ray transients with known values of $\dot P$ (Nova Muscae 1991, XTE J1118+480, A0620–00, see \citealt{Gonzales1702} and references therein) was possible, but not encouraging. These Roche--lobe filling systems are all currently in the faint quiescent state. Under the assumption of conservative evolution,  their values of $\dot P$ would instead predict that all three of them should currently have high mass transfer rates. Their accretion discs should therefore be steady rather than transient, putting them stably in bright states -- indeed Nova Muscae 1991 would actually be a ULX. 
The subsequent publication \citep{Bachetti2208} does not answer these points.

\section{Conclusions}
\label{sec:conclusions}

Table \ref{tab:modul} summarises how various models of ULX behaviour compare with observations. 
The case for disc--wind beaming appears fairly strong, and
is so far compatible with a large body of data. Given that the photon--bubble model does not appear promising, this is the only model which can straightforwardly apply to both black--hole and neutron--star accretors.

Other models come into consideration if we allow for the idea that ULXs might not be a single homogenous population, but
instead consist of various different types of systems (as in the example of long and short gamma--ray bursts). IMBH clearly cannot
account for PULXs or Be--star ULXs, and have not been widely tested in other aspects. Unless their masses are $> 3\times 10^4\msun$ they cannot account for the striking universal 
$L_{\rm soft} \propto T^{-4}$ behaviour of ULX soft X--ray components with
luminosities $> 3 \times 10^{39}\,{\rm erg\, s^{-1}}$. 

\begin{table*}[h!]
\caption{Models of ULXs}
\vskip 5.0pt
{
\setlength{\tabcolsep}{0.8pt}
\label{tab:modul}
{\small
\hfill{}
\begin{tabular}{ |l|c|c|c|c|c|c|c|c|} 
 \hline\hline
 model 
 & blackbody $L\propto T^{-4}$ & winds \& nebulae & evolution & luminosity fn  
 & $L >3 \times 10^{40}\, \ergs$  &  pulsing                  & BeX 
 & BH \& NS                         \\
 \hline\hline
 disc-wind beaming & yes & yes &  yes & yes & yes&  yes & yes & yes\\
 \hline
 magnetic beaming & no & ? &  yes & ? & yes&  yes & no & no \\
 \hline
 IMBH 
 & needs $M > 3\times 10^4\msun$ & ? &  ? & ? 
 &   yes & no     & no    & no     \\
 \hline
 magnetars 
 & no & no & fieldstrength? & no 
  & no & yes & no & no \\
 \hline
 photon bubble 
 & no & no & ? & ? 
  & no & no    & no   & yes       \\
 \hline 
 oriented jets 
 & ? & ? & ? & ? 
 & no & ?    & no       & ?    \\ 
\hline\hline
\end{tabular}}
}
\hfill{}
\vskip 0.2truecm 
\end{table*}

Observations of Be -- X--ray binaries pose a particular problem for models invoking magnetic beaming, as they make regular
transitions between {    normal Be X--ray binary states to PULXs, and back again, when their mass supply rates are increased or decreased by Kozai--Lidov cycles. 
None of the Be--star ULXs is observationally distinct from other Be X--ray systems when it is in its usual Be X--ray binary state, and in the ULX state these 
systems have all the same properties as other PULXs, including a markedly more rapid spin up rate (cf Fig {\ref{fig:dotnul}}. None of the Be X--ray binaries appears to have a very strong magnetic field.
 This makes it inescapable that very strong dipole magnetic fields are not {\it required} for making a PULX.}
{Highly--magnetic 
PULXs might then at most 
constitute some fraction of the PULX population, but there is so far no clinching demonstration that {\it any} PULX contains a neutron star with this kind of field. Even then there remains the difficulty that this 
model requires us to believe that strong--field neutron stars in binaries are unobservable unless the companion star supplies them with mass at a  super--Eddington rate.}
\\

\bigskip
\noindent
{\bf Acknowledgments}

\bigskip
JPL was supported by a grant from the French Space Agency CNES. MM thanks G. Israel, S. Carpano, M. Bachetti, F. Fuerst, S. Tsygankov, R. Sathyaprakash and G. Rodriguez for providing the data for Figure 7.

\appendix

\section{``Standard'' X-ray pulsars}
\label{sec:appdx}

To illustrate the fact that PULXs differ from standard X-ray pulsars not only in luminosity $> 10^{39}\,{\rm erg\, s}^{-1}$,  but also in their very large spin-up rates 
$\dot \nu > 10^{-10}\,{\rm s}^{-2}$,  we used the XRP sample with known luminosities and spin-up rates (Table \ref{tab:XRP}) as compiled by \citet[][]{Ziolkowski0185}.

\begin{table*}[ht]
{
\setlength{\tabcolsep}{1.2pt}
\caption{Observed properties of `classical' X-ray pulsars}
\vskip 5pt
{\small
\hfill{}
\begin{tabular}{ ||l|c|l|c|} 
 \hline\hline
Name & $L_X (\rm max)$  [erg\,s$^{-1}$] & $P_s $ [s] &  $\dot \nu$ [s$^{-2}$]   \\
\hline\hline
SMC X-1 & $5.0 \times 10^{38}$  & 0.71 & $2.6 \times 10^{-11}$ \\
\hline
Cen X-3 & $5.0 \times 10^{37}$& 4.84 & $1.9 \times 10^{-12}$ \\
\hline
GX 1+4 & $ 4.0 \times 10^{37}$ & 122 & $5.2 \times 10^{-12}$\\
\hline 
4U 0115+63 &$ 3.4 \times 10^{37}$ & 3.61 & $2.2 \times 10^{-12}$ \\
\hline
A 0535+26 & $ 1.0 \times 10^{37}$  & 104 & $3.0 \times 10^{-12}$ \\
\hline
Her X-1 & $ 1.0 \times 10^{37}$ & 1.24 & $8.5 \times 10^{-14}$ \\
\hline
Vela X-1 & $ 1.5 \times 10^{36}$ & 283 & $5.6 \times 10^{-13}$ \\
\hline
GX 301-2 & $ 1.0 \times 10^{36}$ & 696 & $3.8 \times 10^{-13}$ \\
\hline
X-Per &  $ 3.9 \times 10^{33}$ & 835 & $6.4 \times 10^{-15}$ \\
\hline\hline
\end{tabular}}
}
\hfill{}
\label{tab:XRP}\\
\vskip 0.2truecm 
\end{table*}

\newpage

\bibliographystyle{elsarticle-harv}
\biboptions{authoryear}
\bibliography{bib_merged}


%
%
%
\end{document}